\documentclass[prb,aps,twocolumn,eqsecnum,showpacs]{revtex4}
\usepackage{graphicx}

\usepackage{epsfig}
\usepackage{graphicx}
\usepackage{dcolumn}
\usepackage{bm}

\begin{document}

\title{       Frustration and entanglement in the $t_{2g}$
              spin--orbital model on a triangular lattice: \\
              valence--bond and generalized liquid states }

\author{      Bruce Normand }
\affiliation{ D\'epartement de Physique, Universit\'e de Fribourg,
              CH--1700 Fribourg, Switzerland \\
              Theoretische Physik, ETH--H\"onggerberg,
              CH--8093 Z\"urich, Switzerland }

\author{      Andrzej M. Ole\'s }
\affiliation{ Marian Smoluchowski Institute of Physics,
              Jagellonian University, Reymonta 4, PL--30059 Krak\'ow,
              Poland \\
              Max-Planck-Institut f\"ur Festk\"orperforschung,
              Heisenbergstrasse 1, D-70569 Stuttgart, Germany }

\date{\today}

\begin{abstract}

We consider the spin--orbital model for a magnetic system with singly
occupied but triply degenerate $t_{2g}$ orbitals coupled into a planar,
triangular lattice, as would be exemplified by NaTiO$_2$. We investigate
the ground states of the model for interactions which interpolate between
the limits of pure superexchange and purely direct exchange interactions.
By considering ordered and dimerized states at the mean--field level, and
by interpreting the results from exact diagonalization calculations on
selected finite systems, we demonstrate that orbital interactions are
always frustrated, and that orbital correlations are dictated by the
spin state, manifesting an intrinsic entanglement of these degrees of
freedom. In the absence of Hund coupling, the ground state changes from
a highly resonating, dimer--based, symmetry--restored spin and orbital
liquid phase, to one based on completely static, spin--singlet valence
bonds. The generic properties of frustration and entanglement survive
even when spins and orbitals are nominally decoupled in the ferromagnetic
phases stabilized by a strong Hund coupling. By considering the same model
on other lattices, we discuss the extent to which frustration is attributable
separately to geometry and to interaction effects.

\end{abstract}


\maketitle

\section{Introduction}

Frustration in magnetic systems may be of geometrical origin, or may
arise due to competing exchange interactions, or indeed both.\cite{Diep}
For quantum spins, frustration acts to enhance the effects of quantum
fluctuations, leading to a number of different types of magnetically
disordered state, among which some of the more familiar are static and
resonating valence--bond (VB) phases. A further form of solution in
systems with frustrated spin interactions is the emergence of novel
ordered states from a highly degenerate manifold of disordered states,
and the mechanism for their stabilization has become known simply as
``order--by--disorder''.\cite{Diep,Faz99} Many materials are now known
whose physical properties could be understood only by employing
microscopic models with frustrated spin interactions in which some
of these theoretical concepts operate.

A different and still richer situation occurs in the class of
transition--metal oxides or fluorides with partly filled $3d$ orbitals
and near--degeneracy of active orbital degrees of freedom. In undoped
systems, large Coulomb interactions on the transition--metal ions
localize the electrons, and the low--energy physics is that of a Mott
(or charge--transfer\cite{Zsa85}) insulator. Their magnetic properties
are described by superexchange spin--orbital models, derived directly
from the real electronic structure and containing linearly independent
but strongly coupled spin and orbital operators.\cite{Ole05} Such models
emerge from the charge excitations which involve various multiplet
states,\cite{Kug73,Cas78} in which ferromagnetic (FM) and
antiferromagnetic (AF) interactions, as well the tendencies towards
ferro--orbital (FO) and alternating orbital (AO) order, compete with
each other. This leads to a profound, intrinsic frustration of
spin--orbital exchange interactions, which occurs even in case of only
nearest--neighbor interactions for lattices with unfrustrated geometry,
such as the square and cubic lattices.\cite{Fei97} The underlying
physics is formulated in the Goodenough--Kanamori rules,\cite{Goode}
which imply that the two types of order are complementary in typical
situations: AO order favors a FM state while FO order coexists with AF
spin order. Only recently have exceptions to these rules been noticed,
\cite{Ole06} and the search for such exceptions, and thus for more
complex types of spin--orbital order or disorder, have become the
topic of much active research.

A case study for frustration in coupled spin--orbital systems is
provided by the one--dimensional (1D) SU(4) model.\cite{Li98} One
expects {\it a priori\/} no frustration in one dimension and with only
nearest--neighbor interactions. However, spin and orbital interactions,
the latter formulated in terms of pseudospin operators, appear on a
completely symmetrical footing for every bond, and favor respectively
AF and AO ordering tendencies, which compete with each other. In fact
a low--energy but magnetically disordered spin state also frustrates
the analogous pseudospin--disordered state, and conversely. This
competition results in strong, combined spin--orbital quantum
fluctuations which make it impossible to separate the two subsystems,
and it is necessary to treat explicitly entangled spin--pseudospin
states.\cite{Ole06,Pss07} While in one sense this may be considered
as a textbook example of frustration and entanglement, the symmetry
of the entangled sectors is so high that joint spin--pseudospin
operators are as fundamental as the separate spin and pseudospin
operators, forming parts of a larger group of elementary (and
disentangled) generators. The fact that the 1D SU(4) model is exactly
solvable also results in fundamental symmetries between the intersite
correlation functions for the spin and orbital (and spin--orbital)
sectors.\cite{Fri99} We return below to a more detailed discussion
of entanglement and its consequences. Although indicative of the rich
underlying physics (indeed, unconventional behavior has been identified
for the SU(4) Hamiltonian on the triangular lattice,\cite{Li98,rpmfm})
the implications of this model are rather limited because it does not
correspond to the structure of superexchange interactions in real
correlated materials.

Realistic superexchange models for perovskite transition--metal
oxides with orbital degrees of freedom have been known for more than
three decades,\cite{Kug73,Cas78} but the intrinsic frustrating effects
of spin--orbital interactions have been investigated only in recent
years.\cite{Fei97,Kha00} A primary reason for this delay was the
complexity of the models and the related quantum phenomena, which
require advanced theoretical methods beyond a straightforward
mean--field theory. The structure of spin--orbital superexchange
involves interactions between SU(2)--symmetric spins $\{{\vec S}_i,
{\vec S}_j\}$ on two nearest--neighbor transition--metal ions $\{i,j\}$,
each coupled to orbital operators $\{{\vec T}_i,{\vec T}_j\}$ which
obey only much lower symmetry (at most cubic for a cubic lattice),
and its general form is\cite{Ole05}
\begin{equation}
\label{som1}
{\cal H}_J = J \sum_{\langle ij \rangle \parallel \gamma} \left\{
{\hat J}_{ij}^{(\gamma)} \left( {\vec S}_i \cdot {\vec S}_j \right) +
{\hat K}_{ij}^{(\gamma)} \right\}.
\end{equation}
The energy scale $J$ is determined (Sec.~II) by the interaction terms
and effective hopping matrix elements between pairs of directional $e_g$
orbitals [$(dd\sigma)$ element] or $t_{2g}$ orbitals [$(dd\pi)$ element]
The orbital operators ${\hat J}_{ij}^{(\gamma)}$ and ${\hat K}_{ij}
^{(\gamma)}$ specify the orbitals on each bond $\langle ij \rangle
\parallel \gamma$, which participate in $d^n_i d^n_j \rightleftharpoons
d^{n+1}_i d^{n-1}_j$ virtual excitations, and thus have the symmetry of
the lattice. The form of the orbital operators depends on the valence
$n$, on the type ($e_{g}$ or $t_{2g}$) of the orbitals and, crucially,
on the bond direction in real space.\cite{Bri04} It is clear from
Eq.~(\ref{som1}) that individual terms in the Hamiltonian ${\cal
H}_J$ can be minimized for particularly chosen spin and orbital
configurations,\cite{Ole05} but in general the structure of the
orbital operators ensures a competition between the different bonds.

This directional nature is the microscopic origin of the intrinsic
frustration mentioned above, which is present even in the absence
of geometrical frustration. Both spin and orbital interactions are
frustrated, making long--range order more difficult to realize
in either sector, and enhancing the effects of quantum fluctuations.
Quite generally, because insufficient (potential) energy is available
from spin or orbital order, instead the system is driven to gain
(kinetic) energy from resonance processes, promoting phases with
short--range dynamical correlations and leading naturally to spin
and/or orbital disorder. Disordering tendencies are particularly
strong in highly symmetric systems, which for crystalline materials
means cubic and hexagonal structures. Among possible magnetically
disordered phases for spin systems, tendencies towards dimer formation
are common in the regime of predominantly AF spin interactions, and new
phases with VB correlations occur. This type of physics was discussed
first for $e_g$ orbitals on the cubic lattice,\cite{Fei97} and, in the
context of BaVS$_3$, for one version of the problem of $t_{2g}$ orbitals
on a triangular lattice.\cite{rmkzmpfbf} The same generic behavior has
since been found for $t_{2g}$ orbitals on the cubic lattice,\cite{Hor03}
$e_g$--orbital systems on the triangular lattice,\cite{Ver04,Mil07} and
for $t_{2g}$ orbitals in the pyrochlore geometry.\cite{Mat04,Mat05}
By analogy with spin liquids, the orbital--liquid phase\cite{Diep} has
been introduced for systems with both $e_g$\cite{Fei97,Fei05} and
$t_{2g}$\cite{Kha00,Kha05} orbital degrees of freedom. The orbital
liquid is a phase in which strong orbital fluctuations restore the
symmetry of the orbital sector, in the sense that the instantaneous
orbital state of any site is pure, but the time average is a uniform
occupation of all available orbital states. We note that in the
discussion of orbital liquids in $t_{2g}$ systems,\cite{Kha00,Kha05}
it was argued that the spin sector would be ordered. To date little is
known concerning the behavior of orbital correlations in an orbital
liquid, the possible instabilities of the orbital liquid towards
dimerized or VB phases, or its interplay with lattice degrees of
freedom.

One possible mechanism for the formation of an orbital liquid state
is the positional resonance of VBs. There has been considerable recent
discussion of spin--orbital models in the continuing search for a
realistic system realizing such a resonating VB (RVB) state,\cite{Mil07}
including in a number of the references cited in the previous paragraph.
While the RVB state was first proposed for the $S = \frac12$ Heisenberg
model on a triangular lattice,\cite{rfa} extensive analysis of spin--only
models has not yet revealed a convincing candidate system, although the
nearest--neighbor dimer basis has been shown to deliver a very good
description of the low--energy sector for the $S = \frac12$ Heisenberg
model on a kagome lattice.\cite{rmm} To date, the only rigorous proof
for RVB states has been obtained in rather idealized quantum dimer
models (QDMs),\cite{Rok88} most notably on the triangular
lattice.\cite{Moe01} The insight gained from this type of study can,
however, be used\cite{Mil07} to formulate some qualitative criteria
for the emergence of an RVB ground state. These combine energetic and
topological requirements, both of which are essential: the energetics
of the system must establish a proclivity for dimer formation, a high
quasi--degeneracy of basis states in the candidate ground manifold,
and additional energy gains from dimer resonance; exact degeneracy
between topological sectors (determined by a non--local order parameter
related to winding of wave functions around the system) is a prerequisite
to remove the competing possibility of a ``solid'' phase with dimer,
plaquette or other ``crystalline'' order.\cite{rrfbim}

We comment here that the ``problem'' of frustration, and the resulting
highly degenerate manifolds of states which may promote resonance
phenomena, is often solved by interactions with the lattice. Lattice
deformations act to lift degeneracies and to stabilize particular
patterns of spin and orbital order, the most familiar situation being
that in colossal--magnetoresistance manganites.\cite{Wei04} The same
physics is also dominant in a number of spinels, where electron--lattice
interactions are responsible both for the Verwey transition in
magnetite\cite{Pie06} and for $t_{2g}$ orbital order below it,
as well as for inducing the Peierls state in CuIr$_2$S$_4$ and
MgTi$_2$O$_4$.\cite{Kho05} Similar phenomena are also expected\cite{Kho05}
to play a role in NaTiO$_2$. Here, however, we will not introduce a
coupling to phonon degrees of freedom, and focus only on purely electronic
interactions whose frustration is not quenched by the lattice.

The spin--orbital interactions on a triangular lattice are particularly
intriguing. This lattice occurs for edge--sharing MO$_6$ octahedra in
structures such as NaNiO$_2$ or LiNiO$_2$, where the consecutive $\langle
111 \rangle$ planes of Ni$^{3+}$ ions are well separated. These two
$e_g$--electron systems behave quite differently: while NaNiO$_2$
undergoes a cooperative Jahn--Teller structural transition followed by
a magnetic transition at low temperatures ($T_{\rm N} = 20$ K), both
transitions are absent in LiNiO$_2$.\cite{Hol04} Possible reasons for
this remarkable contrast were discussed in Ref.~\onlinecite{Rei05}, where
the authors noted in particular that realistic spin--orbital superexchange
neither has an SU(2)$\otimes$SU(2) structure,\cite{Ver04} nor can it ever
be reduced only to the consideration of FM spin terms.\cite{Mos02} These
studies showed in addition that LiNiO$_2$ is not a spin--orbital liquid,
and that the reasons for the observed disordered state are subtle, as
spins and orbitals are thought likely to order in a strictly
two--dimensional (2D) spin--orbital model.\cite{Rei05}

\begin{figure}[t!]
\mbox{\includegraphics[width=4.1cm]{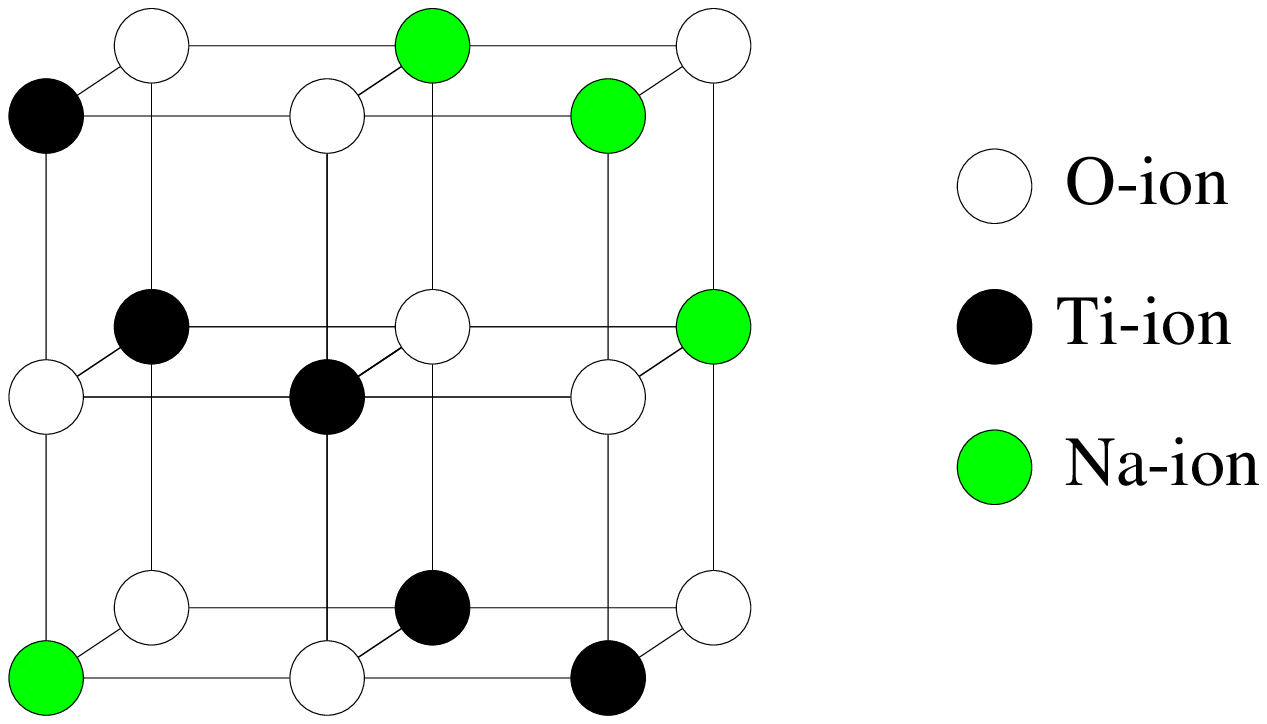}}
\hskip .3cm
\mbox{\includegraphics[width=3.6cm]{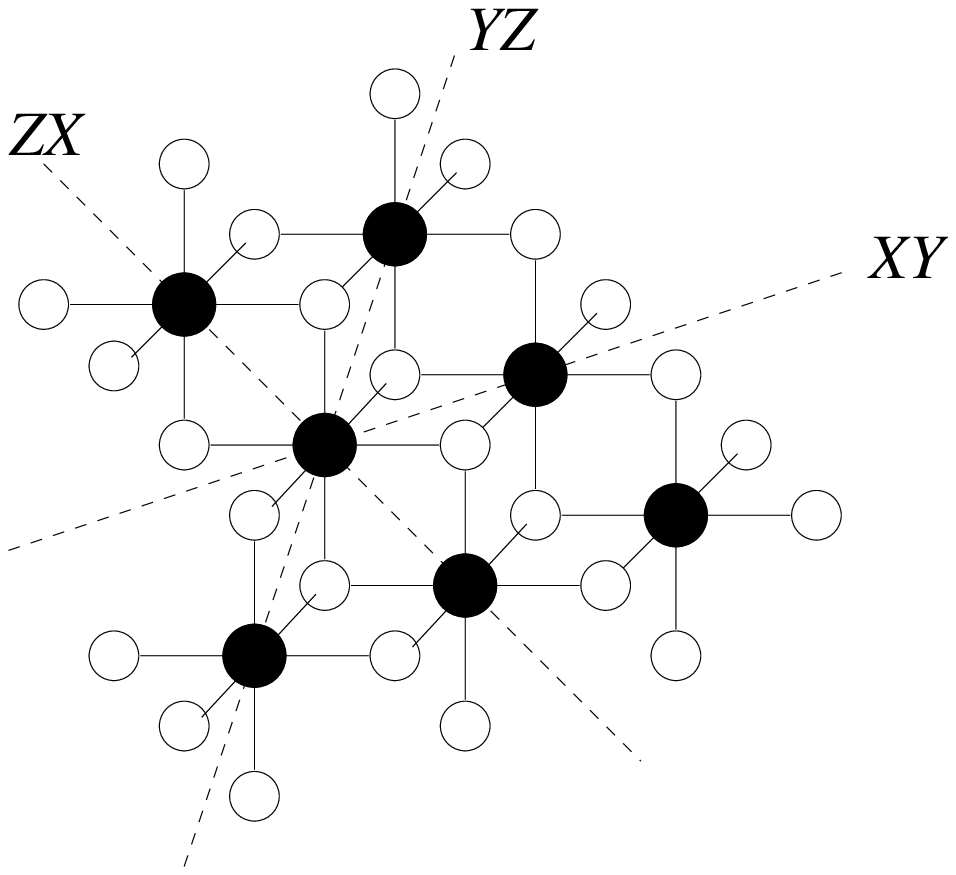}}
\vskip .3cm
\centerline{(a) \hskip 3.5cm (b)}
\caption{(Color online)
Structure of the transition--metal oxide with edge--sharing octahedra
realized for NaTiO$_2$:
(a) fragment of crystal structure, with Ti and Na ions shown respectively
    by black and green (grey) circles separated by O ions (open circles);
(b) titanium $\langle 111 \rangle$ plane with adjacent oxygen layers,
    showing each Ti$^{3+}$ ion coordinated by six oxygen atoms (open
    circles). The directions of the Ti--Ti bonds are labeled as $XY$,
    $YZ$, and $ZX$, corresponding to the plane spanned by the connecting
    Ti--O bonds.
This figure is reproduced from Ref.~\onlinecite{Rei05}, where it served
to explain the structure of LiNiO$_2$.}
\label{structure}
\end{figure}

The possibilities offered for exotic phases in this type of model and
geometry motivate the investigation of a realistic spin--orbital model
with active $t_{2g}$ orbitals, focusing first on $3d^1$ electronic
configurations. The threefold degeneracy of the orbitals is maintained,
although, as noted above, this condition may be hard to maintain in real
materials at low temperatures. A material which should exemplify this
system is NaTiO$_2$ (Fig.~\ref{structure}), which is composed of Ti$^{3+}$
ions in $t_{2g}^1$ configuration, but has to date had rather limited
experimental\cite{Hir85,Tak92} and theoretical\cite{Pen97} attention.
Considerably more familiar is the set of triangular cobaltates best
known for superconductivity in Na$_x$CoO$_2$: here the Co$^{4+}$ ions
have $t_{2g}^5$ configuration and are expected to be analogous to the
$d^1$ case by particle--hole symmetry. The effects of doping have
recently been removed by the synthesis of the insulating end--member
CoO$_2$.\cite{Mot08} Another system for which the same spin--orbital
model could be applied is Sr$_2$VO$_4$, where the V$^{4+}$ ions occupy
the sites of a square lattice.\cite{Ito91}

The model with hopping processes of pure superexchange type was
considered in the context of doped cobaltates by Koshibae and
Maekawa.\cite{Kos03} These authors noted that, like the cubic
system, two $t_{2g}$ orbitals are active for each bond direction
in the triangular lattice, but that the superexchange interactions
are very different from the cubic case because the effective
hopping interchanges the active orbitals. Here we focus only on
insulating systems, whose entire low--energy physics is described
by a spin--orbital model. In addition to superexchange processes
mediated by the oxygen ions, on the triangular lattice it is
possible to have direct--exchange interactions, which result from
charge excitations due to direct $d-d$ hopping between those
$t_{2g}$ orbitals which do not participate in the superexchange.
The ratio of these two types of interaction ($\alpha$, defined in
Sec.~II) is a key parameter of the model. Further, in
transition--metal ions\cite{Ole05} the coefficients of the
different microscopic processes depend on the Hund exchange $J_H$
arising from the multiplet structure of the excited intermediate
$d^2$ state,\cite{Gri71} and we introduce
\begin{equation}
\label{eta}
\eta = \frac{J_H}{U},
\end{equation}
as the second parameter of the model. The aim of this investigation
is to establish the general properties of the phase diagram in the
$(\alpha,\eta)$ plane.

We conclude our introductory remarks by returning to the question
of entanglement. In the analysis to follow we will show that the
presence of conflicting ordering tendencies driven by different
components of the frustrated intersite interactions can be related
to the entanglement of spin and orbital interactions. By ``entanglement''
we mean that the correlations in the ground state involve simultaneous
fluctuations of the spin and orbital components of the wave function
which cannot be factorized. We will introduce an intersite spin--orbital
correlation function to identify and quantify this type of entanglement
in different regimes of the phase diagram. It has been shown\cite{Ole06}
that such spin--orbital entanglement is present in cubic titanates or
vanadates for small values of the Hund exchange $\eta$. Here we will
find entanglement to be a generic feature of the model for all exchange
interactions, even in the absence of dimer resonance, and that only the
FM regime at sufficiently high $\eta$, which is fully factorizable,
provides a counterpoint where the entanglement vanishes.

The paper is organized as follows. In Sec.~\ref{sec:som} we derive
the spin--orbital model for magnetic ions with the $d^1$ electronic
configuration (Ti$^{3+}$ or V$^{4+}$) on a triangular lattice.
The derivation proceeds from the degenerate Hubbard model, and the
resulting Hamiltonian contains both superexchange and direct exchange
interactions. We begin our analysis of the model, which covers the full
range of physical parameters, in Sec.~\ref{sec:mfa} by considering
patterns of long--ranged spin and orbital order representative of all
competitive possibilities. These states compete with magnetically or
orbitally disordered phases dominated by VB correlations on the bonds,
which are investigated in Sec.~\ref{sec:dim}. The analysis suggests
strongly that all long--range order is indeed destabilized by quantum
fluctuations, leading over much of the phase diagram to liquid phases
based on fluctuating dimers, with spin correlations of only the
shortest range. In Sec.~\ref{sec:ed} we present the results of exact
diagonalization calculations performed for small clusters with three,
four, and six bonds, which reinforce these conclusions and provide
detailed information about the local physical processes leading to the
dominance of resonating dimer phases. In each of Secs.~\ref{sec:mfa},
\ref{sec:dim}, and \ref{sec:ed}, we conclude with a short summary of
the primary results, and the reader who is more interested in an
overview, rather than in detailed energetic comparisons and actual
correlation functions for the different phases, may wish to read only
these. Some insight into the competition and collaboration between
frustration effects of different origin can be obtained by varying
the geometry of the system, and Sec.~\ref{sec:rhk} discusses the
properties of the model on related lattices. A discussion and
concluding summary are presented in Sec.~\ref{sec:summa}.

\section{Spin--orbital model}
\label{sec:som}

\subsection{Hubbard model for $t_{2g}$ electrons}
\label{sec:Hub}

We consider the spin--orbital model on the triangular lattice which
follows from the degenerate Hubbard--like model for $t_{2g}$ electrons.
It contains the electron kinetic energy and electronic interactions for
transition--metal ions arranged on the $\langle 111 \rangle$ planes of
a compound with local cubic symmetry and with the $d^1$ ionic
configuration, and as such is applicable to Ti$^{3+}$ or V$^{4+}$
[Fig.~1(a)]. The kinetic energy is given by
\begin{eqnarray}
\label{hkin}
H_{t} & = & - \!\! \sum_{\langle ij \rangle{\parallel} \gamma,\mu\nu,
\sigma} \!\! t^{(\gamma)}_{\mu\nu} \! \left( d^{\dagger}_{i\mu\sigma}
d_{j\nu\sigma} + d^{\dagger}_{j\nu\sigma} d_{i\mu\sigma} \right) \! ,
\end{eqnarray}
where $d^{\dagger}_{i\mu\sigma}$ are creation operators for an
electron with spin $\sigma = \uparrow,\downarrow$ and orbital
``color'' $\mu$ at site $i$, and the sum is made over all the
bonds $\langle ij \rangle {\parallel} \gamma$ spanning the three
directions, $\gamma = a,b,c$, of the triangular lattice. This
notation is adopted from the situation encountered in a cubic
array of magnetic ions, where only two of the three $t_{2g}$
orbitals are active on any one bond $\langle ij \rangle$, and
contribute $t^{(\gamma)}_{\mu\nu}$ to the kinetic energy, while
the third lies in the plane perpendicular to the $\gamma$ axis and
thus hopping processes involving the $2p_{\pi}$ oxygen orbitals is
forbidden by symmetry.\cite{Kha01,Har03} We introduce the labels
$a \equiv yz$, $b \equiv xz$, and $c \equiv xy$ also for the three
orbital colors, and in the figures to follow their respective
spectral colors will be red, green, and blue.

\begin{figure}[t!]
\begin{center}
\includegraphics[width=7.5cm]{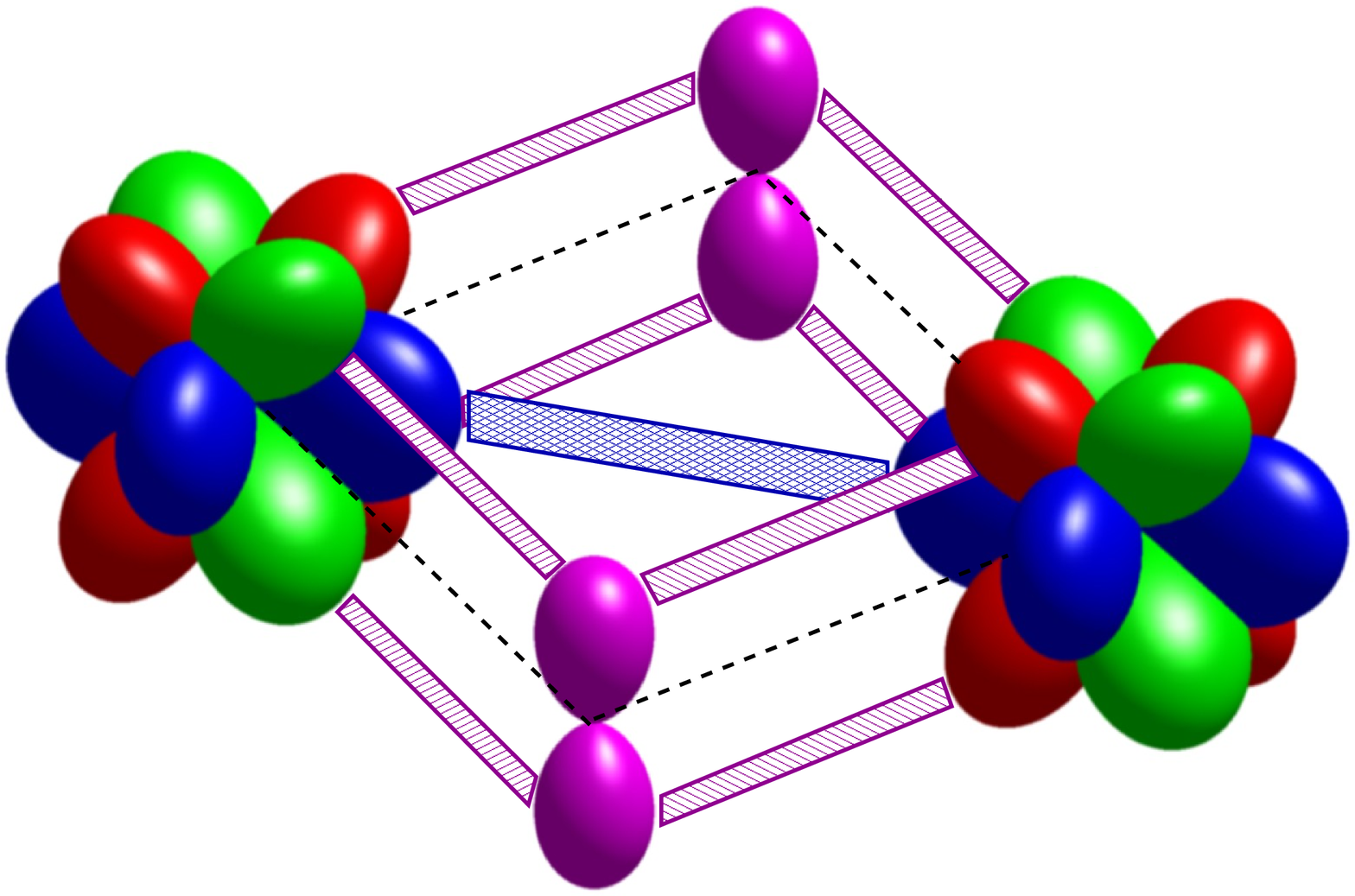}
\centerline{(a)} \vskip .1cm
\includegraphics[width=7.7cm]{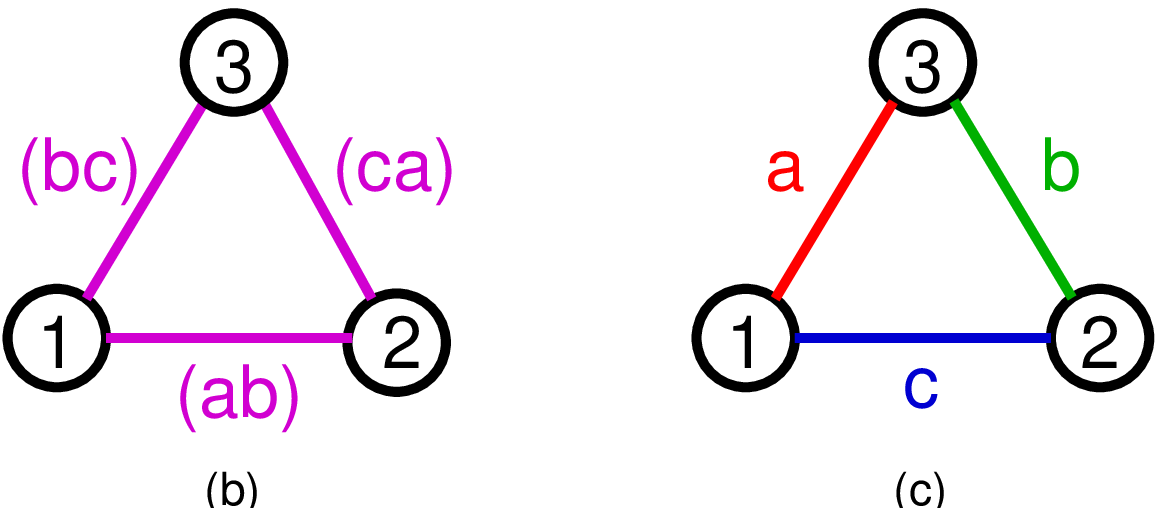}
\end{center}
\caption{(Color online) (a) Schematic representation of the
hopping processes in Eq.~(\ref{hkin}) which contribute to magnetic
interactions on a representative bond $\langle ij \rangle$ along
the $c$--axis in the triangular lattice. The $t_{2g}$ orbitals are
represented by different colors (greyscale intensities).
Superexchange processes involve O $2p_z$ orbitals (violet), and
couple pairs of $a$ and $b$ orbitals (red, green) with effective
hopping elements $t$, interchanging their orbital color. Direct
exchange couples $c$ orbitals (blue) with hopping strength $t'$.
(b) Pairs of $t_{2g}$ orbitals active in superexchange and (c)
single orbitals active in direct exchange; horizontal bonds
correspond to the situation depicted in panel (a). }
\label{fig:hops}
\end{figure}

For the triangular lattice formed by the ions on the $\langle 111
\rangle$ planes of transition--metal oxides (Fig.~\ref{structure})
it is also the case that only two $t_{2g}$ orbitals participate in
(superexchange) hopping processes via the oxygen sites. However,
unlike the cubic lattice, where the orbital color is conserved,
here any one active orbital color is exchanged for the other one
[Fig.~\ref{fig:hops}(a)]. Using the same convention, that each
direction in the triangular lattice is labeled by its inactive
orbital color\cite{Kha04} $\gamma = a, b, c$, the hopping elements
for a bond oriented (for example) along the $c$--axis in Eq.~(\ref{hkin})
are $t^{(c)}_{ab} = t^{(c)}_{ba} = t$, while $t^{(c)}_{aa} = t^{(c)}_{bb}
 = 0$. In addition, and also in contrast to the cubic system, for the
triangular geometry a direct hopping from one $c$ orbital to the other,
{\it i.e.} without involving the oxygen orbitals, is also permitted on
this bond (Fig.~\ref{fig:hops}), and this element is denoted by $t'
 = t^{(c)}_{cc}$. We will also refer to these hopping processes as
off--diagonal and diagonal. We stress that while the lattice structure of
magnetic ions is triangular, the system under consideration retains local
cubic symmetry of the metal--oxygen octahedra, which is crucial to ensure
that the degeneracy of the three $t_{2g}$ orbitals is preserved.

The electron--electron interactions are described by the on--site
terms\cite{Ole83}
\begin{eqnarray}
\label{hint}
H_{int}\! & = &\! U \sum_{i\mu} n_{i\mu\uparrow} n_{i\mu\downarrow}
  + \left(U - \frac{5}{2} J_H\right)\!\sum_{i,\mu < \nu,\sigma\sigma'}\!
                      n_{i\mu\sigma} n_{i\nu\sigma'}        \nonumber \\
 &-&\!2 J_H \! \sum_{i,\mu < \nu} \! {\vec S}_{i\mu}\!\cdot\!{\vec S}_{i\nu}
     \! + \! J_H \!\! \sum_{i,\mu \neq \nu}\!
             d^{\dagger}_{i\mu   \uparrow} d^{\dagger}_{i\mu\downarrow}
             d^{       }_{i\nu \downarrow }d^{       }_{i\nu  \uparrow},
\end{eqnarray}
where $U$ and $J_H$ represent respectively the intraorbital
Coulomb and on--site Hund exchange interactions. Each pair of
orbitals $\{\mu,\nu\}$ is included only once in the interaction
terms. The Hamiltonian (\ref{hint}) describes rigorously the
multiplet structure of $d^2$ ions within the $t_{2g}$ subspace,
and is rotationally invariant in the orbital space.\cite{Ole83}

When the Coulomb interaction is large compared with the hopping
elements ($U \gg t, t'$), the system is a Mott insulator with one
$d$ electron per site in the $t_{2g}$ orbitals, whence the local
constraint in the strongly correlated regime is
\begin{equation}
\label{n=1}
n_{ia} + n_{ib} + n_{ic} = 1,
\end{equation}
where $n_{i\gamma} = n_{i\gamma\uparrow} + n_{i\gamma\downarrow}$.
The operators act in the restricted space $n_{i\gamma} = 0,1$. The
low--energy Hamiltonian may be obtained by second--order perturbation
theory, and consists of a superposition of terms which follow from
virtual $d_i^1 d_j^1 \rightleftharpoons d_i^2 d_j^0$ excitations.
Because each hopping process may be of either off--diagonal ($t$)
[Fig.~\ref{fig:hops}(b)] or diagonal ($t'$) type
[Fig.~\ref{fig:hops}(c)], the Hamiltonian consists of several
contributions which are proportional to three coupling constants,
\begin{equation}
\label{allj}
J_s = \frac{4t^2}{U},    \hskip .5cm
J_d = \frac{4t'^2}{U},   \hskip .5cm
J_m = \frac{4tt'}{U}.
\end{equation}
These represent in turn the superexchange term, the direct exchange
term, and mixed interactions which arise from one diagonal and one
off--diagonal hopping process.

We choose to parameterize the Hamiltonian by the single variable
\begin{equation}
\label{alpha}
\alpha = \sin^2\theta,
\end{equation}
with
\begin{equation}
\label{angle}
\tan\theta = \frac{t'}{t},
\end{equation}
which gives $J_s = J \cos^2 \theta$, $J_m = J \sin \theta \cos \theta$,
and $J_d = J \sin^2 \theta$; $J$ is the energy unit, which specifies
respectively the superexchange ($J = J_s$) and direct--exchange ($J =
J_d$) constants in the two limits $\alpha = 0$ and $\alpha = 1$. The
Hamiltonian
\begin{equation}
\label{som}
{\cal H} = J \left\{ (1 - \alpha) \; {\cal H}_s
                + \sqrt{(1 - \alpha) \alpha} \; {\cal H}_m
                + \alpha \; {\cal H}_d \right\}
\end{equation}
consists of three terms which follow from the processes described by
the exchange elements in Eqs.~(\ref{allj}), each of which contains
contributions from both high-- and low--spin excitations.

\subsection{Superexchange}
\label{sec:sex}

Superexchange contributions to ${\cal H}$ can be expressed in the form
\begin{eqnarray}
\label{Hs}
{\cal H}_s \! & = & \! \frac{1}{2} \sum_{\langle ij \rangle \parallel \gamma}
\Big\{ r_1\Big( \vec{S}_i \! \cdot \! \vec{S}_j + \frac{3}{4} \Big)
  \Big[ A_{ij}^{(\gamma)} \! + \frac{1}{2} (n_{i\gamma} + n_{j\gamma})
  - 1 \Big] \nonumber \\
 & & + r_2 \Big( \vec{S}_i \! \cdot \! \vec{S}_j - \frac{1}{4}\Big)
\Big[ A_{ij}^{(\gamma)} \! - \frac{1}{2} (n_{i\gamma} + n_{j\gamma}) + 1
\Big] \nonumber \\ & & - \frac{2}{3} (r_2 - r_3) \Big( \vec{S}_i \!
\cdot \! \vec{S}_j - \frac{1}{4} \Big) B_{ij}^{(\gamma)} \Big\},
\end{eqnarray}
where one recognizes a structure similar to that for superexchange
in cubic vanadates,\cite{Kha00,Ole05} with separation into a spin
projection operator on the triplet state, $(\vec{S}_i \cdot \vec{S}_j
 + \frac{3}{4})$, and an operator $(\vec{S}_i \cdot \vec{S}_j
 - \frac{1}{4})$ which is finite only for low--spin excitations.
These operators are accompanied by coefficients $(r_1,r_2,r_3)$
which depend on the Hund exchange parameter (\ref{eta}), and are
given from the multiplet structure of $d^2$ ions\cite{Gri71} by
\begin{equation}
\label{rr}
r_1 = \frac{1}{1 - 3 \eta},  \hskip 0.5cm
r_2 = \frac{1}{1 - \eta},    \hskip 0.5cm
r_3 = \frac{1}{1 + 2 \eta}.
\end{equation}
The Coulomb and Hund exchange elements deduced from the spectroscopic
data of Zaanen and Sawatzky\cite{Zaa90} are $U = 4.35$ eV and $J_H = 0.59$
eV, giving a realistic value of $\eta \simeq 0.136$ for Ti$^{2+}$ ions.
For V$^{2+}$ one finds\cite{Zaa90} $U = 4.98$ eV and $J_H = 0.64$ eV,
whence $\eta \simeq 0.13$, and the values for V$^{3+}$ ions are expected
to be very similar. Finally, for Co$^{3+}$ ions,\cite{Miz96} $U = 6.4$ eV
and $J_H = 0.84$ eV, giving again $\eta \simeq 0.13$. The value $\eta
 = 0.13$ therefore appears to be quite representative for
transition--metal oxides with partly filled $t_{2g}$ orbitals,
whereas somewhat larger values have been found for systems with
active $e_g$ orbitals due to a stronger Hund exchange.\cite{Ole05}

The orbital operators $A_{ij}$ and $B_{ij}$ in Eq.~(\ref{Hs}) depend on
the bond direction $\gamma$ and involve two active orbital colors,
\begin{eqnarray}
\label{som1abn}
\hskip -.7cm
A_{ij}^{(\gamma)} \!\! & = & \!\! \Big( T_{i\gamma}^+ T_{j\gamma}^+ \!
 + T_{i\gamma}^- T_{j\gamma}^- \Big) \! - \! 2T_{i\gamma}^z T_{j\gamma}^z
 + \! \frac{1}{2} n_i^{(\gamma)} n_j^{(\gamma)},\\
\hskip -.7cm
B_{ij}^{(\gamma)} \! & = & \!\!
\Big( T_{i\gamma}^+ T_{j\gamma}^- \! + T_{i\gamma}^- T_{j\gamma}^+ \Big)
\! - \! 2T_{i\gamma}^z T_{j\gamma}^z + \! \frac{1}{2} n_i^{(\gamma)}
n_j^{(\gamma)}.
\end{eqnarray}
For illustration, in the case $\gamma = c$ ($\langle ij \rangle \parallel
c$), the orbitals $a$ and $b$ at site $i$ are interchanged (off--diagonal
hopping) at site $j$, and the electron number operator is $n_i^{(\gamma)}
= n_{ia} + n_{ib}$. The quantity $n_{i\gamma}$ in Eq.~(\ref{Hs}) is the
number operator for electrons on the site in orbitals inactive for hopping
on bond $\gamma$, $n_{i\gamma} = 1 - n_i^{(\gamma)}$, or $n_{ic}$ in this
example.

For a single bond, the orbital operators in Eq.~(\ref{som1abn}) may be
written in a very suggestive form by performing a local transformation
in which the active orbitals are exchanged on one bond site, specifically
$|a \rangle \rightarrow |b \rangle$ and $|b \rangle \rightarrow |a
\rangle$ on bond $\gamma = c$.\cite{Kos03} Then
\begin{eqnarray}
\label{som2abn}
\hskip -.7cm
A_{ij}^{(\gamma)} \!\! & = & \!\! 2 \Big( {\vec T}_{i\gamma} \cdot {\vec
T}_{j\gamma} \! + \! \frac{1}{4} n_i^{(\gamma)} n_j^{(\gamma)} \Big),\\
\hskip -.7cm
B_{ij}^{(\gamma)} \! & = & \!\! 2 \Big( \vec{T}_{i\gamma} \times
\vec{T}_{j\gamma} + \frac{1}{4} n_i^{(\gamma)} n_j^{(\gamma)} \Big),
\end{eqnarray}
where the scalar product in $A_{ij}$ is the conventional expression
for pseudospin--1/2 variables, and the cross product in $B_{ij}$
is defined as
\begin{equation}
\vec{T}_{i\gamma} \times \vec{T}_{j\gamma} = \frac12 (
T_{i\gamma}^ + T_{j\gamma}^+ + T_{i\gamma}^- T_{j\gamma}^- ) +
T_{i\gamma}^z T_{j\gamma}^z.
\end{equation}
Equations (\ref{Hs}) and (\ref{som2abn}) make it clear that for a
single superexchange bond, the minimal energy is obtained either by
forming an orbital singlet, in which case the optimal spin state is
a triplet, or by forming a spin singlet, in which case the preferred
orbital state is a triplet; we refer to these bond wavefunctions
respectively as (os/st) and (ss/ot). The two states are degenerate
for $\eta = 0$, while for finite Hund exchange
\begin{eqnarray}
\label{eosst}
   E_{\rm (os/st)} & = & - J r_1,            \\
\label{essot}
   E_{\rm (ss/ot)} & = & - \frac{1}{3} J \left( 2 r_2 + r_3 \right),
\end{eqnarray}
and the (os/st) state is favored. This propensity for singlet formation
in the $\alpha = 0$ limit will drive much of the physics to be analyzed
in what follows.

Because of the off--diagonal nature of the hopping term, in the
original electronic basis (before the local transformation) the
orbital singlet is the state
\begin{equation}
|\psi_{os} \rangle = \frac{1}{\sqrt{2}} \left( |aa \rangle - |bb
\rangle \right),
\end{equation}
while the orbital triplet states are
\begin{eqnarray}
|\psi_{ot+} \rangle & = & |ab \rangle,                           \\
|\psi_{ot0} \rangle & = & \frac{1}{\sqrt{2}}
                     \left( |aa \rangle + |bb \rangle\right),    \\
|\psi_{ot-} \rangle & = & |ba \rangle.
\end{eqnarray}
The locally transformed basis then gives a clear analogy which can
be used for single bonds and dimer phases in combination with all
of the understanding gained for the Heisenberg model. However, we
stress here that the local transformation fails for systems with
more than 1 bond in the absence of static dimer formation. This
arises due to frustration, and can be shown explicitly in numerical
calculations, but we will not enter into this point in more detail
here. However, we take the liberty of retaining the notation of the
local transformation, particularly in Sec.~\ref{sec:dim} when
considering dimers. Because the transformation interchanges the
definitions of FO and AO configurations, we will state clearly in
each section the basis in which the notation is chosen.

\subsection{Direct Exchange}
\label{sec:dex}

The direct exchange part is obtained by considering virtual excitations
of active $\gamma$ orbitals on a bond $\langle ij \rangle \parallel
\gamma$, which yield
\begin{eqnarray}
\label{Hd}
{\cal H}_d \! & = & \! \frac{1}{4} \sum_{\langle ij \rangle \parallel \gamma}
\Big\{ \Big[ -r_1 \Big( \vec{S}_i \! \cdot \! \vec{S}_j + \frac{3}{4}\Big)
 + r_2 \Big( \vec{S}_i \! \cdot \! \vec{S}_j - \frac{1}{4} \Big) \Big]
\nonumber \\ & & \hskip 1.3cm \times \Big[ n_{i\gamma} (1 - n_{j\gamma})
 + (1 - n_{i\gamma}) n_{j\gamma} \Big] \nonumber \\ & & + \frac{1}{3}
\left( 2r_2 + r_3 \right) \Big( \vec{S}_i \! \cdot \! \vec{S}_j - \frac{1}{4}
\Big) \; 4 n_{i\gamma} n_{j\gamma} \Big\}.
\end{eqnarray}
Here there are no orbital operators, but only number operators
which select electrons of color $\gamma$ on bonds oriented along
the $\gamma$--axis. When only only one active orbital is occupied
[$n_{i\gamma} (1 - n_{j\gamma})$], this electron can gain energy
$- \frac14 J$ from virtual hopping at $\eta = 0$, a number which
has only a weak dependence on the bond spin state at $\eta > 0$.
When both active orbitals are occupied ($n_{i\gamma}
n_{j\gamma}$), placing the two electrons in a spin singlet yields
the far lower bond energy $- J$, and thus again one may expect
much of the discussion to follow to center on dimer--based states
of the extended system. Again the triplet $d^2$ spin excitation
corresponds to the lowest energy, $(U - 3J_H)$, and only the lower
two excitations involve spin singlets which could minimize the
bond energy. The structure of these terms is the same as in the 1D
$e_g$ spin--orbital model,\cite{Dag04} or the case of the spinel
MgTi$_2$O$_4$.\cite{Mat04} A simplified model for the
triangular--lattice model in this limit, using a lowest--order
expansion in $\eta$ for the spin but not for orbital interactions,
was introduced in Ref.~\onlinecite{Jac07}.

\subsection{Mixed Exchange}
\label{sec:mex}

Finally, the two different types of hopping channel may also contribute
to two--step, virtual $d_i^1 d_j^1 \rightleftharpoons d_i^2d_j^0$
excitations with one off--diagonal ($t$) and one diagonal ($t'$) process.
The occupied orbitals are changed at both sites (Fig.~\ref{fig:hops}),
and as for the superexchange term the resulting effective interaction
may be expressed in terms of orbital fluctuation operators. To avoid a
more general but complicated notation, we write this term only for
$c$--axis bonds,
\begin{eqnarray}
\label{Hm}
{\cal H}_m^{(c)} \! & = & \! - \frac{1}{4} \sum_{\langle ij \rangle
\parallel c} \Big[ r_1 \Big( \vec{S}_i \! \cdot \! \vec{S}_j
 + \frac{3}{4} \Big) - r_2 \Big( \vec{S}_i \! \cdot \! \vec{S}_j
 - \frac{1}{4} \Big) \Big] \nonumber \\ & & \times \Big( T_{ia}^+
T_{jb}^+ + T_{ib}^- T_{ja}^- + T_{ib}^+ T_{ja}^+ + T_{ia}^- T_{jb}^- \Big),
\end{eqnarray}
where the orbital operators are
\begin{eqnarray}
\label{TaTb}
T_{ia}^+ & = b_i^{\dagger}c_i^{}, \hskip .7cm T_{ib}^+ = c_i^{\dagger} a_i^{},
\nonumber \\ T_{ia}^- & = c_i^{\dagger} b_i^{}, \hskip .7cm T_{ib}^- =
a_i^{\dagger} c_i^{}.
\end{eqnarray}
These definitions are selected to correspond to the $\uparrow$--pseudospin
components of both operators being $|b_i \rangle$ for $T_{ia}^z$ and $|c_i
\rangle$ for $T_{ib}^z$. The form of the ${\cal H}_m^{(a)}$ and ${\cal
H}_m^{(b)}$ terms is obtained from Eq.~(\ref{Hm}) by a cyclic permutation
of the orbital indices. By inspection, this type of term is finite only
for bonds whose sites are occupied by linear superpositions of different
orbital colors, and creates no strong preference for the spin configuration
at small $\eta$.

\subsection{Limit of vanishing Hund exchange}
\label{sec:eta0}

In the subsequent sections we will give extensive consideration to
the model of Eq.~(\ref{som}) at $\eta = 0$. In this special case the
multiplet structure collapses (spin singlet and triplet excitations
are degenerate), one finds a single charge excitation of energy
$U$, and the Hamiltonian reduces to the form
\begin{widetext}
\begin{eqnarray}
\label{som0}
{\cal H}(\eta = 0) & = &
J \sum_{\langle ij \rangle \parallel \gamma}
\Big\{ (1 - \alpha) \Big[ 2 \Big( \vec{S}_i \! \cdot \! \vec{S}_j
 + \frac{1}{4} \Big) \Big( \vec{T}_{i\gamma} \! \cdot \! \vec{T}_{j\gamma}
 + \frac{1}{4} n_i^{(\gamma)} n_j^{(\gamma)} \Big)
 + \frac{1}{2} (n_{i\gamma} + n_{j\gamma}) - 1 \Big] \nonumber \\
& & \hskip 1.3cm + \alpha \Big[ \Big( \vec{S}_i \! \cdot \! \vec{S}_j
 - \frac{1}{4} \Big) n_{i\gamma} n_{j\gamma} - \! \frac{1}{4} \Big(
n_{i\gamma} (1 - n_{j\gamma}) + (1 - n_{i\gamma}) n_{j\gamma} \Big)
\Big] \nonumber \\
& & \hskip 1.3cm - \frac{1}{4} \sqrt{\alpha (1 - \alpha)} \; \Big(
T_{i\bar{\gamma}}^+ T_{j{\tilde \gamma}}^+ + T_{i{\tilde \gamma}}^-
T_{j\bar{\gamma}}^- + T_{i{\tilde \gamma}}^+ T_{j\bar{\gamma}}^+
+ T_{i\bar{\gamma}}^- T_{j{\tilde \gamma}}^- \Big)
\Big\},
\end{eqnarray}
\end{widetext}
which depends only on the ratio of superexchange to direct exchange
($0 \le \alpha \le 1$). The first line of Eq.~(\ref{som0}) makes
explicit the fact that the spin and orbital sectors are completely
equivalent and symmetrical at $\alpha = 0$, at least at the level of
a single bond. However, we will show that this equivalence is broken
when more bonds are considered, and no higher symmetry emerges because
of the color changes involved for different bond directions, which
change the SU(2) orbital subsector. The second line of Eq.~(\ref{som0})
emphasizes the importance of bond occupation and singlet formation at
$\alpha = 1$ (Sec.~IIC).

In the third line of Eq.~(\ref{som0}), the labels $\bar{\gamma} \ne
{\tilde \gamma}$ refer to the two mixed orbital operators on each
bond [Eq.~(\ref{TaTb})]. Orbital fluctuations are the only processes
contributing to the mixed terms in this limit, where the spin state
of the bond has no effect. We draw the attention of the reader to the
fact that for the parameter choice $\alpha = 0.5$, an electron of
any color at any site has the same matrix element to hop in any
direction. However, because of the different color changes involved
in these processes, again the spin--orbital Hamiltonian does not
exhibit a higher symmetry at this point, a result reflected in the
different operator structures of superexchange and direct exchange
components.

\section{Long--range--ordered states}
\label{sec:mfa}

In this Section we study possible ordered or partially ordered states
for the Hamiltonian of Eq.~(\ref{som}). As explained in Sec.~II, the
parameters of the problem are the ratio of the direct and superexchange
interactions, $\alpha$ (\ref{alpha}), and the strength of the Hund
exchange interaction, $\eta$ (\ref{eta}). Regarding the latter, we
will discuss briefly the transition to ferromagnetic (FM) spin order
for increasing $\eta$ in this framework.

The first necessary step in any analysis of such an interacting system
is to establish the energies of different (magnetically and orbitally)
ordered states. The high connectivity of the triangular--lattice system
suggests that ordered states will dominate, and claims of more exotic
ground states are justifiable only when these are shown to be
uncompetitive. The calculations in this Section will be performed for
static orbital and spin configurations, with the virtual processes
responsible for (super)exchange as the only fluctuations. In the language
of the discussion in Sec.~I, fully ordered states gain only potential
energy at the cost of sacrificing the kinetic (resonance) energy from
fluctuation processes, which we will show in Secs.~IV and V is of
crucial importance here.

\subsection{Possible orbital configurations }
\label{sec:mforb}

The results to follow will be obtained by first fixing the orbital
configuration, either on every site or on particular bonds, and then
computing the spin interaction and optimizing the spin state accordingly.
While this is equivalent to the converse, the procedure is more
transparent and offers more insight into the candidate phases. We limit
the number of states to ordered phases with small unit cells, and the
orbital states to be considered are enumerated in this subsection.
For clarity we adopt the convention of Fig.~\ref{fig:hops}(c) that
horizontal ($c$) bonds have diagonal (direct exchange) hopping of $c$
orbitals, which are shown in blue, and off--diagonal (superexchange)
hopping processes for $a$ and $b$ orbitals [Fig.~\ref{fig:hops}(b)],
respectively red and green; up--slanting ($a$) bonds have diagonal
hopping for $a$ orbitals and off--diagonal hopping between $b$ and $c$
orbitals; down--slanting ($b$) bonds have diagonal hopping for $b$
orbitals and off--diagonal hopping between $a$ and $c$ orbitals. All
Hamiltonians and energies are functions of $\alpha$ and $\eta$, as
given by Eqs.~(\ref{som}), (\ref{Hs}), (\ref{Hd}), and (\ref{Hm}).
To minimize additional notation, they will be quoted in this and
in the next section as functions of the single argument $\alpha$,
with implicit $\eta$--dependence contained in the parameters
$(r_1, r_2, r_3)$. The orbital bond index $\gamma$ will also be
suppressed here and in Sec.~IV.

\begin{figure}[t!]
\mbox{\includegraphics[width=3.5cm]{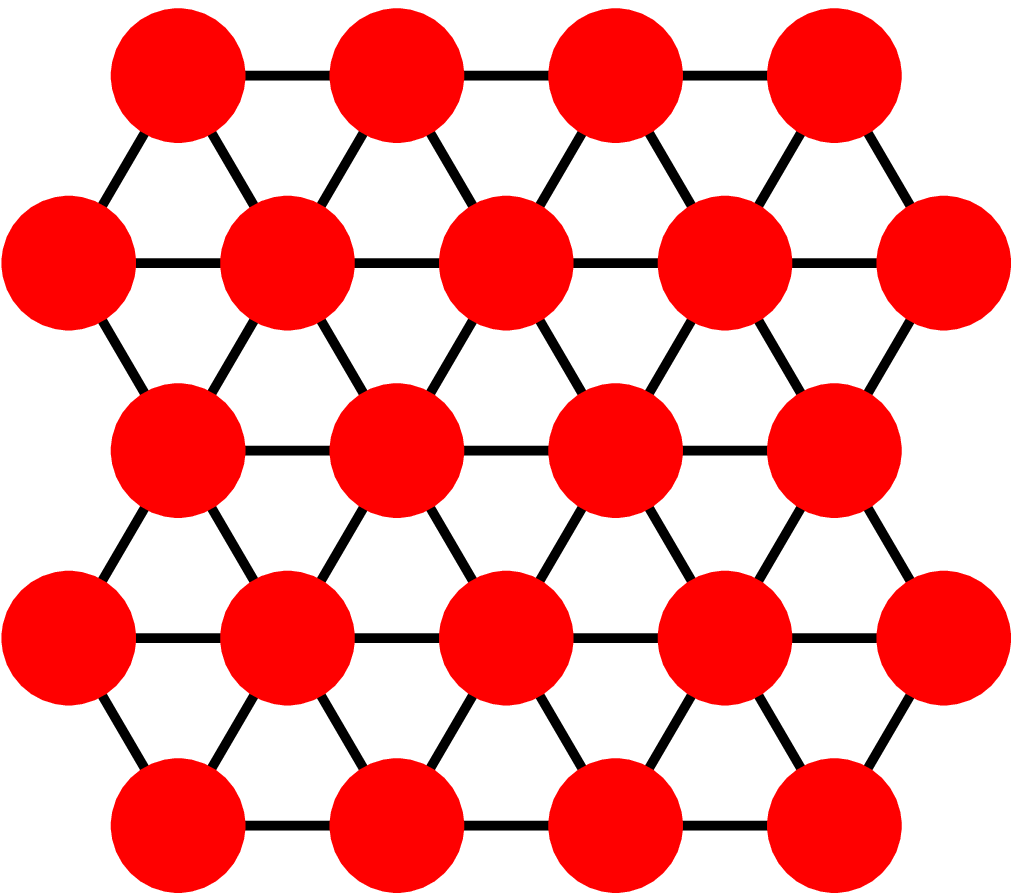}} \hskip .5cm
\mbox{\includegraphics[width=3.5cm]{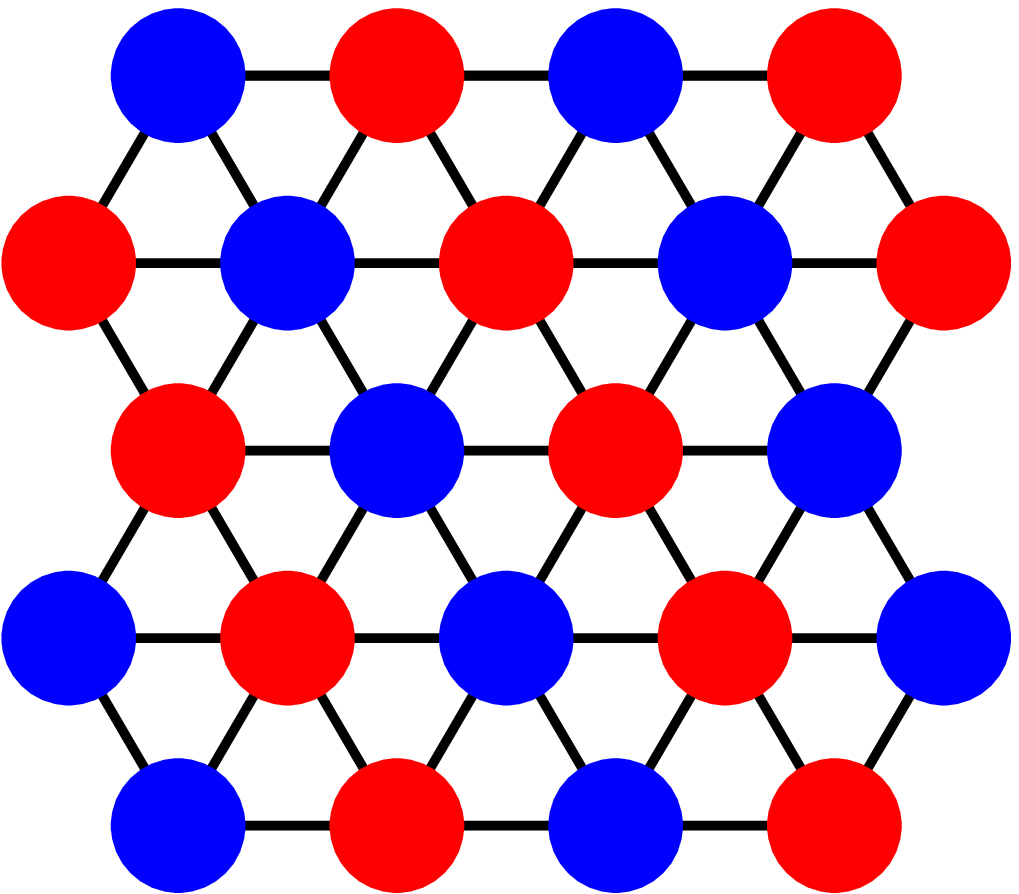}} \centerline{(a)
\hskip 3.5cm (b)} \vskip .2cm
\mbox{\includegraphics[width=3.5cm]{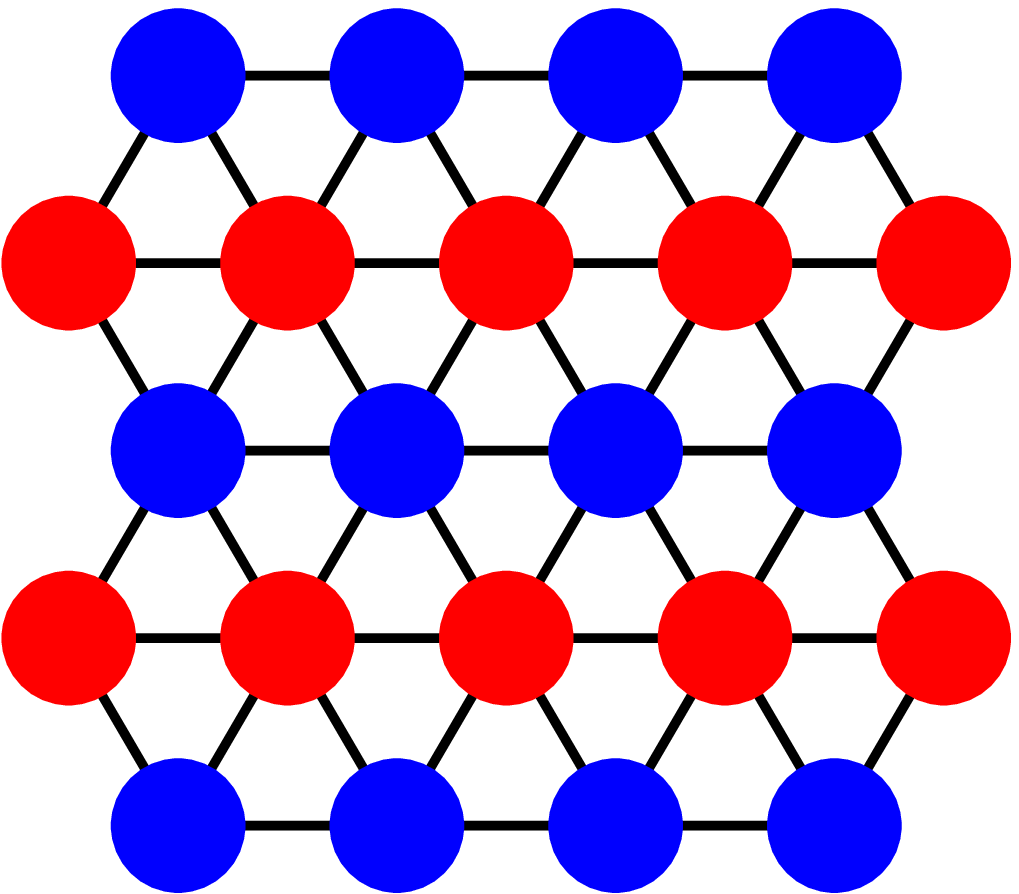}} \hskip .5cm
\mbox{\includegraphics[width=3.5cm]{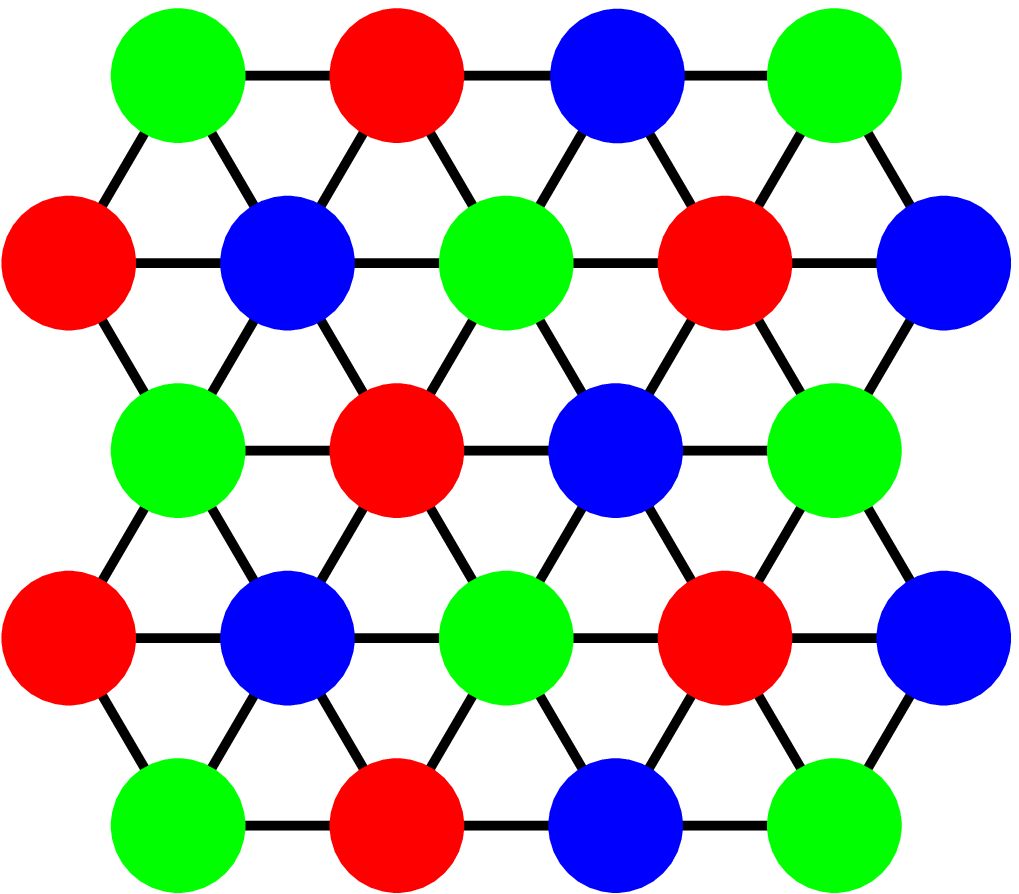}} \centerline{(c)
\hskip 3.5cm (d)} \vskip .2cm
\mbox{\includegraphics[width=3.5cm]{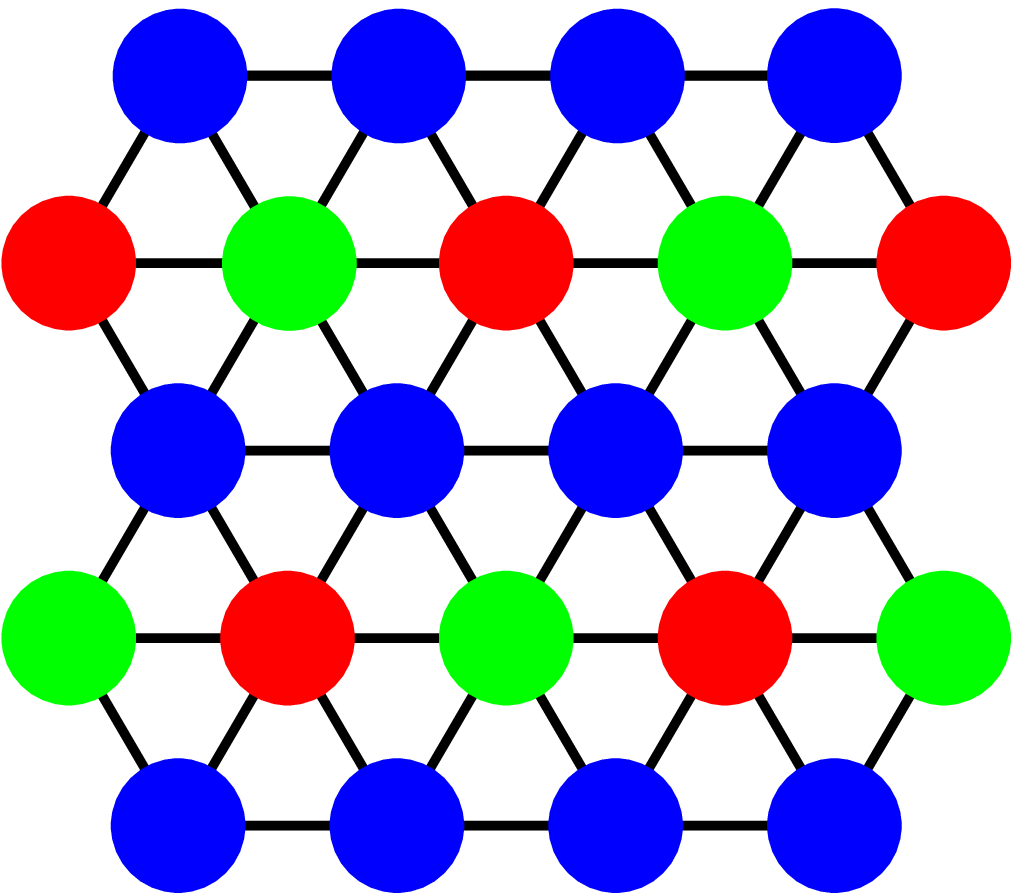}} \hskip .5cm
\mbox{\includegraphics[width=3.5cm]{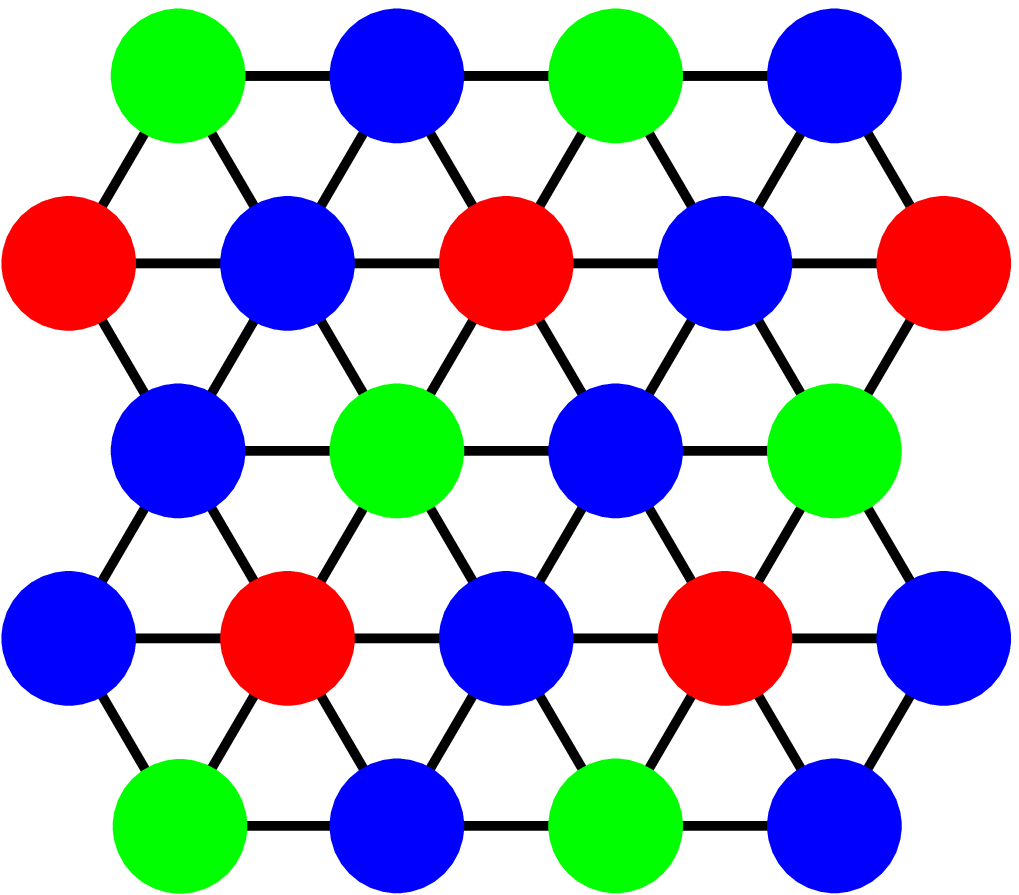}} \centerline{(e)
\hskip 3.5cm (f)} \caption{(Color online) Schematic representation
of possible orbital states with a single color on each site of the
triangular lattice: (a) one--color state; (b) and (c) two
inequivalent two--color states; (d) three--sublattice three--color
state; (e) and (f) two inequivalent three--color states. The
latter two configurations are degenerate with similar states where
the lines of occupied $a$ and $b$ orbitals repeat rather than
being staggered along the direction perpendicular to the lines of
occupied $c$ orbitals. The three--sublattice state (3d) is
nondegenerate ($d = 1$), states (3a), (3b), and (3e) have
degeneracy $d = 3$, and states (3c) and (3f) have degeneracy $d =
6$. } \label{fig:pure}
\end{figure}

We continue to refer to the orbital type as a ``color'', and begin by
listing symmetry--inequivalent states where each site has a unique color.
If the same orbital is occupied at every site [Fig.~\ref{fig:pure}(a)],
the three states with $a$, $b$, or $c$ orbitals occupied are physically
equivalent (degeneracy is $d = 3$). When lines of the same occupied
orbitals alternate along the perpendicular direction there are two
basic possibilities, which are shown in Figs.~\ref{fig:pure}(b) and
\ref{fig:pure}(c). These two--color states differ in their numbers of
active superexchange or direct--exchange bonds, which depend on how the
monocolored lines are oriented relative to the active hopping direction(s)
of the orbital color. There is only one three--color configuration with
equal occupations, which is shown in Fig.~\ref{fig:pure}(d).

\begin{figure}[t!]
{\includegraphics[width=3.5cm]{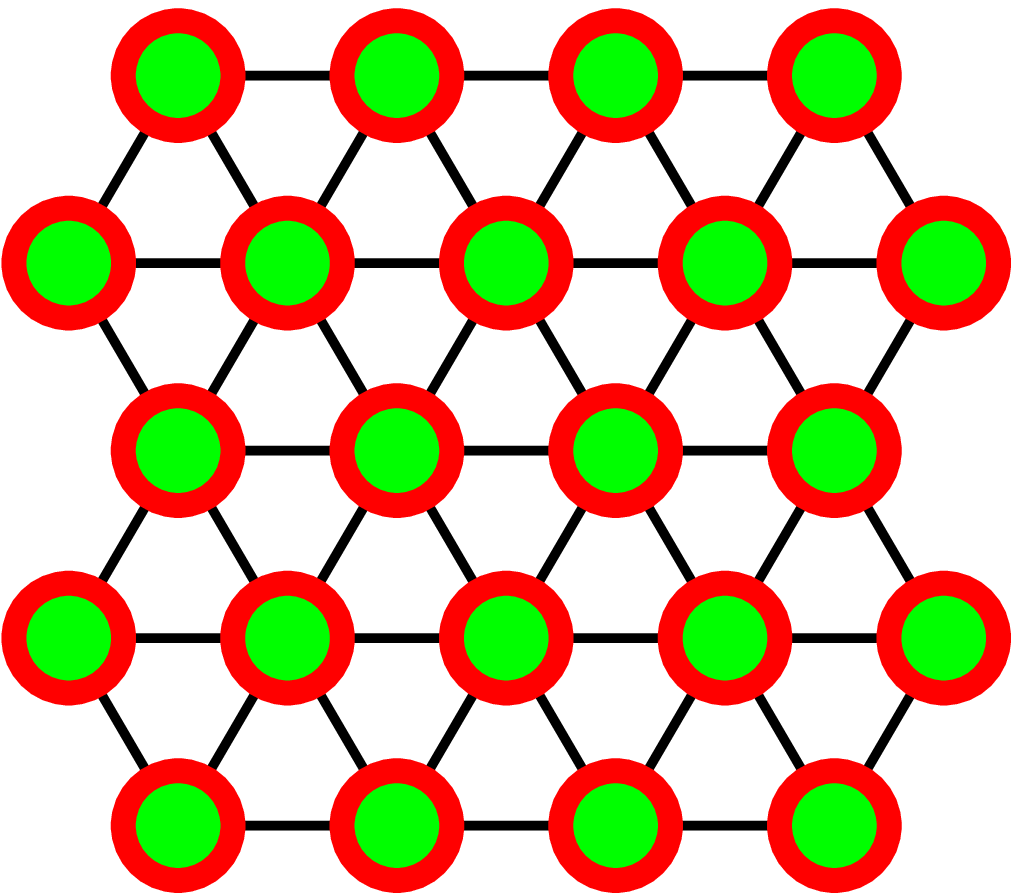} \centerline{(a)}
\mbox{\includegraphics[width=3.5cm]{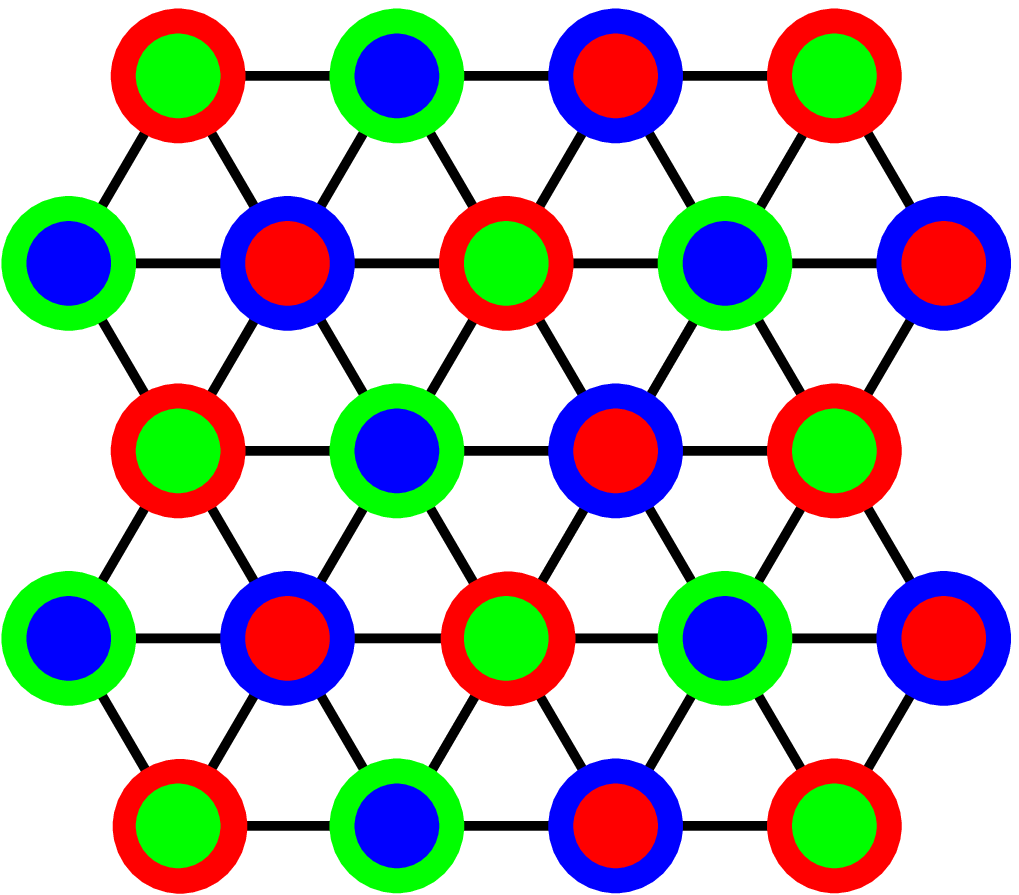}} \hskip .5cm
\mbox{\includegraphics[width=3.5cm]{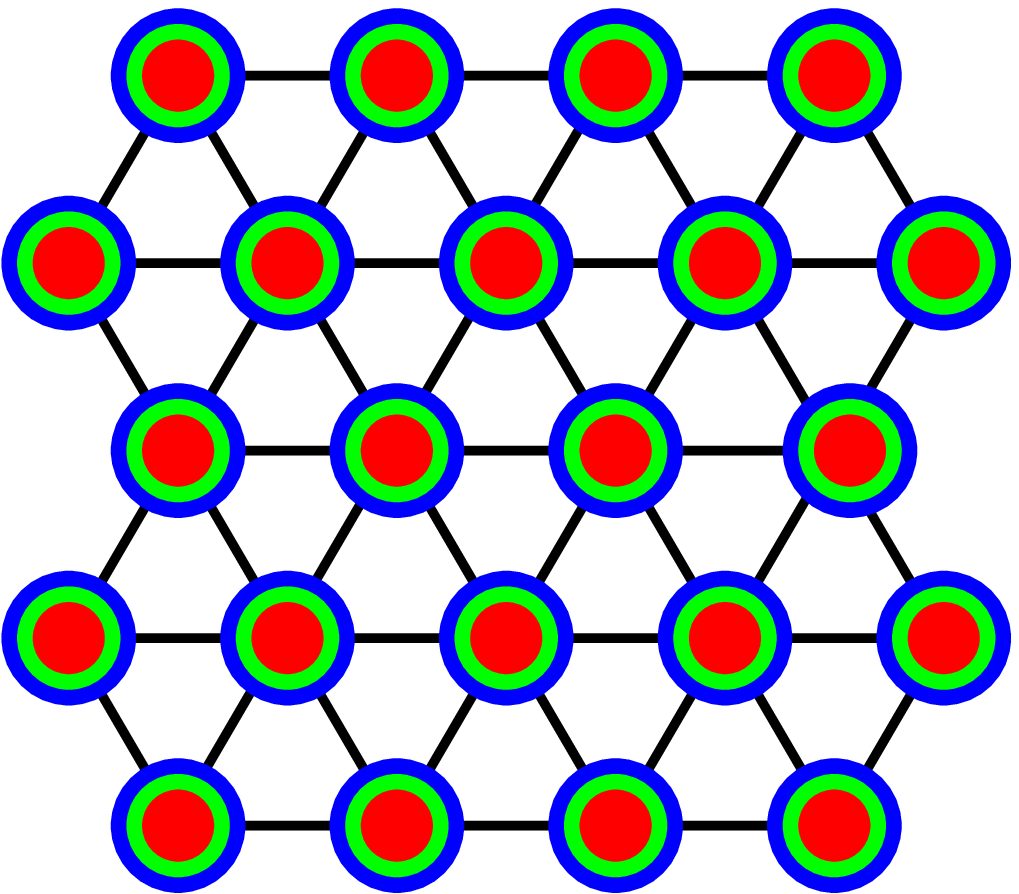}} \centerline{(b)
\hskip 3.5cm  (c)} \vskip .2cm
\mbox{\includegraphics[width=3.5cm]{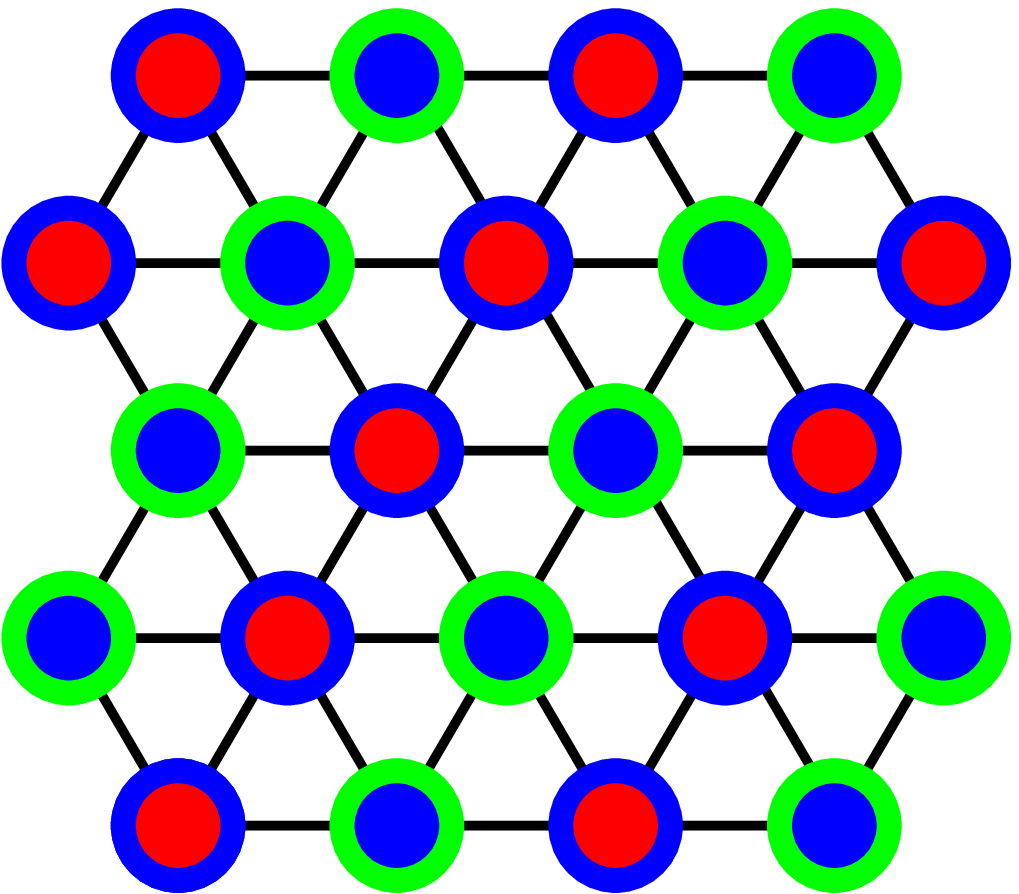}} \hskip .5cm
\mbox{\includegraphics[width=3.5cm]{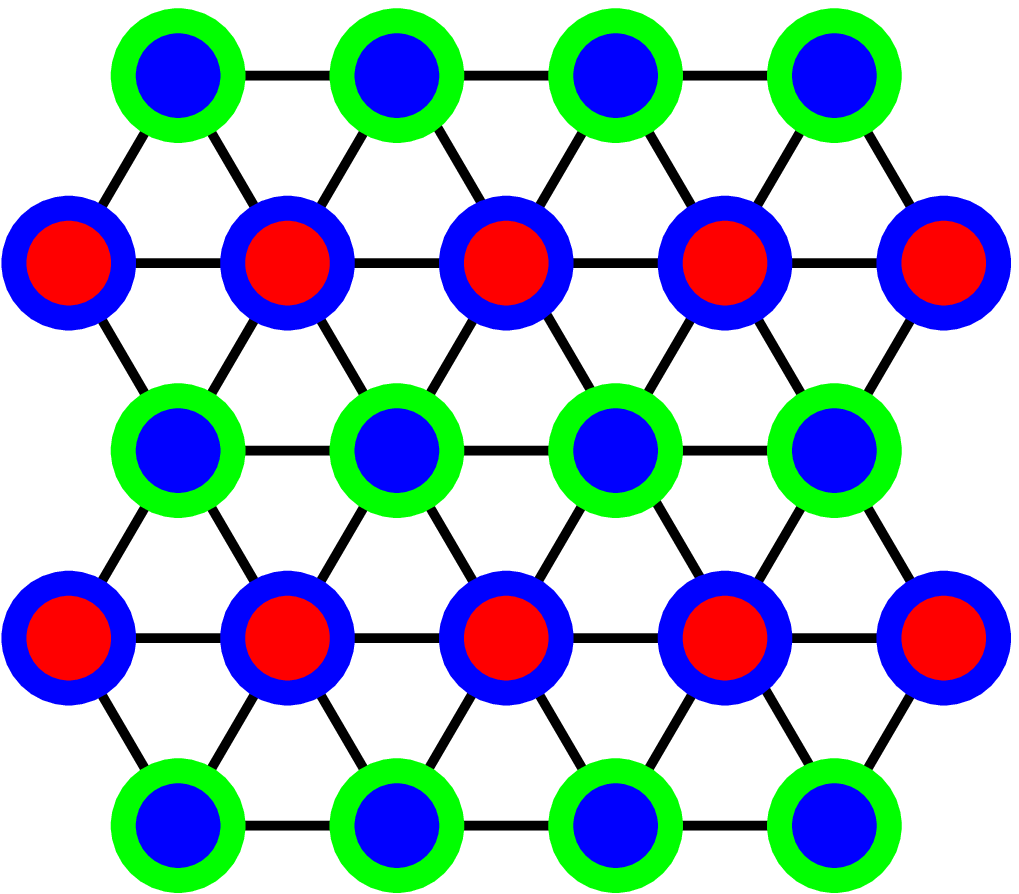}} \centerline{(d)
\hskip 3.5cm  (e)}} \caption{(Color online) Schematic
representation of possible orbital configurations with
superpositions of (a) two orbitals in a two--color state, (b)
three orbitals, (c) two orbitals with equal net weight, and (d)
and (e) two orbitals with differing net weights of all three
orbitals. State (a) has degeneracy $d = 3$, states (b) and (c)
have $d = 1$, and the degeneracies of states (d) and (e) are $d =
6$ and $d = 3$. } \label{fig:mix}
\end{figure}

Turning to orbital states with unequal occupations, motivated by the
tendency of ${\cal H}$ to favor dimer formation in certain limits we
extend our considerations to the possibility of a four--site unit cell
[Figs.~\ref{fig:pure}(e) and \ref{fig:pure}(f)]. More elaborate
three--color unit cells are not considered. In this case the same state
is obtained when the fourth site is occupied by electrons whose orbital
color is any of the other three. Again this state, which breaks rotational
symmetry, differs depending on its orientation relative to the active
hopping axes.

States involving a superposition of either two or three orbitals at
each site can be expected to allow a significantly greater variety of
hopping processes. When either two or three orbital states are partially
occupied at each site (we stress that the condition of Eq.~(\ref{n=1}) is
always obeyed rigorously), one finds the two uniform states represented
in Figs.~\ref{fig:mix}(a) and \ref{fig:mix}(b). These denote the symmetric
wavefunctions $|\psi_2 \rangle = (|\phi_a \rangle + |\phi_b
\rangle)/\sqrt{2}$ and $|\psi_3 \rangle = (|\phi_a \rangle + |\phi_b
\rangle + |\phi_c \rangle)/\sqrt{3}$ at every site, where $|\phi_{\gamma}
\rangle = \gamma^{\dagger} |0 \rangle$. The remaining states shown in
Fig.~\ref{fig:mix} involve only two orbitals per site, but with all
three orbitals partly occupied in the lattice. The average electron
density per site and per orbital is $1/3$ in the state of
Fig.~\ref{fig:mix}(c), while in Figs.~\ref{fig:mix}(d) and \ref{fig:mix}(e)
it is $n_c = \frac12$, $n_a = n_b = \frac14$. The latter two states are
neither unique nor (for general interactions) equivalent to each other,
and represent two classes of states with respective degeneracies 3 and 6.

\subsection{Ordered--state energies: superexchange}
\label{sec:mfsex}

Before analyzing the different possible ordered states for any of the
model parameters, we stress that the spin interactions on a given bond
depend strongly on the orbital occupation of that bond. We begin with
the pure superexchange model $H_s$ (\ref{Hs}), meaning $\alpha = 0$,
for which the question of spin and orbital singlets was addressed in
Sec.~\ref{sec:sex}. Here the spin and orbital scalar products $\langle
{\vec S}_i \cdot {\vec S}_j \rangle$ and $\langle {\vec T}_i \cdot
{\vec T}_j \rangle$ may take only values consistent with long--range
order throughout the system and thus vary between $-1/4$ and $+1/4$.

For a bond on which both electrons occupy active orbitals, one has the
possibility of either FO or AO states. For the FO state, $\langle {\vec
T}_i \cdot {\vec T}_j \rangle = 1/4 = \langle {\vec T}_i \times {\vec
T}_j \rangle$ and $\langle A_{ij} \rangle = \langle B_{ij} \rangle = 1$,
whence the terms of ${\cal H}_s$ can be separated into the physically
transparent form
\begin{eqnarray}
H_{1}^{\rm (FO)} (0) & = & \frac12 J r_1 \left( \frac12 \langle
n_{i\gamma} + n_{j\gamma} \rangle \right) \left( {\vec S}_i \cdot
{\vec S}_j + \frac34 \right) = 0, \nonumber \\ H_{2}^{\rm (FO)}
(0) & = & \frac12 J r_2 \left( 2 - \frac12 \langle n_{i\gamma} +
n_{j\gamma} \rangle \right) \left( {\vec S}_i \cdot {\vec S}_j -
\frac14 \right) \nonumber \\ & = & J r_2 \left( {\vec S}_i \cdot
{\vec S}_j - \frac14 \right), \\ H_{3}^{\rm (FO)} & = & \frac13
J (r_3 - r_2) \left( {\vec S}_i \cdot {\vec S}_j - \frac14 \right),
\nonumber
\label{ensifo}
\end{eqnarray}
specifying a net spin interaction which, because $n_{i\gamma} = 0$, must
be AF if any hopping processes are to occur. In the AO case, $\langle
{\vec T}_i \cdot {\vec T}_j \rangle = - 1/4 = \langle {\vec T}_i \times
{\vec T}_j \rangle$ and $\langle A_{ij}\rangle = \langle B_{ij} \rangle
 = 0$, giving
\begin{eqnarray}
H_1^{\rm (AO)} (0) & = & - \frac12 J r_1 \left( {\vec S}_i \cdot
{\vec S}_j + \frac34 \right), \nonumber \\
H_2^{\rm (AO)} (0) & = & \frac12 J r_2 \left( {\vec S}_i \cdot
{\vec S}_j - \frac14 \right), \\
H_3^{\rm (AO)} (0) & = & 0, \nonumber
\label{ensiao}
\end{eqnarray}
and the spin interaction is constant at $\eta = 0$, with only a weak FM
preference emerging at finite $\eta$. We remind the reader here that
the designations FO and AO continue to be based on the conventional
notation\cite{Fei05} obtained by a local transformation on one bond
site, and in the basis of the original orbitals correspond respectively
to opposite active orbitals and to equal active orbitals. Cases where
only one orbital is active on a bond are by definition AO, but do
contribute a finite spin interaction
\begin{eqnarray}
H_1^1 (0) & = & - \frac14 J r_1 \left( {\vec S}_i \cdot {\vec S}_j
 + \frac34 \right), \nonumber \\
H_2^1 (0) & = &   \frac14 J r_2 \left( {\vec S}_i \cdot {\vec S}_j
 - \frac14 \right), \\
H_3^1 (0) & = &      0, \nonumber
\label{ensi1o}
\end{eqnarray}
which again has only a weak FM tendency at $\eta > 0$. Clearly, when
neither electron may hop, the bond does not contribute a finite energy.

We begin with the uniform, one--color orbital state of
Fig.~\ref{fig:pure}(a), meaning that all bonds are AO by the
definition of the previous paragraph. In two directions both
electrons are active, while in the third none are. The energy
per bond is
\begin{equation}
\label{e1afm}
E_{\rm FM}^{(3a)}(0) =  - \frac{1}{3} J r_1.
\end{equation}
and the spin configuration is FM. However, an antiferromagnetic (AF)
spin configuration on the square lattice defined by the active hopping
directions has energy
\begin{equation}
\label{e1aaf}
E_{\rm AF}^{(3a)}(0) = - \frac{1}{6} J \left( r_1 + r_2 \right),
\end{equation}
from which one observes that all spin states are degenerate at $\eta
 = 0$. The ordered spin state spin is then FM for any finite $\eta$.
We note in passing that the energy per bond for a square lattice
would have the significantly lower value $-\frac12 J$ for the same
${\cal H}_s$ convention, by which is meant the presence of the constants
$+\frac{3}{4}$ and $- \frac14$ in Eq.~(\ref{Hs}). This result is a direct
reflection of the geometrical frustration of the triangular lattice,
an issue to which we return in Sec.~\ref{sec:rhk}.

\begin{figure}[t!]
\mbox{\includegraphics[width=3.7cm]{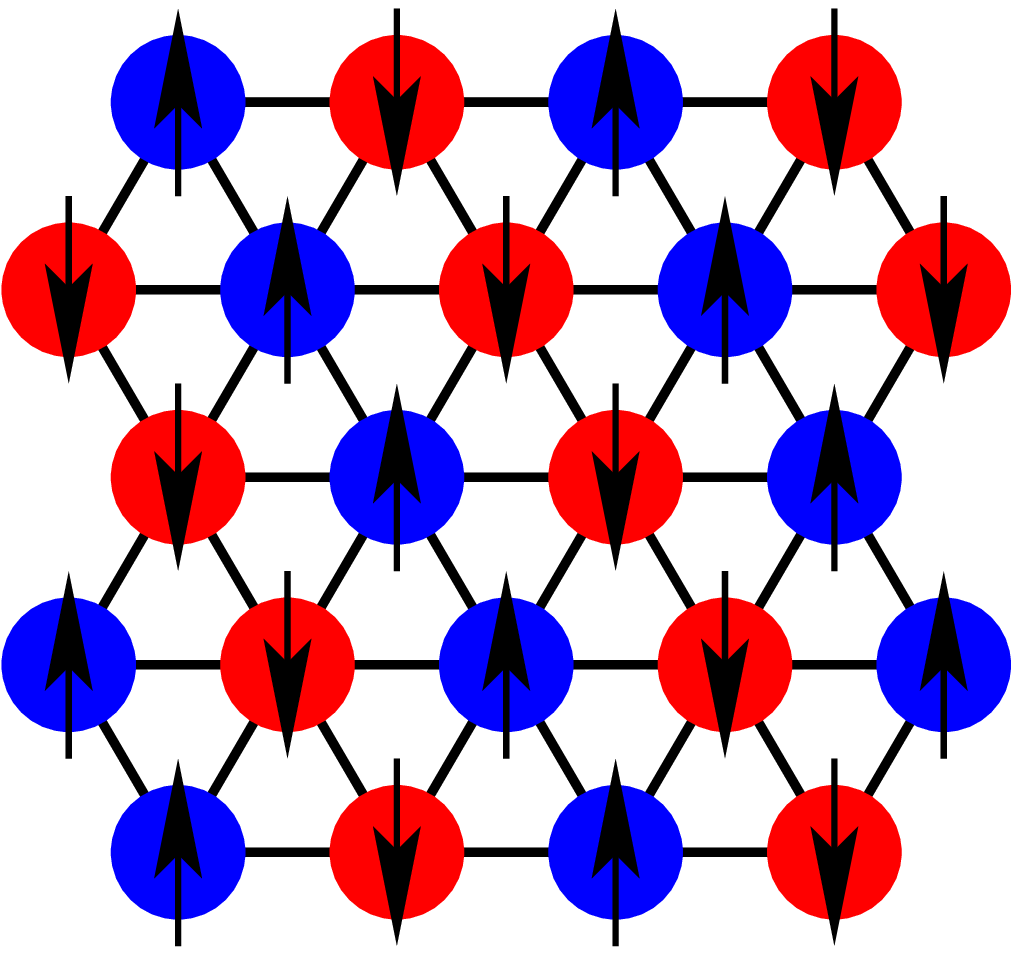}}
\hskip .5cm
\mbox{\includegraphics[width=3.7cm]{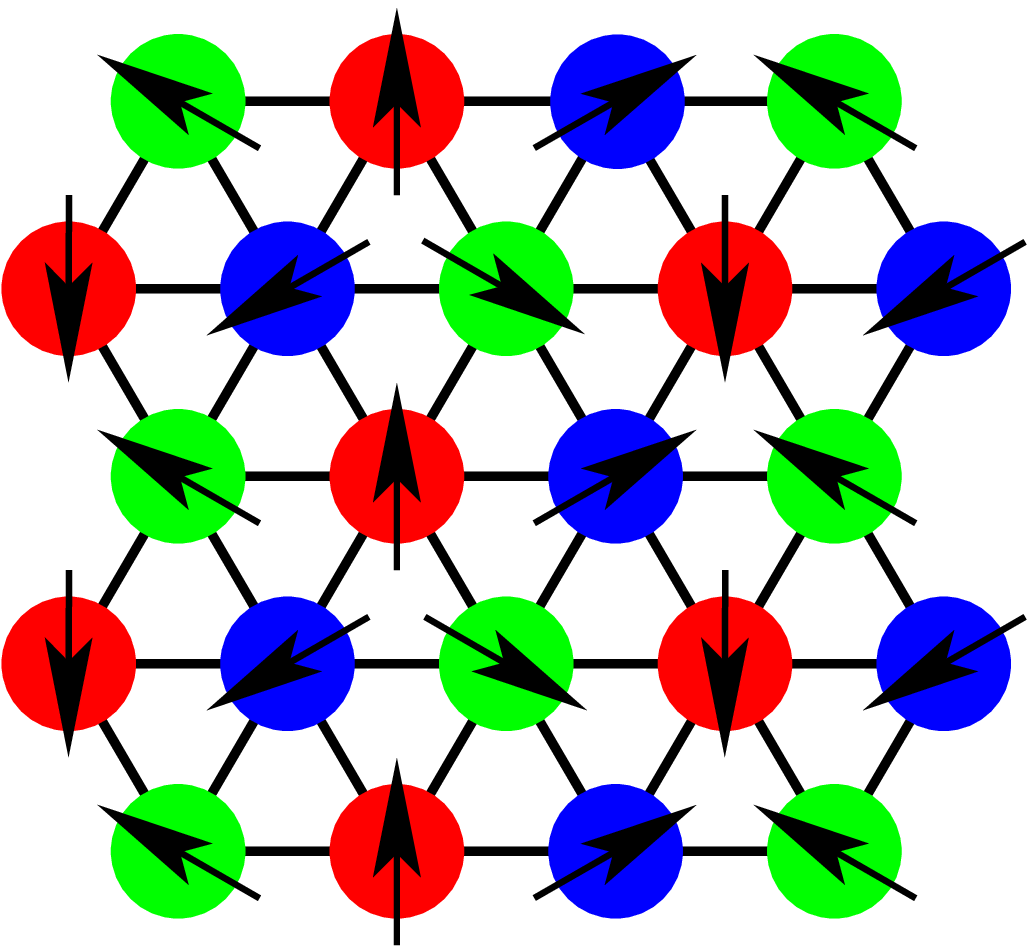}}
\centerline{(a) \hskip 3.5cm  (b)}
\caption{(Color online)
Spin configurations minimizing the total energy of the superexchange
Hamiltonian ${\cal H}_s$ ($\alpha = 0$) for given fixed patterns of
orbital order:
(a) AFF state for the orbital ordering pattern of Fig.~3(c), showing
    how the FM line is selected by the direction (here $b$) giving zero
    frustration;
(b) 60--120$^{\circ}$ ordered spin configuration minimizing the total
    energy for the orbital ordering pattern of Fig.~3(d).}
\label{osc}
\end{figure}

The state of Fig.~\ref{fig:pure}(b) involves one set of (alternating) AO
lines with two active orbitals and two sets of (AO) lines each with one
active orbital. All sets of lines favor FM order at finite $\eta$, with
\begin{equation}
\label{e1bfm}
    E_{\rm FM}^{(3b)}(0) = - \frac{1}{3} J r_1.
\end{equation}
Here the square--lattice state which becomes degenerate at $\eta = 0$, with
\begin{equation}
\label{e1baf}
    E_{\rm AF}^{(3b)}(0) = - \frac{1}{6} J \left( r_1 + r_2 \right),
\end{equation}
is more accurately described as one with two lines of AF spins and one
of FM spins [Fig.~\ref{osc}(a)], and will be denoted henceforth as AFF.

The state of Fig.~\ref{fig:pure}(c) involves one set of FO lines with
two active orbitals, one set of lines with one active orbital, one half
set of AO lines with two active orbitals and one half set of inactive
lines. The two--active FO lines will favor AF order, while the AO and
the one--active lines will favor FM order only at $\eta > 0$, giving
\begin{equation}
\label{e1caf}
E_{\rm AFF}^{(3c)}(0) = - \frac{1}{72}J\; ( 9 r_1 + 11 r_2 + 4 r_3 )
\end{equation}
from the AFF configuration, but with 2 equivalent directions for the
FM line. At $\eta = 0$ the energy is again $- \frac{1}{3}J$. Both
$E_{\rm AF}^{(3b)} (0)$ and $E_{\rm AFF}^{(3c)} (0)$ can be regarded
as the energy of an unfrustrated system, in the sense that the spin
order enforced in any one direction by the orbital configuration at
no time denies the system the ability to adopt the energy--minimizing
configuration in other directions. However, at finite $\eta$ the
configurations shown in Figs.~\ref{fig:pure}(b) and \ref{fig:pure}(c)
will be penalized relative to the uniform (AO) order of
Fig.~\ref{fig:pure}(a) due to the presence of AF bonds.

We insert here an important observation: the orbital state of
Fig.~\ref{fig:pure}(c) also admits the formation of 1D AF
Heisenberg spin chains on the FO ($b$--axis) lines. The energy
per bond of such a state includes constant interchain contributions
which are independent of the spin state ($\langle {\vec S}_i \cdot
{\vec S}_j \rangle = 0$) on these bonds. Of these interchain bonds,
1/4 are FO with two active orbitals and 1/2 have one active orbital.
One finds
\begin{equation}
\label{e1c1d}
E_{\rm 1D}^{(3c)}(0) = - \frac{1}{9}J \ln2\; ( 2r_2 + r_3 )
                       - \frac{1}{24} J (3 r_1 + r_2) ,
\end{equation}
which gives $E_{\rm 1D}^{(3c)}(0) = - 0.3977 J$ at $\eta = 0$. This
energy is significantly lower than that of an ordered magnetic state,
a result showing that the kinetic energy gained from resonance processes
on the chains is far more significant than minimal potential energy gain
obtainable from an ordering of the magnetic moments on the active
interchain bonds which are active, and thus provides strong evidence
in favor of the hypothesis that any ordered state will ``melt'' to a
quantum disordered one in this system. We will return to this issue
below.

For the two--color superposition [Fig.~\ref{fig:mix}(a)], one set of
bonds always has two active orbitals, but with equal probability of
being FO or AO, while the other two sets of bonds have a 1/4 probability
of having two active orbitals, which are FO, or a 1/2 probability of
having one active orbital (and a 1/4 probability of having none). Under
these circumstances, the net system Hamiltonian can be expressed by
summing over all the possible orbital states, although this is not
necessarily a useful exercise when the spin state may not be isotropic.
By inserting the three most obvious ordered spin states, FM, AF
(meaning here the AF state of the triangular lattice with 120$^{\circ}$
bond angles and $\langle{\vec S}_i \cdot {\vec S}_j\rangle =
- \frac{1}{8}$) and AFF, the candidate energies are
\begin{eqnarray}
      E_{\rm FM}^{(4a)}(0) & = &  - \frac{1}{6} J r_1,   \\
      E_{\rm AF}^{(4a)}(0) & = & E_{\rm AFF}^{(4a)}(0) \; = \;
       - \frac{1}{48} J\; (5 r_1 + 7 r_2 + 2 r_3). \nonumber
\end{eqnarray}
The coincidence for the results for the AF and AFF ordered states
in this case is an accidental degeneracy. The final energy $E_{\rm
AF(F)}^{(4a)} = -\frac{7}{24}J$ at $\eta = 0$ shows that both states
are compromises, and it is not possible to put all bonds in their
optimal spin state simultaneously. This arises because of the presence
of two--active FO components in all three lattice directions, and will
emerge as a quite generic feature of superposition states, albeit not
one without exceptions.

In general there is no compelling reason (given by ${\cal H}$ for any
value of $\alpha$) to expect that two--color superpositions of this
type may be favorable. While the 120$^{\circ}$ state of a
triangular--lattice antiferromagnet is one compromise within a space of
SU(2) operators, this type of symmetry--breaking is not relevant within
the orbital sector, where there are three colors and the two--color
subsector of active orbitals in the $\alpha = 0$ limit changes as a
function of the bond orientation.

In the equally weighted three--color state [Fig.~\ref{fig:pure}(d)],
all bonds are FO and it is easy to show that 1/3 of them (arranged as
isolated triangles) have two active orbitals while the other 2/3 have
one active orbital. The two--active bonds favor AF order while the
one--active bonds have only a weak preference for FM order at finite
$\eta$. In this case the problem becomes frustrated and is best
resolved by a kind of AF state on the triangular lattice where the
strong triangles have 120$^{\circ}$ angles and alternating triangles
have spins either all pointing in or all pointing out
[Fig.~\ref{osc}(b)]; then 2/3 of the intertriangle bonds have
60$^{\circ}$ angles while the other 1/3 have 120$^{\circ}$ angles.
The energy of this state is
\begin{equation}
\label{afm1d}
    E^{(3d)}(0) = - \frac{1}{144} J\; (19 r_1 + 17 r_2 + 6 r_3),
\end{equation}
and $E^{(3d)}(0) = - \frac{7}{24} J$ at $\eta = 0$, a value again inferior
to the optimal energy due to the manifest spin frustration.

In the state of Fig.~\ref{fig:pure}(e), the only AO bonds (1/6 of the
total) contain inactive orbitals. Of the remaining bonds, 3/6 have two
active FO orbitals (in all three directions) and 2/6 have one active
orbital. Once again the system is composed of strongly coupled triangles,
but this time in a square array and with strong coupling in their basal
direction by one set of two--active FO bonds. Possible competitive
spin--ordered states would be AF or AFF, with energies
\begin{eqnarray}
E_{\rm AF}^{(3e)}(0) & = & -  \frac{1}{96} J\; (5 r_1 + 15 r_2 + 6 r_3), \\
E_{\rm AFF}^{(3e)}(0) & = & - \frac{1}{288} J\; (15 r_1 + 35 r_2 + 16 r_3).
\nonumber
\end{eqnarray}
The lowest energy is obtained for 120$^{\circ}$ AF order, with the
frustrated value $E_{\rm AF}^{(3e)}(0) = - \frac{13}{48}J$ for $\eta = 0$.

For the state in Fig.~\ref{fig:pure}(f) the FO bonds (1/6) and only 1/6 of
the AO bonds have two active orbitals, while the other 2/3 of the bonds
have one active orbital. In this case
\begin{eqnarray}
    E_{\rm FM}^{(3f)}(0) & = & - \frac14 J r_1, \nonumber \\
    E_{\rm AF}^{(3f)}(0) & = & - \frac{1}{96} J\; (15 r_1 + 13 r_2 + 2 r_3), \\
    E_{\rm AFF}^{(3f)}(0) & = & -\frac{1}{192} J\; (36 r_1 + 17 r_2 + 5 r_3),
    \nonumber
\end{eqnarray}
leading again to an AF spin state. At $\eta = 0$ one has
$E_{\rm AF}^{(3f)}(0) = - \frac{5}{16}J$, {\it i.e.}~relatively
weaker frustration.

Turning now to three--color superpositions, the ``uniform'' orbital
state [Fig.~\ref{fig:mix}(b)] is one in which on every bond there is
a probability 2/9 of having two active FO orbitals, 2/9 for two active
AO orbitals, 4/9 of one active orbital and 1/9 of no active orbitals.
The appropriately weighted bond interaction strengths may be summed to
give the net interaction, which for the three spin states considered
results in the energies
\begin{eqnarray}
E_{\rm FM}^{(4b)}(0) & = &  - \frac{2}{9} J r_1, \nonumber    \\
E_{\rm AF}^{(4b)}(0) & = & - \frac{1}{36} J\; (5 r_1 + 5 r_2 + r_3), \\
E_{\rm AFF}^{(4b)}(0) & = & - \frac{1}{81} J\; (12 r_1 + 10 r_2 + 2 r_3),
\nonumber
\end{eqnarray}
and thus the AF state is lowest, with the value $E_{\rm AF}^{(4b)}(0)
 = - \frac{11}{36} J$ at $\eta = 0$. While this orbital configuration
does not attain the minimal energy of $- \frac{1}{3} J$, it is a close
competitor: although it involves every bond, the fractional probabilities
of each being in a two--active state mean that it cannot maximize
individual bond contributions. However, we will see in Sec.~IIID that
state (4b) lies lowest over much of the phase diagram ($0 < \alpha < 1$)
as a result of the contributions from mixed terms.

For states with unequal site occupations, in Fig.~\ref{fig:mix}(c) one has
a situation where on 1/3 of the bonds (arranged in separate triangles)
there is a 1/4 probability of two active FO orbitals and a 1/2 probability
of one active orbital, while on the remaining 2/3 of the bonds there is a
1/4 probability of two active AO orbitals, 1/4 of two active FO orbitals
and 1/2 of having one active orbital. On computing the net energies for
the three standard spin configurations, one obtains
\begin{eqnarray}
E_{\rm FM}^{(4c)}(0) & = & -  \frac{5}{24} J r_1 , \nonumber              \\
E_{\rm AF}^{(4c)}(0) & = & - \frac{1}{192} J\; (25 r_1 + 25 r_2 + 8 r_3), \\
E_{\rm AFF}^{(4c)}(0) & = & - \frac{1}{216} J\; (30 r_1 + 25 r_2 + 8 r_3),
\nonumber
\end{eqnarray}
where the AF state with $E_{\rm AF}^{(4c)}(0) = -\frac{29}{96} J$ is the
lowest at $\eta = 0$. However, this state is also manifestly frustrated.

In the unequally weighted state of Fig.~\ref{fig:mix}(d), the problem is
best considered once again as lines of different bond types. Here 1/6 of
the lines have two active orbitals (1/2 FO and 1/2 AO), 1/6 of the lines
have probability 1/4 of two active orbitals (AO) and 1/2 of one active
orbital, 1/3 of the lines have probability 1/4 of two active FO orbitals,
1/4 of two active AO orbitals and 1/2 of one active orbital, and the
remaining 1/3 of the lines have probability 1/4 of two active orbitals
(FO) and 1/2 of one active orbital. The ordered spin states yield the
energies
\begin{eqnarray}
E_{\rm FM}^{(4d)}(0) & = & -  \frac{5}{24} J r_1, \nonumber \\
E_{\rm AF}^{(4d)}(0) & = & -  \frac{1}{192} J (25 r_1 + 27 r_2 + 6 r_3), \\
E_{\rm AFF}^{(4d)}(0) & = & - \frac{1}{144} J (21 r_1 + 17 r_2 + 4 r_3),
\nonumber
\end{eqnarray}
whence it is again the AF state, with a small degree of unrelieved
frustration in its energy $E_{\rm AF}^{(4d)}(0) = - \frac{29}{96} J$,
which lies lowest at $\eta = 0$.

Finally, the state of Fig.~\ref{fig:mix}(e) has the orbital pattern of
Fig.~\ref{fig:mix}(d) rotated in such a way that the number of active
orbitals in different bond directions is changed. Now 1/3 of the bonds
have probabilities 1/4 of two active orbitals (AO) and 1/2 of one
active orbital, while the remaining 2/3 have probabilities 1/4 of two
active orbitals (FO), 1/4 of two active orbitals (AO) and 1/2 of one
active orbital. The ordered--state energies are
\begin{eqnarray}
E_{\rm FM}^{(4e)}(0) & = & -  \frac{1}{4} J r_1, \nonumber \\
E_{\rm AF}^{(4e)}(0) & = & - \frac{1}{96} J (15 r_1 + 13 r_2 + 2 r_3),  \\
E_{\rm AFF}^{(4e)}(0) & = & - \frac{1}{36} J (6 r_1 + 5 r_2 + r_3),
\nonumber
\end{eqnarray}
of which the AFF states lies lowest at $\eta = 0$, achieving the
unfrustrated value $E_{\rm AFF}^{(4e)}(0)= - \frac{1}{3}J$. That it is
possible to obtain this energy in an orbital superposition is because
of the absence of FO bond contributions in one direction, which can
then be chosen to be FM.

The results of this section and the conclusions one may draw from them
are summarized in Subsec.~\ref{sec:mfasum} below.

\subsection{ Ordered--state energies: direct exchange }
\label{sec:mfdex}

In the limit of only direct exchange, the analysis is somewhat simpler.
The Hamiltonian is ${\cal H}_d$ of Eq.~(\ref{Hd}), and in this case a
particle on any site is active in only one direction, which leads to
the immediate observation that in a static orbital configuration it is
never possible to have, on average, active exchange processes on more
than 2/3 of the bonds. For simplicity we repeat the Hamiltonian for the
two cases of AO order between sites, in which case by definition at most
one of the orbitals is active, and FO order between sites, which is
restricted to the case where neighboring sites have the same orbital
color and the correct bond orientation. We stress that in this
subsection the definitions FO and AO are entirely conventional, as
the local transformation of Sec.~\ref{sec:sex} is not relevant at
$\alpha= 1$, and thus the designation FO implies orbitals of the same
color, and AO orbitals of different colors. One obtains the expressions
\begin{eqnarray}
\label{dexabe}
{\cal H}^{\rm (AO)} (1) & = & \frac14 J \left[ - r_1
\left( {\vec S}_i \cdot {\vec S}_j + \frac34 \right) \right. \nonumber \\
& & \hskip .7cm + r_2 \left. \left(
{\vec S}_i \cdot{\vec S}_j - \frac14 \right) \right], \\
{\cal H}^{\rm (FO)} (1) & = & \frac{1}{3} J\; (2 r_2 + r_3)
\left( {\vec S}_i\cdot {\vec S}_j - \frac14 \right),
\end{eqnarray}
which in the $\eta = 0$ limit reduce to the forms
\begin{eqnarray}
\label{dexabf}
{\cal H}^{\rm (AO)} (1) & = & - \frac14 J, \\
{\cal H}^{\rm (FO)} (1) & = & J \left( {\vec S}_i \cdot {\vec S}_j
- \frac14 \right).
\end{eqnarray}

It is clear (Sec.~II) that for a single bond, the most favorable state
is a spin singlet, which would contribute energy $-J$, but at the possible
expense of placing all of the neighboring bonds in suboptimal states. The
very strong preference for such singlet bonds means that any mean--field
study of the minimal energy is incomplete without the consideration of
dimerized (or valence--bond) states (Sec.~\ref{sec:dim}). The analysis of
this section can be considered as elucidating the optimal energies to be
gained from long--ranged magnetic and orbital order on these bonds, where
the optimal energy of any one is $-\frac12 J$. Also as noted in Sec.~II,
any active AO bond gains an exchange energy ($-\frac14 J$) simply
because it does not prevent one of the two particles from performing
virtual hopping processes, and this we term ``avoided blocking''. In
the limit of zero Hund exchange, these will give a highly degenerate
manifold of all possible spin states, from which FM states are
selected at finite $\eta$.

We begin again with one--color state of Fig.~\ref{fig:pure}(a),
which we denote henceforth as (3a). Only one set of lattice bonds
has finite interactions, all FO, and therefore the system behaves
as a set of AF Heisenberg spin chains with energy per bond
\begin{equation}
\label{af2a1}
    E_{\rm AF1D}^{(3a)}(1) = - \frac{1}{9} J \ln 2\; (2r_2 + r_3),
\end{equation}
whence $E_{\rm AF1D}^{(3a)}(1) = - 0.2310 J$ at $\eta = 0$.

In state (3b), the FO lines do not correspond to active hopping
directions. The remaining two directions then form an AO square
lattice with
\begin{equation}
\label{fm2b1}
    E_{\rm FM}^{(3b)}(1) = - \frac{1}{6} J r_1.
\end{equation}
This can be called a ``pure avoided--blocking'' energy. The spins are
unpolarized at $\eta = 0$, where all bond spin states are equivalent,
but any finite $\eta$ will select FM order (hence the notation). We
will see in the remainder of this section that $E = - \frac{1}{6} J$
is the optimal energy obtainable by a 2D ordered state in the
direct--exchange limit ($\alpha = 1$), where the net energy is
generically higher than at $\alpha = 0$ quite simply because there
are half as many hopping channels. Thus the ``melting'' of such
ordered states into quasi--1D states becomes clear from the outset,
and can be understood due to the very low connectivity of the
active hopping network on the triangular lattice.

In state (3c), one of the FO lines is active, and forms AF Heisenberg
spin chains. Electrons in the other FO line are active only in a
cross--chain direction, where their bonds are AO, and gain
avoided--blocking energy, whence
\begin{equation}
\label{af2c1}
   E_{\rm AF}^{(3c)}(1) = - \frac{1}{12} J (2 \ln 2 + 1) \;\; = \;\;
   - 0.1988 J
\end{equation}
at $\eta = 0$. As in the preceding subsection, the coherent state of
each Heisenberg chain is not altered by the presence of additional
electrons from other chains executing virtual hopping processes onto
empty orbitals of individual sites. The spin chains remain uncorrelated
and only quasi--long--range--ordered until a finite value of $\eta$,
where FM spin polarization and a long--range--ordered state are favored.

In the two--color superposition (4a), 1/3 of the bonds are inactive,
while on the other 2/3 one has probability 1/4 of two active electrons
(FO), 1/2 of one active (AO) and 1/4 of two inactive electrons. In this
case, one obtains an effective square lattice on which an AF spin
configuration is favored by the FO processes, with
\begin{equation}
\label{af3a1}
    E_{\rm AF}^{(4a)}(1) = - \frac{1}{72}J\; (3 r_1 + 7 r_2 + 2 r_3),
\end{equation}
so again $E_{\rm AF}^{(4a)}(1) = - \frac{1}{6}J$ at $\eta = 0$.

The uniform three--color state (3d) maximizes AO bonds, but 1/3 of the
bonds on the lattice remain inactive. Thus
\begin{equation}
\label{fm2d1}
    E_{\rm FM}^{(3d)}(1) = - \frac{1}{6} J r_1,
\end{equation}
and Hund exchange will select the FM spin state.

The three--color state (3e) has FO lines oriented in their active
direction and will, as in state (3c), form Heisenberg chains linked by
bonds with AO order. While the geometry of the interchain coupling can
differ depending on the orbital alignment in the inactive chains, it
does not create a frustrated spin configuration and the net energy is
$E_{\rm AF}^{(3e)}(1) = E_{\rm AF}^{(3c)}(1)$. The state (3f) has only
inactive FO lines and so gains only avoided--blocking energy, from 2/3
of the bonds in the system, whence $E_{\rm FM}^{(3f)}(1)
 = E_{\rm FM}^{(3d)}(1)$.

In the uniform three--color superposition (4b), every bond has probability
1/9 of containing two active electrons (FO), 4/9 of one active electron
and 4/9 of remaining inactive. For the three different ordered spin
configurations considered in Subsec.~\ref{sec:mfsex} the energies are
\begin{eqnarray}
E_{\rm FM}^{(4b)}(1)  & = & -   \frac{1}{9} J r_1, \nonumber            \\
E_{\rm AF}^{(4b)}(1)  & = & -  \frac{1}{72} J\; (5 r_1 + 5 r_2 + r_3),     \\
E_{\rm AFF}^{(4b)}(1) & = & -  \frac{1}{81} J\; (6 r_1 + 5 r_2 + r_3),
\nonumber
\end{eqnarray}
and one finds the energy $E_{\rm AF}^{(4b)}(1) = - \frac{11}{72} J$ for
the 120$^0$ AF state at $\eta = 0$.

The three--color state (4c) is one in which 1/3 of the bonds (arranged
on isolated triangles) have probability 1/4 of being in a state with two
active electrons and 1/2 of containing one active electron, while on the
other 2/3 of the bonds there is simply a 1/2 probability of one active
orbital. The respective energies are
\begin{eqnarray}
E_{\rm FM}^{(4c)}(1)  & = & - \frac{1}{8} J r_1, \nonumber                \\
E_{\rm AF}^{(4c)}(1)  & = & - \frac{1}{192} J\; (15 r_1 + 13 r_2 + 2 r_3),  \\
E_{\rm AFF}^{(4c)}(1) & = & - \frac{1}{216} J\; (18 r_1 + 13 r_2 + 2 r_3).
\nonumber
\end{eqnarray}
At $\eta = 0$, the energy $E_{\rm FM}^{(4c)}(1) = - \frac{5}{32} J$
is minimized by a 120$^{\circ}$ state on the triangles, which are also
isolated magnetically in this limit. Finite values of $\eta$ result in
FM interactions between the triangles, and a frustrated problem in the
spin sector which by inspection is resolved in favor of a net FM
configuration only at large $\eta$ ($\eta > 0.23)$.

Finally, the three--color states (4d) and (4e) yield two possibilities
in the $\alpha = 1$ limit, namely where one of the minority colors is
aligned with its active direction and where neither is. In the former
case,
\begin{eqnarray}
E_{\rm FM}^{(4d)}(1)  & = & -  \frac{5}{48} J r_1,  \nonumber             \\
E_{\rm AF}^{(4d)}(1)  & = & - \frac{1}{384} J\; (25 r_1 + 27 r_2 + 6 r_3),\\
E_{\rm AFF}^{(4d)}(1) & = & -  \frac{1}{96} J\; (7 r_1 + 7 r_2 + 2 r_3),
\nonumber
\end{eqnarray}
and the lowest energy $E_{\rm AFF}^{(4d)}(1)= - \frac{1}{6} J$ at
$\eta = 0$ is given by the directionally anisotropic AFF spin
configuration. This is because 1/2 of the lines, in two of the three
directions, have some AF preference from their 1/4 probability of
containing two active orbitals, while the third direction has no
preference at $\eta = 0$, and in any case favors FM spins at $\eta > 0$.
In the latter case, the only AF tendencies arise along lines in a single
direction, but avoided--blocking energy is sufficient to exclude the
possibility of a Heisenberg chain state. Here
\begin{eqnarray}
E_{\rm FM}^{(4e)}(1)  & = & -   \frac{1}{8} J r_1,  \nonumber              \\
E_{\rm AF}^{(4e)}(1)  & = & - \frac{1}{192} J\; (15 r_1 + 13 r_2 + 2 r_3), \\
E_{\rm AFF}^{(4e)}(1) & = & -  \frac{1}{72} J\; (6 r_1 + 5 r_2 + r_3),
\nonumber
\end{eqnarray}
whence $E_{\rm AFF}^{(4e)}(1)= - \frac{1}{6} J$ at $\eta = 0$, in fact
with two degenerate possibilities for the orientation of the FM line.

\subsection{ Ordered--state energies: $\alpha = 0.5$ }
\label{sec:mfa05}

To illustrate the properties of the model in the presence of
finite direct and superexchange contributions, {\it i.e.}~at
intermediate values of $\alpha$, we consider the point $\alpha =
0.5$. As shown in Sec.~II, there is no special symmetry at this
point, because the contributions from diagonal and off--diagonal
hopping remain intrinsically different. States with long--ranged
orbital (and spin) order at $\alpha = 0.5$ are mostly very easy
to characterize, because all virtual processes, of both types,
allowed by the given configuration are able to contribute in full
to the net energy. For the many of the states considered in this
section, the energetic calculation for $\alpha = 0.5$ is merely
an exercise in adding the $\alpha = 0$ and $\alpha = 1$ results
with equal weight. Exceptions occur for superposition states
gaining energy from processes contained in $H_m$ [Eq.~(\ref{Hm})],
and are in fact decisive here. Because these terms involve explicitly
a finite density of orbitals of all three colors on the bond in
question, with the active diagonal color represented on both
sites, only for states (4b), (4c), and (4d), but not (4e)
[Figs.~\ref{fig:mix}(b--e)], will it be necessary to consider
this contribution.

For state (3a), in two directions both electrons are active by
off--diagonal hopping, while in the third both may hop diagonally.
Diagonal hopping favors an AF spin configuration, while the
off--diagonal hopping bonds have only a weak preference (by Hund
exchange) for FM order. The ordered--state spin solution is then
a doubly degenerate AFF state with energy per bond
\begin{equation}
E^{(3a)}(0.5) = - \frac{1}{72} J\; (9 r_1 + 7 r_2 + 2 r_3),
\label{e2ap5}
\end{equation}
giving $E^{(3a)}(0.5) = - \frac14 J$ at $\eta = 0$. We remind the
reader that the prefactor of the superexchange and direct exchange
contributions is only half as large as in Subsecs.~\ref{sec:mfsex}
and \ref{sec:mfdex} [Eq.~(\ref{som})], so the overall effect of
additional hopping processes in this state is in fact an
unfrustrated energy summation. We also comment that, exactly at
$\eta = 0$, there is no obvious preference for any magnetic order
between the diagonal--hopping chains. Only at unrealistically
large values of $\eta$ would the system sacrifice this
diagonal--hopping energy to establish a square--lattice FM state.
At finite $\eta$, the one--color orbital state represents a compromise
between competing spin states preferred by the two types of hopping
contribution.

State (3b) has no diagonal--hopping chains, and these processes
therefore enforce only a weak preference for a FM square lattice.
Because the off--diagonal hopping processes also favor FM order at
finite $\eta$ (Subsec.~\ref{sec:mfsex}), the two types of contribution
cooperate and one obtains
\begin{equation}
E^{(3b)}(0.5) = - \frac14 J r_1.
\label{e2bp5}
\end{equation}

State (3c) contains one half set of diagonal--hopping chains,
which fall along one of the directions which in the spin state
favored by the off--diagonal hopping processes could be FM or AF;
this degeneracy will therefore be broken. The other half set of
chains will gain only avoided--blocking energy from diagonal
processes, which will take place in the FM direction and thus
cause no frustration even at finite $\eta$. One obtains
\begin{equation}
E^{(3c)}(0.5) = - \frac{1}{144} J \left( 3 r_1 + 7 r_2 + 2 r_3
\right),
\label{e2cp5}
\end{equation}
and thus $E^{(3c)}(0.5) = - \frac14 J$ at $\eta = 0$ from this AFF
configuration. The additive contributions from superexchange and
direct exchange remove the possibility that Heisenberg--chain states
in either of the directions favored separately by off--diagonal
(Sec.~IIIB) or diagonal (Sec.~IIIC) hopping could result in an
overall lowering of energy.

As in Subsec.~\ref{sec:mfdex}, in the two--color superposition (4a) the
diagonal hopping processes are optimized by an AFF spin configuration.
Although this is one of the degenerate states minimizing the
off--diagonal Hamiltonian, the directions of the FM lines do
not match. Insertion of the four possible spin states yields
\begin{eqnarray}
E_{\rm FM}^{(4a)}(0.5) & = & - \frac{1}{8} J r_1, \nonumber \\
E_{\rm AF}^{(4a)}(0.5) & = & - \frac{1}{96} J\; (8 r_1 + 10 r_2 + 3
r_3), \nonumber \\
 E_{\rm AFF(0)}^{(4a)}(0.5) & = & - \frac{1}{72} J\; (6 r_1 + 7
r_2 + r_3), \\
E_{\rm AFF(1)}^{(4a)}(0.5) & = & -\frac{1}{144} J\; (12 r_1
 + 14 r_2 + 3 r_3), \nonumber
\label{e3ap5}
\end{eqnarray}
whence the lowest final energy is $E_{\rm AF}^{(4a)} (0.5) = - \frac{7}
{32} J$ at $\eta = 0$. As noted in the previous sections for this spin
configuration, the optimal energy for all bonds is not attainable
within the off--diagonal hopping sector, and the addition of the
(small) diagonal--hopping contribution causes little overall change.

The equally weighted three--color state (3d) has no lines of
diagonal--hopping bonds, and in fact these contribute only
avoided--blocking energy on the bonds between the strong triangles
defined by the off--diagonal problem, adding to the weak propensity
for FM intertriangle bonds arising only from the Hund exchange. The
diagonal processes can be taken only to alter this energy, and not to
promote any tendency towards an alteration of the spin state, whose
energy is then
\begin{equation}
E^{(3d)}(0.5) = - \frac{1}{144} J\; (19 r_1 + 11 r_2 + 3 r_3),
\label{e2dp5}
\end{equation}
with $E^{(3d)}(0.5) = - \frac{11}{48} J$ at $\eta = 0$.

State (3e) is already frustrated in the off--diagonal sector, and
diagonal--hopping processes contribute primarily on otherwise inactive
bonds without changing the frustration conditions. For the two candidate
spin configurations
\begin{eqnarray}
E_{\rm AF}^{(3e)}(0.5) & = & - \frac{1}{96} J\; (5 r_1 + 12 r_2 + 4 r_3),
\nonumber \\ E_{\rm AFF}^{(3e)}(0.5) & = & - \frac{1}{192} J\; (11 r_1
 + 19 r_2 +  8 r_3),
\label{e2ep5}
\end{eqnarray}
a competition won by the 120$^{\circ}$ AF--ordered state with
$E_{\rm AF}^{(3e)}(0.5) = - \frac{7}{32} J$ at $\eta = 0$.

State (3f) lacks active lines of diagonal--hopping processes, and
thus the avoided--blocking energy may be added simply to the results
for the off--diagonal sector, giving
\begin{eqnarray}
E_{\rm FM}^{(3f)}(0.5) & = & - \frac{5}{24} J r_1, \nonumber \\
E_{\rm AF}^{(3f)}(0.5) & = & - \frac{1}{192} J\; (25 r_1 + 19 r_2 + 2 r_3), \\
E_{\rm AFF}^{(3f)}(0.5) & = & - \frac{5}{384} J\; (12 r_1 + 5 r_2 + r_3),
\nonumber
\label{e2fp5}
\end{eqnarray}
or a minimum of $E_{\rm AF}^{(3f)}(0.5) = - \frac{23}{96} J$ at $\eta = 0$.

In the uniform three--color superposition (4b), on every bond there is
a probability 4/9 of having only off--diagonal hopping processes, 2/9
for 2 active FO orbitals and 2/9 for two active AO orbitals, a probability
1/9 of having only diagonal hopping processes, and a probability 4/9 of
other processes. These last include the contributions from one active
diagonal or off--diagonal electron, and mixed processes contained in the
Hamiltonian ${\cal H}_m$ (\ref{Hm}); none of these three possibilities
favors any given bond spin configuration other than a FM orientation at
finite $\eta$. The net energy contributions are
\begin{eqnarray}
E_{\rm FM}^{(4b)}(0.5) & = & - \frac{2}{9} J r_1, \nonumber \\
E_{\rm AF}^{(4b)}(0.5) & = & - \frac{1}{144} J \; (20 r_1 + 18 r_2 + 3 r_3), \\
E_{\rm AFF}^{(4b)}(0.5) & = & - \frac{1}{54} J \; (8 r_1 + 6 r_2 + r_3),
\nonumber
\label{e3bp5}
\end{eqnarray}
and thus the AF state is lowest, with $E_{\rm AF}^{(4b)}(0.5) =
 - \frac{41}{144} J$ at $\eta = 0$. While this energy differs from that
for the AFF spin configuration by only $\frac{1}{144} J$, its crucial
property is that it lies below the value $- \frac14 J$ obtained by
direct summation of the superexchange and direct--exchange contributions.

For this orbital configuration, all three spin states gain a net energy
of $- \frac{1}{18} J$ at $\eta = 0$ from mixed processes, and these are
sufficient, as we shall see, to reduce the otherwise partially frustrated
ordered--state energy to the global minimum for this value of $\alpha$.
By a small extension of the calculation, the energy of the 120$^0$ AF
spin state may be deduced at $\eta = 0$ for all values of $\alpha$, and
is given by
\begin{equation}
E_{\rm AF}^{(4b)}(\alpha) = - \frac{1}{72} J \; \big( 22 - 11 \alpha +
8 \sqrt{\alpha ( 1 - \alpha)} \big).
\label{e3baa}
\end{equation}
Comparison with the value obtained by direct summation, $E = - \frac{1}{6}
(2 - \alpha)$, reveals that state (4b) is the lowest--lying fully spin
and orbitally ordered configuration in the region $0.063 < \alpha < 0.983$.
That this state dominates over the majority of the phase diagram is a
direct consequence of its ability to gain energy from mixed processes.

The non--uniform three--color state (4c) also presents a delicate
competition between spin configurations of very similar energies. From
the preceding subsections, it is clear that in this case diagonal and
off--diagonal processes favor different ground states, while there will
also be a mixed contribution from 1/3 of the bonds. The energies of the
three standard spin configurations are
\begin{eqnarray}
E_{\rm FM}^{(4c)}(0.5) & = & - \frac{1}{12} J r_1, \nonumber \\
E_{\rm AF}^{(4c)}(0.5) & = & - \frac{1}{384} J\; (45 r_1 + 41 r_2 + 10 r_3), \\
E_{\rm AFF}^{(4c)}(0.5) & = & - \frac{1}{432} J\; (54 r_1 + 41 r_2 + 10 r_3),
\nonumber
\label{e3cp5}
\end{eqnarray}
where the AF state, obtaining $E_{\rm AF}^{(4c)}(0.5) = - \frac14 J$
is the lowest at $\eta = 0$.

Finally, in the three--color states (4d) and (4e), which are composed of
lines of two--color sites, this delicate balance between different spin
configurations persists. For configuration (4d), an AFF state with the
same orientation of the FM line (along the $b$--axis) is both favored
by diagonal hopping processes and competitive for off--diagonal processes.
With inclusion of a small contribution due to mixed processes, the three
ordered spin states have energies
\begin{eqnarray}
E_{\rm FM}^{(4d)}(0.5) & = & - \frac{51}{288} J r_1, \nonumber \\
E_{\rm AF}^{(4d)}(0.5) & = & - \frac{1}{768} J\; (85 r_1 + 84 r_2 +
18 r_3), \\
E_{\rm AFF}^{(4d)}(0.5) & = & - \frac{1}{576} J\; (71 r_1 + 59 r_2 + 14 r_3),
\nonumber
\label{e3dp5}
\end{eqnarray}
from which the AFF state minimizes the energy at $\eta = 0$ with
$E_{\rm AFF}^{(4d)}(0.5) = - \frac14 J$.

For state (4e), which has no mixed contribution, the orientations of the
FM lines in the optimal AFF states do not match, and it is necessary, as
above, to consider both possibilities when performing a full comparison.
These four ordered spin states yield the energies
\begin{eqnarray}
E_{\rm FM}^{(4e)}(0.5) & = & - \frac{3}{16} J r_1, \nonumber \\
E_{\rm AF}^{(4e)}(0.5) & = & - \frac{1}{128} J\; (15 r_1 + 13 r_2 +
2 r_3), \nonumber \\
E_{\rm AFF(0)}^{(4e)}(0.5) & = & -\frac{1}{144} J\; (18 r_1 + 13 r_2
+ 2 r_3), \nonumber \\
E_{\rm AFF(1)}^{(4e)}(0.5) & = &
 - \frac{1}{48} J\; (6 r_1 + 4 r_2 + r_3),
\label{e3ep5}
\end{eqnarray}
among which the AF state in fact lies lowest at $\eta = 0$, achieving
the weakly frustrated value $ E_{\rm AF}^{(4e)}(0.5) = - \frac{15}{64} J$.

\subsection{Summary}
\label{sec:mfasum}

Here we summarize the results of this section in a concise form. For
the superexchange model ($\alpha = 0$), a considerable number of 2D
ordered orbital and spin states exist which return the energy
$- \frac13 J$ at $\eta = 0$. This degeneracy is lifted at any finite
Hund exchange in favor of orbital states [(3a), (3b)] permitting a fully
FM spin alignment. Most other orbital configurations introduce a
frustration in the spin sector at small $\eta$, while some offer the
possibility of a change of ground--state spin configuration at finite
$\eta$, where $r_1$ exceeds the $r_2$ and $r_3$ contributions and
begins to favor states with more FM bonds.

However, the value $E = - \frac13 J$ per bond remains a rather poor
minimum for a system as highly connected as the triangular lattice,
even if, as in the superexchange limit, active hopping channels exist
only in two of the three lattice directions for each orbital color.
Indeed, the limitations of the available ordering (potential) energy are
clearly visible from the fact that a significantly lower overall energy
is attained in systems which abandon spin order in favor of the resonance
(kinetic) energy gains available in one lattice direction. The result
$E_{\rm 1D}^{(3c)}(0) = - 0.3977 J$ is the single most important obtained
in this section, and in a sense obviates all of the considerations made
here for fully ordered states, mandating the full consideration of 2D
magnetically and orbitally disordered phases.

\begin{figure}[t!]
\includegraphics[width=7.7cm]{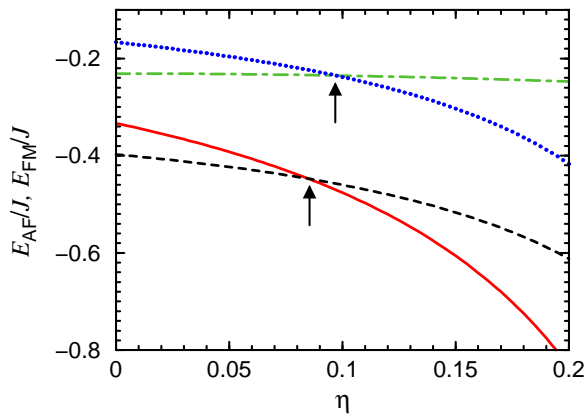}
\caption{(Color online)
Minimum energies per bond obtained for orbitally ordered phases,
showing a transition as a function of Hund exchange $\eta$ from
quasi--1D, AF--correlated to FM ordered spin states. For the
superexchange Hamiltonian ${\cal H_s}$ of Sec.~II ($\alpha = 0$), the
transition is from the quasi--1D spin state on orbital configuration
(3c) [black, dashed line from Eq.~(\ref{e1c1d})] to the one--color
orbital state (3a) [red, solid line from Eq.~(\ref{e1afm})]. For the
direct--exchange Hamiltonian ${\cal H}_d$ ($\alpha = 1$), the transition
is from the purely 1d spin state on the one--color orbital state (3a)
[green, dot--dashed line from Eq.~(\ref{af2a1})] to the two--colour,
avoided--blocking state (3b) [blue, dotted line from Eq.~(\ref{fm2b1})].
The transitions to FM order as obtained from the mean--field
considerations of this section are marked by arrows.}
\label{fig:fm}
\end{figure}

In the study of ordered states, it becomes clear that the Hund
exchange acts to favor FM spin alignments at high $\eta$. Because
the ``low--spin'' states of minimal energy are in fact stabilized by
quantum corrections due to AF spin fluctuations, the lowest energies
at $\eta = 0$ are never obtained for FM states, and therefore
increasing $\eta$ drives a phase transition between states of
differing spin and orbital order. We show in Fig.~\ref{fig:fm} the
transitions from quasi--1D AF--correlated states at low $\eta$, for
both $\alpha = 0$ and $\alpha = 1$, to FM states of fixed orbital
and spin order (3b). The transitions occur at the values $\eta_c (0)
 = 0.085$ and $\eta_c (1) = 0.097$, indicating that FM ordered states
may well compete in the physical parameter regime. We note again that
the energies in the superexchange limit are lower by approximately a
factor of two compared to the direct--exchange limit simply because
of the number of available hopping channels.

We note also that there is never a situation in which the spin
Hamiltonian becomes that of a Heisenberg model on a triangular lattice.
This demonstrates again the inherent frustration introduced by the orbital
sector. However, the fact that the ordered--state energy can never be
lowered to the value $E_{\rm HAF} = - \frac{3}{8}J$, which might be
expected for a two--active FO situation on every bond, far less the value
$-\frac12 J$ which could be achieved if it were possible to optimize
every bond in some ordered configuration, can be taken as a qualitative
reflection of the fact that on the triangular lattice the orbital
degeneracy ``enhances'' rather than relieves the (geometrical)
frustration of superexchange interactions (Sec.~\ref{sec:rhk}).

The limit of direct exchange ($\alpha = 1$) is found to be quite
different: the very strong tendency to favor spin singlet states, and
the inherent one--dimensionality of the model in this limit (one active
hopping direction per orbital color), combine to yield no competitive
states with long--ranged magnetic order. Their optimal energy is very
poor because of the restricted number of hopping channels, and coincides
with the (``avoided--blocking'') value for the model with only AO bonds,
$E = - \frac{1}{6} J$. Thus these states form part of a manifold with
very high degeneracy. However, even at this level it is clear that more
energy, meaning kinetic (from resonance processes) rather than potential,
may be gained by forming quasi--1D Heisenberg--chain states with little
or no interchain coupling and only quasi--long--ranged magnetic order.
Studies of orbital configurations permitting dimerized states are
clearly required (Sec.~IV). Finite Hund exchange acts to favor ordered
FM configurations, which will take over from chain--like states at
sufficiently high values of $\eta$ (Fig.~\ref{fig:fm}).

Finally, ordered states of the mixed model show a number of compromises.
At $\alpha = 0.5$, where the coefficients of superexchange and
direct--exchange are equal, some configurations are able to return the
unfrustrated sum of the optimal states in each sector when considered
separately, namely $- \frac14 J$. However, superposition states, which
are not optimal in either limit, can redeem enough energy from mixed
processes to surpass this value, and in fact the maximally superposed
configuration (4b) is found to minimize the energy over the bulk of the
phase diagram. Still, the net energy of such states remains small compared
to expectations for a highly connected state with three available hopping
channels per orbital color. Because of the directional mismatch between
the diagonal and off--diagonal hopping sectors, no quasi--1D states with
only chain--like correlations are able to lower the ordered--state
energy in the intermediate regime.

\section{Dimer states}
\label{sec:dim}

As shown in Sec.~II, the spin--orbital model on a single bond favors
spin or orbital dimer formation in the superexchange limit, and spin
dimer formation in the direct--exchange limit. The physical
mechanism responsible for this behavior is, as always, the fluctuation
energy gain from the highly symmetric singlet state. On the basis of
this result, combined with our failure to find any stable, energetically
competitive states with long--ranged spin and orbital order in either
limit of the model (Sec.~\ref{sec:mfa}), we proceed to examine states
based on dimers. Given the high connectivity of the triangular lattice,
dimer--based states are not expected {\it a priori\/} to be capable of
attaining lower energies than ordered ones, and if found to be true it
would be a consequence of the high frustration, which as noted in Sec.~I
has its origin in both the interactions and the geometry. Here we consider
static dimer coverings of the lattice, and compute the energies they gain
due to inter--singlet correlations. The tendency towards the formation of
singlet dimer states will be supported by the numerical results in
Sec.~\ref{sec:ed}, which will also address the question of resonant
dimer states.

\subsection{ Superexchange model }
\label{sec:dimsex}

Motivated by the fact that the spin and orbital sectors in ${\cal H}_s$
(\ref{Hs}) are not symmetrical, we proceed with a simple decoupling of
spin and orbital operators. Extensive research on spin--orbital models
has shown that this procedure is unlikely to capture the majority of
the physical processes contributing to the final energy, particularly
in the vicinity of highly symmetric points of the general Hamiltonian.
The results to follow are therefore to be treated as a preliminary guide,
and a basis from which to consider a more accurate calculation of the
missing energetic contributions. We remind the reader that the notation
FO and AO used in this subsection is again that obtained by performing
a local transformation on one site of every dimer. As noted in
Sec.~\ref{sec:sex}, this procedure is valid for the discussion of
states based on individual dimerized bonds, where it represents merely
a notational convenience. For FO configurations, which in the original
basis have different orbital colors, one might in principle expect that,
because of the color degeneracy, there should be more ways to realize
these without frustration than there are to realize AF spin
configurations; however, because of the directional dependence of
the hopping, we will find that this is not necessarily the case (below).

The basic premise of the spin--orbital decoupling is that if the spin
(orbital) degrees of freedom on a dimer bond form a singlet state, their
expectation value $\langle {\vec S}_i \cdot {\vec S}_j \rangle$ ($\langle
{\vec T}_i \cdot {\vec T}_j \rangle$) on the neighboring interdimer bonds
will be precisely zero. The optimal orbital (spin) state of the interdimer
bond may then be deduced from the effective bond Hamiltonian obtained by
decoupling. Because ${\cal H}_s$ depends on the number of electrons on
the sites of a given bond which are in active orbitals, and this number
is well defined only for the dimer bonds, the effective Hamiltonian will
be obtained by averaging over all occupation probabilities. In contrast
to the pure Heisenberg spin Hamiltonian, here the interdimer bonds
contribute with finite energies, and the dimer distribution must be
optimized. A systematic optimization will not be performed in this
section, where we consider only representative dimer coverings giving
the semi--quantitative level of insight required as a prelude to adding
dimer resonance processes (Sec.~V).

\begin{figure}[t!]
\mbox{\includegraphics[width=2.5cm]{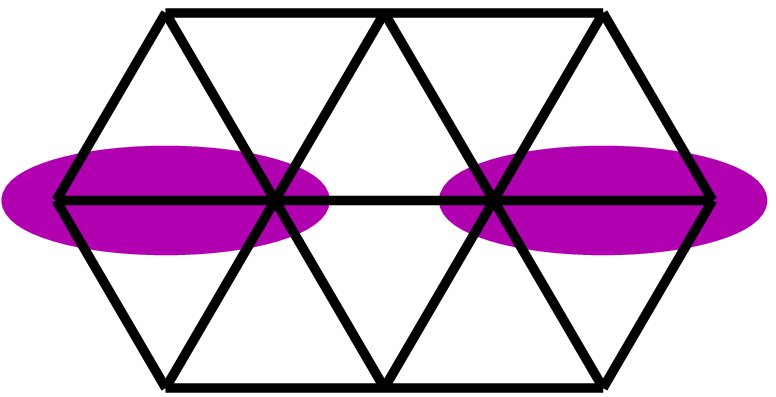}}
\hskip .7cm
\mbox{\includegraphics[width=4.7cm]{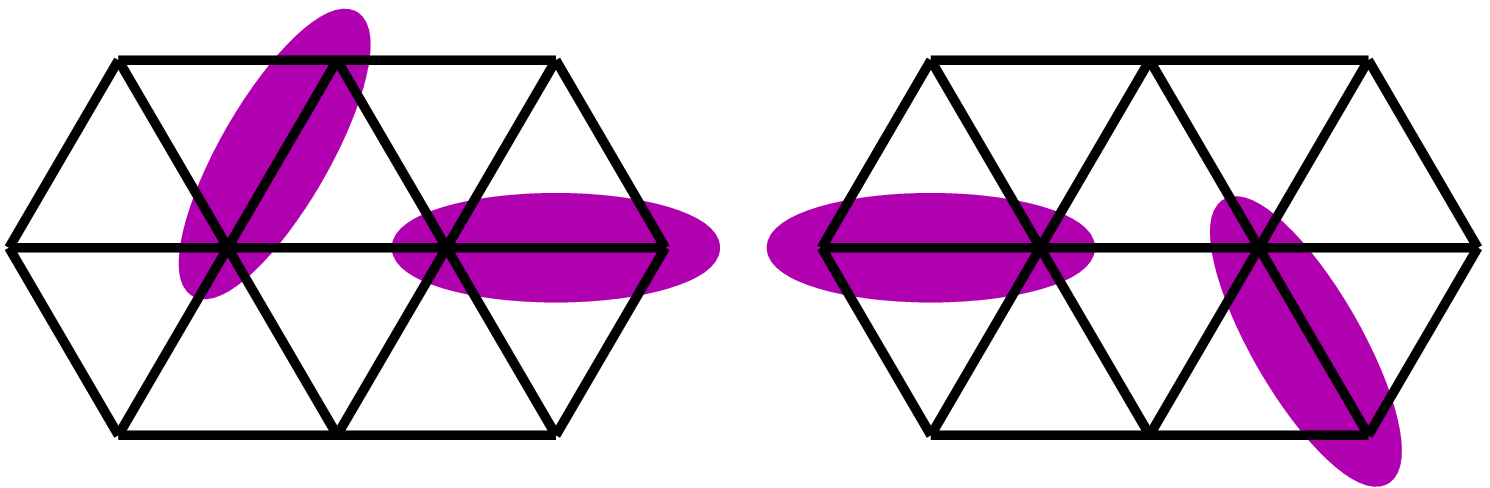}}
\centerline{(a) \hskip 4.0cm  (b) \hskip 0.9cm}
\vskip .3cm
\mbox{\includegraphics[width=7.0cm]{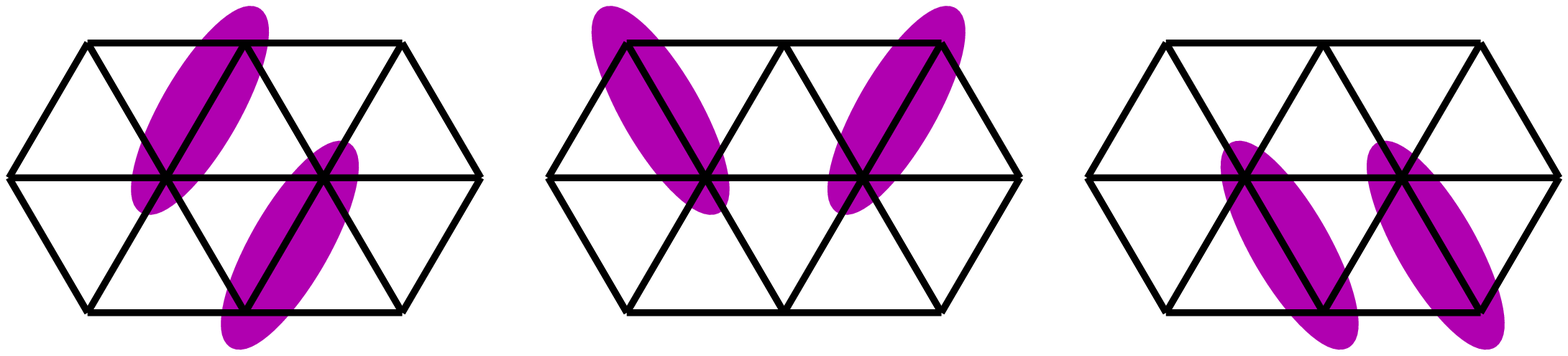}}
\centerline{(c)}
\caption{(Color online)
Types of interdimer bond differing in effective interaction due to
dimer coordination: (a) ``linear'', (b) ``semi--linear'', (c)
``non--linear''.}
\label{fig:bonds}
\end{figure}

On the triangular lattice there are three essentially different types
of interdimer bond, which are shown in Fig.~\ref{fig:bonds}). For a
``linear'' configuration [Fig.~\ref{fig:bonds}(a)], the number of
electrons in active orbitals on the interdimer bond is two; for the
8 possible configurations where one dimer bond is aligned with the
interdimer bond under consideration [Fig.~\ref{fig:bonds}(b)], the
number is one on the corresponding site and one or zero with equal
probability on the other; for the 14 remaining configurations where
neither dimer bond is aligned with the interdimer bond
[Fig.~\ref{fig:bonds}(c)], the number is one or zero for both sites.
The number of electrons in active orbitals is then two for type
(\ref{fig:bonds}a), two or one, each with probability 1/2, for type
(\ref{fig:bonds}b), and two, one or zero with probabilities 1/4, 1/2,
and 1/4 for type (\ref{fig:bonds}c).

The effective interdimer interactions for each type of bond can be
deduced in a manner similar to the treatment of the previous section.
Considering first the situation for a bond of type (\ref{fig:bonds}a)
with (os/st) dimers, setting $\langle {\vec T}_i \cdot {\vec T}_j
\rangle = 0$ yields one high--spin and two low--spin terms which
contribute
\begin{eqnarray}
H_{1}^{\rm (os,7a)} (0) & = & - \frac14 J r_1 \Big( {\vec S}_i
\cdot {\vec S}_j + \frac{3}{4} \Big), \label{ehdt2e} \nonumber   \\
H_{2}^{\rm (os,7a)} (0) & = &   \frac{3}{4} J r_2 \Big( {\vec S}_i
\cdot {\vec S}_j - \frac14 \Big),      \\
H_{3}^{\rm (os,7a)} (0) & = &   \frac{1}{6} J (r_3 - r_2) \Big(
{\vec S}_i \cdot {\vec S}_j - \frac14 \Big). \nonumber
\end{eqnarray}
Clearly $H_{1}^{\rm (os,7a)}$ favors FM (high--spin) interdimer spin
configurations with coefficient $\frac14$, while $H_{2}^{\rm (os,7a)}$
and $H_{3}^{\rm (os,7a)}$ favor AF (low--spin) configurations with
coefficient $\frac{3}{8}$ (both at $\eta = 0$). Because $r_1$ exceeds
$r_2$ and $r_3$ when Hund exchange is finite, one expects a critical
value of $\eta$ where FM configurations will be favored. Simple
algebraic manipulations using all three terms suggest that this
value, which should be relevant for a linear chain of (os/st) dimers,
is $\eta_c = \frac{1}{8}$. In the limit $\eta \rightarrow 0$, the
effective bond Hamiltonian simplifies to
\begin{equation}
H_{\rm eff}^{\rm (os,7a)} (0) = \frac12 J \left( {\vec S}_i \cdot
{\vec S}_j - \frac34 \right).
\label{ehdt2}
\end{equation}

For a bond of type (\ref{fig:bonds}a) with (ss/ot) dimers, setting
$\langle{\vec S}_i \cdot {\vec S}_j \rangle = 0$ on the interdimer
bond yields
\begin{eqnarray}
H_{1}^{\rm (ss,7a)} (0) & = &  \frac{3}{4} J r_1 \left( {\vec T}_i
\cdot {\vec T}_j - \frac14 \right), \label{ehds2e} \nonumber \\
H_{2}^{\rm (ss,7a)} (0) & = & - \frac14 J r_2 \left( {\vec T}_i \cdot
{\vec T}_j + \frac{3}{4} \right), \\
H_{3}^{\rm (ss,7a)} (0) & = & - \frac{1}{6} J (r_3 - r_2) \left(
{\vec T}_i \times {\vec T}_j + \frac14 \right). \nonumber
\end{eqnarray}
Here $H_{1}^{\rm (ss,7a)}$ favors AO configurations with coefficient
$\frac{3}{8}$, while $H_{2}^{\rm (ss,7a)}$ and $H_{3}^{\rm (ss,7a)}$
both favor FO configurations with coefficient $\frac14$ (at $\eta = 0$).
Over the relevant range of Hund exchange coupling, $0 < \eta < 1/3$,
there is no change in sign and AO configurations are always favored.
The effective bond Hamiltonian for $\eta \rightarrow 0$ is
\begin{equation}
H_{\rm eff}^{\rm (ss,7a)} (0) = \frac12 J \left( {\vec T}_i \cdot
{\vec T}_j - \frac34 \right).
\label{ehds2}
\end{equation}

For bonds of type (\ref{fig:bonds}b), when only one electron occupies
an active orbital the corresponding decoupled interdimer bond
Hamiltonians are, for (os/st) dimers,
\begin{eqnarray}
H_{1}^{\rm (os,1)} (0) & = & - \frac14 J r_1 \Big( {\vec S}_i \cdot
{\vec S}_j + \frac{3}{4} \Big), \label{ehdt1e} \nonumber \\
H_{2}^{\rm (os,1)} (0) & = &  \frac14 J r_2 \Big( {\vec S}_i \cdot
{\vec S}_j - \frac14 \Big),    \\
H_{3}^{\rm (os,1)} (0) & = &  0. \nonumber
\end{eqnarray}
The final interdimer interaction is obtained by averaging over these
expressions and those (\ref{ehdt2e}) for two active orbitals per bond,
and takes the rather cumbersome form
\begin{eqnarray}
H_{\rm eff}^{\rm (os,7b)} (0) & = & \frac{1}{12} J\; (r_3 + 5 r_2
 - 3 r_1) \; {\vec S}_i \cdot {\vec S}_j \nonumber \\ & & - \frac{1}{48}
J \; (9 r_1 + 5 r_2 + r_3),
\end{eqnarray}
which reduces in the limit $\eta \rightarrow 0$ to
\begin{equation}
H_{\rm eff}^{\rm (os,7b)} (0) = \frac14 J \left( {\vec S}_i \cdot
{\vec S}_j - \frac{5}{4} \right).
\label{ehdt1}
\end{equation}
For (ss/ot) dimers, the situation cannot be formulated analogously,
because if only one electron on the bond is active, the orbital
state of the other electron has no influence on the hopping
process, {\it i.e.}~${\vec T}_i \cdot {\vec T}_j$ is not a
meaningful quantity. The resulting expressions lead then to
\begin{eqnarray}
\label{ehds1e}
H_{\rm eff}^{\rm (ss,7b)} (0) \!\! & = & \!\! \frac{1}{8} J (3 r_1
\! - \! r_2) {\vec T}_i \cdot {\vec T}_j \! - \! \frac{1}{12} J (r_3
\! - \! r_2) {\vec T}_i \! \times \! {\vec T}_j \nonumber \\ & &
 - \frac{1}{48} J\, (9 r_1 + 5 r_2 + r_3),
\end{eqnarray}
which has the $\eta \rightarrow 0$ limit
\begin{equation}
H_{\rm eff}^{\rm (ss,7b)} (0) = \frac14 J
\left( {\vec T}_i \cdot {\vec T}_j - \frac{5}{4} \right).
\label{ehds1}
\end{equation}

Finally, for a bond of type (\ref{fig:bonds}c), there is no contribution
from interdimer bond states with no electrons in active orbitals, so
the above results [(\ref{ehdt2e}, \ref{ehdt1e}) and (\ref{ehds2e},
\ref{ehds1e})] are already sufficient to perform the necessary
averaging. With (os/st) dimers
\begin{eqnarray}
H_{\rm eff}^{\rm (os,7c)} (0) & = & \frac{1}{48} J \, (2 r_3 + 13 r_2
 - 9 r_1) \; {\vec S}_i \cdot {\vec S}_j \nonumber \\
& & - \frac{1}{192} J\; (27 r_1 + 13 r_2 + 2 r_3),
\end{eqnarray}
which reduces in the limit $\eta \rightarrow 0$ to
\begin{equation}
H_{\rm eff}^{\rm (os,7c)} (0) = \frac{1}{8} J \left( {\vec S}_i \cdot
{\vec S}_j - \frac{7}{4} \right),
\label{ehdt0}
\end{equation}
while for (ss/ot) dimers,
\begin{eqnarray}
H_{\rm eff}^{\rm (ss,7c)} (0) \!\! & = & \!\! \frac{1}{16} J (3 r_1 \!
 - \! r_2) {\vec T}_i \cdot {\vec T}_j \! - \! \frac{1}{24} J (r_3 \!
 - \! r_2) {\vec T}_i \! \times \! {\vec T}_j \nonumber \\ & &
 - \frac{1}{192} J \, (27 r_1 + 13 r_2 + 2 r_3),
\end{eqnarray}
which in the $\eta \rightarrow 0$ limit gives
\begin{equation}
H_{\rm eff}^{\rm (ss,7c)} (0) = \frac{1}{8} J \left( {\vec T}_i \cdot
{\vec T}_j - \frac{7}{4} \right).
\label{ehds0}
\end{equation}

\begin{table}[b!]
\caption{
Occurrence probabilities for bonds of each type for four simple
periodic dimer coverings of the triangular lattice.}
\begin{ruledtabular}
\begin{tabular}{ccccc}

 configuration & dimer &  bond (7a)  &  bond (7b)  &  bond (7c)  \cr

\colrule

Fig.~\ref{pdc}(a) &$\frac{1}{6}$&$\frac{1}{6}$&  0  &$\frac{2}{3}$ \cr
Fig.~\ref{pdc}(b) &$\frac{1}{6}$&  0  &$\frac{1}{3}$&$\frac{1}{2}$ \cr
Fig.~\ref{pdc}(c) &$\frac{1}{6}$&  0  &$\frac{1}{3}$&$\frac{1}{2}$ \cr
Fig.~\ref{pdc}(d) &$\frac{1}{6}$&  0  &$\frac{1}{3}$&$\frac{1}{2}$ \cr

\end{tabular}
\end{ruledtabular}
\label{tab:dim}
\end{table}

These results have clear implications for the nearest--neighbor
correlations in an extended system. By inspection, systems composed
of either type of dimer would favor AF (spin) and AO interdimer
bonds, to the extent allowed by frustration, and ``linear'' [type
(\ref{fig:bonds}a)] bonds over ``semi--linear'' [type (\ref{fig:bonds}b)]
bonds over ``non--linear'' [type (\ref{fig:bonds}c)] bond types in
Fig.~\ref{fig:bonds}, to the extent allowed by geometry. Discussion of
this type of state requires in principle the consideration of all possible
dimer coverings, but will be restricted here to a small number of periodic
arrays which illustrate much of the essential physics of extended dimer
systems within this model.

\begin{figure}[t!]
\mbox{\includegraphics[width=4.1cm]{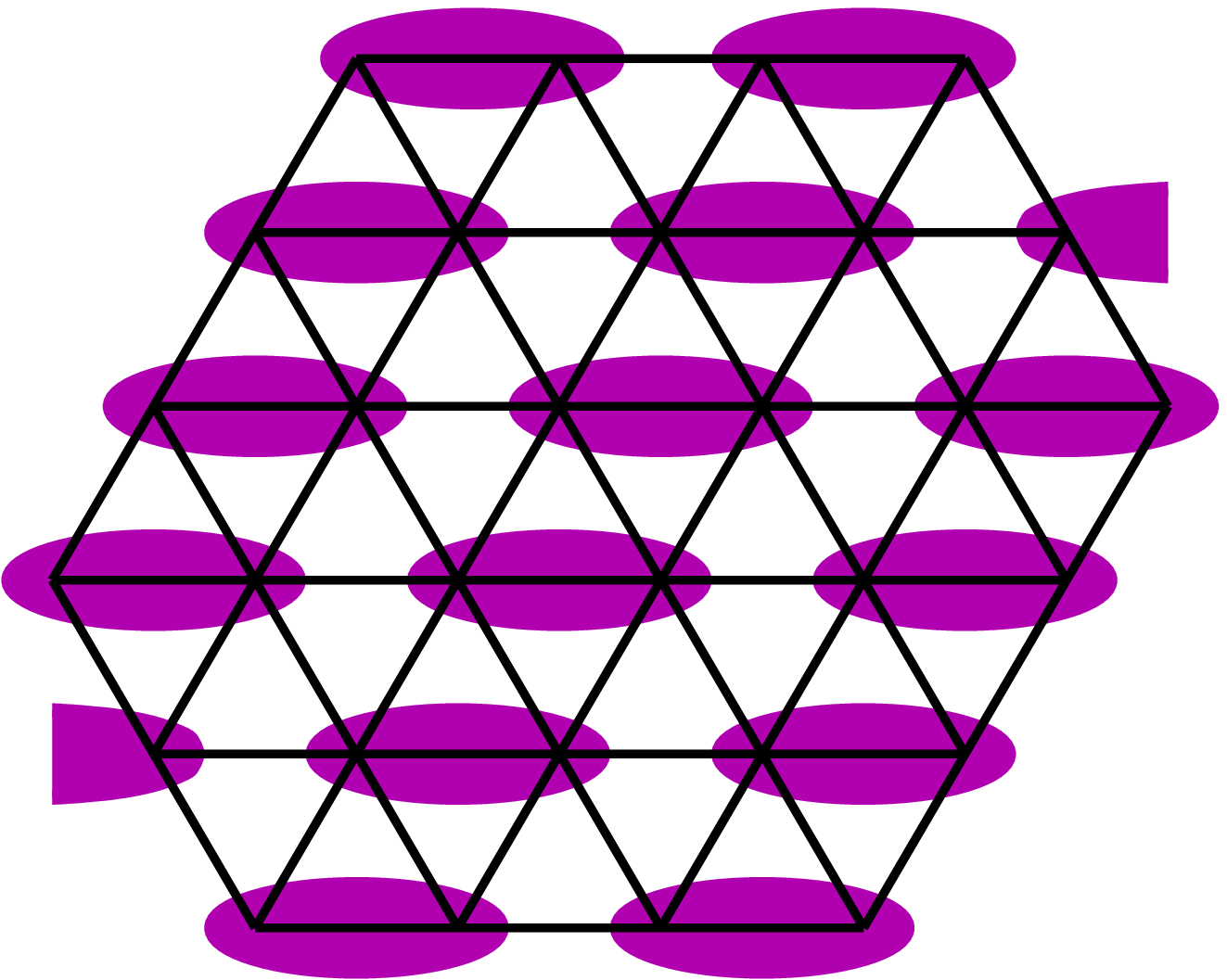}} \hskip .2cm
\mbox{\includegraphics[width=4.1cm]{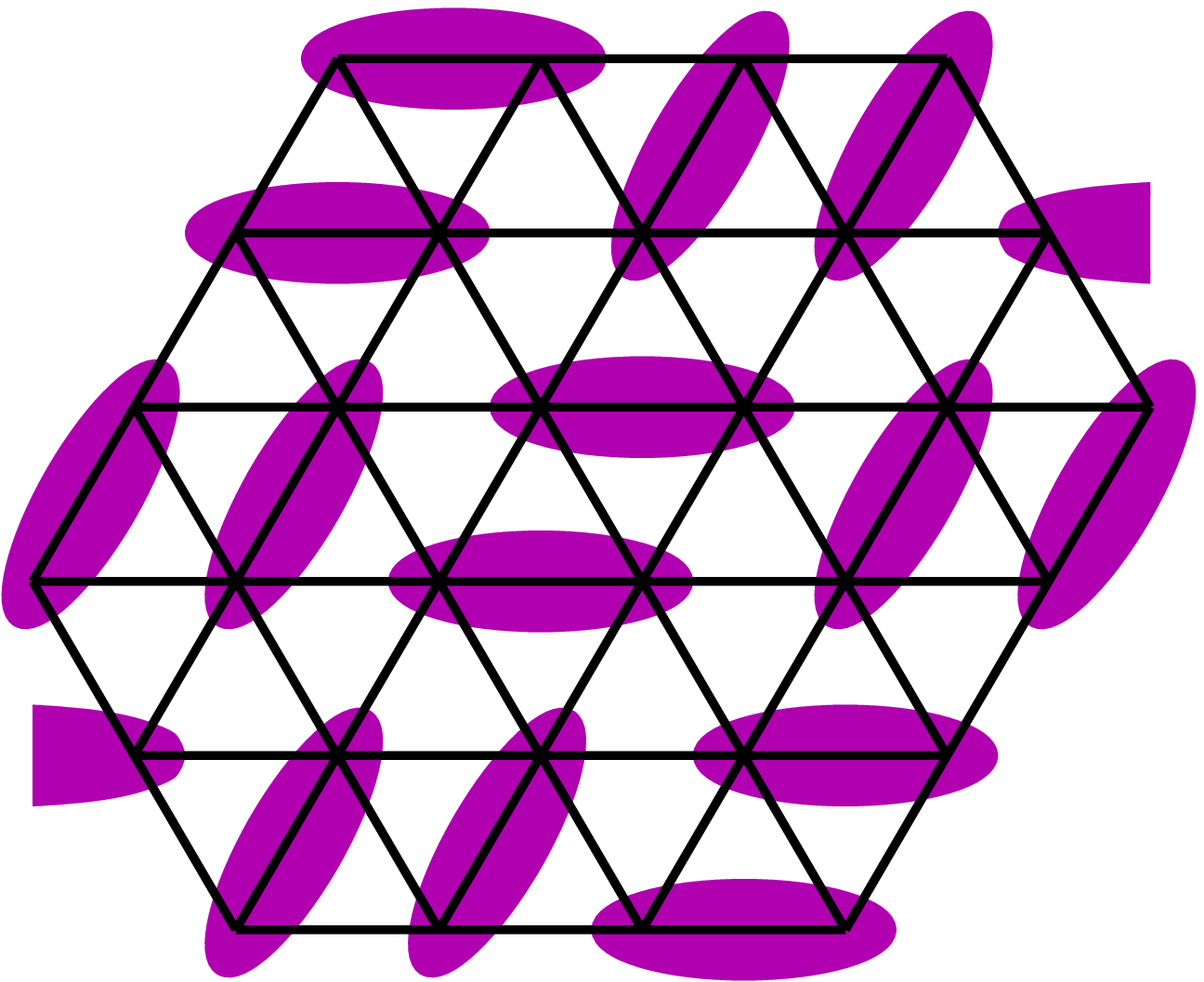}} \centerline{(a)
\hskip 3.5cm  (b)} \vskip .2cm
\mbox{\includegraphics[width=4.1cm]{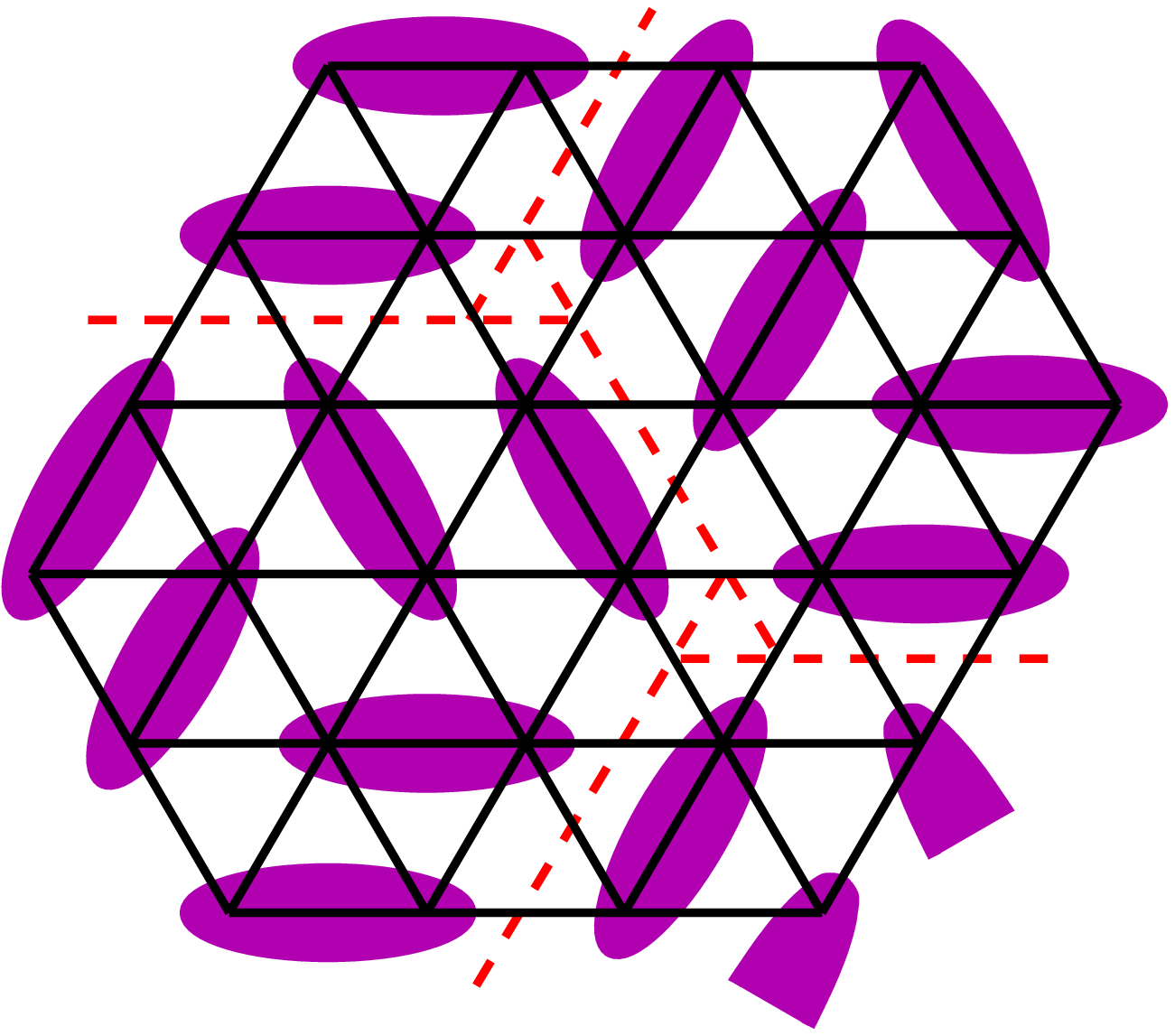}} \hskip .2cm
\mbox{\includegraphics[width=4.1cm]{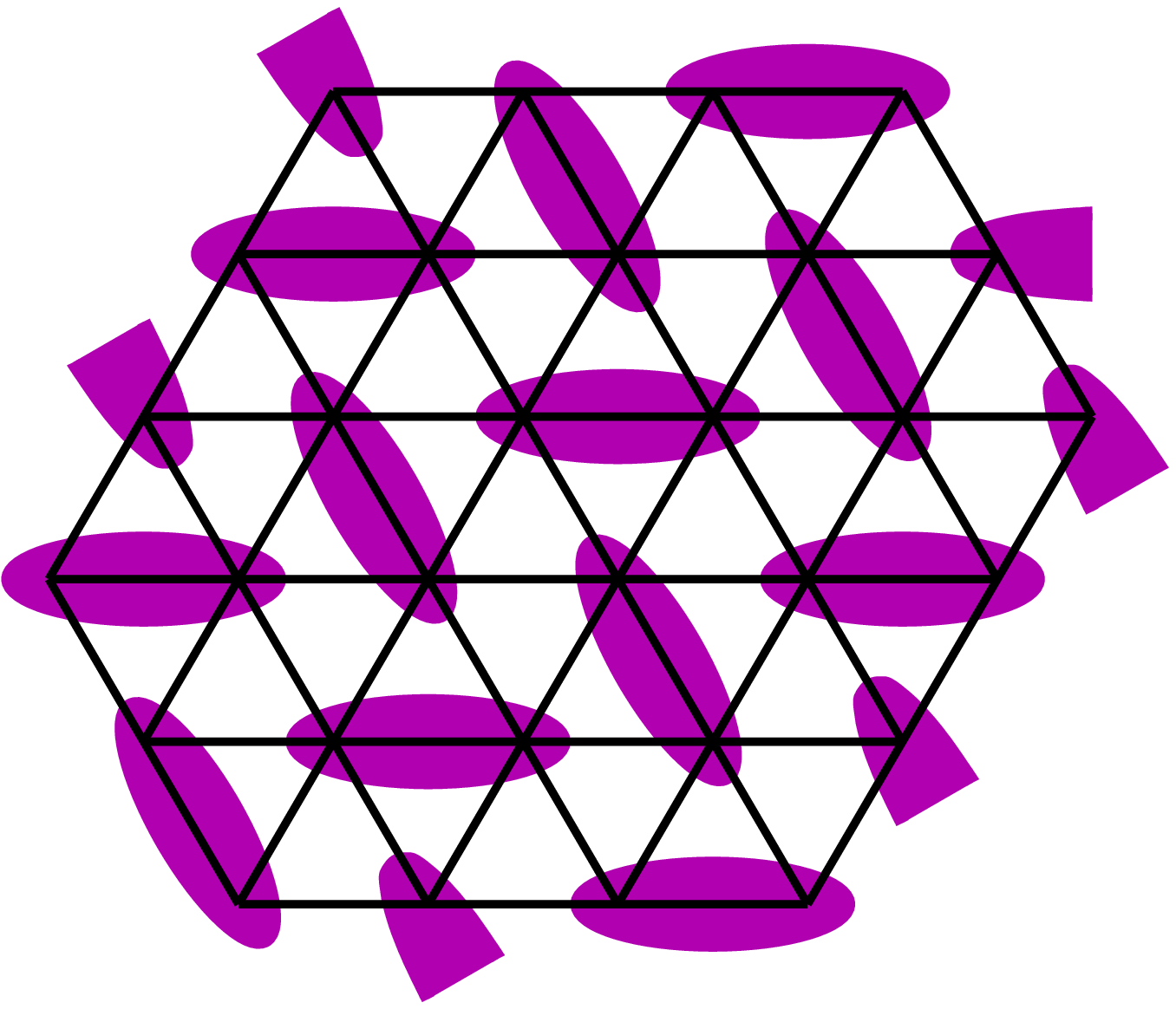}} \centerline{(c)
\hskip 3.5cm  (d)} \caption{(Color online) Periodic dimer
coverings on the triangular lattice, each representative of a
class of coverings: (a) linear; (b) plaquette; (c) 12--site unit
cell; (d) ``zig--zag''. } \label{pdc}
\end{figure}

We begin by considering the periodic covering of Fig.~\ref{pdc}(a),
a fully linear conformation (of ground--state degeneracy 12) whose
interdimer bond types (Table I) maximize the possible number of bonds
of type (\ref{fig:bonds}a). The counterpoint shown in Fig.~\ref{pdc}(b)
consists of pairs of dimer bonds with alternating orientations in two of
the three lattice directions, and constitutes the simplest configuration
minimizing (to zero) the number of type--(\ref{fig:bonds}a) interdimer
bonds. The coverings in Figs.~\ref{pdc}(c) and (d) have the same
property. These configurations exemplify a quite general result,
that any dimer covering in which there are no linear configurations
[type (\ref{fig:bonds}a)] of any pair of dimers will have 1/3
type--(\ref{fig:bonds}b) bonds, and thus the remaining 1/2 of the
bonds must be of type (\ref{fig:bonds}c). The coverings shown in
Figs.~\ref{pdc}(a) and (b, c, d) represent the limiting cases on
numbers of each type of bond, in that any random dimer covering will
have values between these. Indeed, it is straightforward to argue that,
in changes of position of any set of dimers within a covering, the
creation of any two bonds of type (\ref{fig:bonds}b) will destroy
one of type (\ref{fig:bonds}a) and one of type (\ref{fig:bonds}c),
and conversely.

Having established this effective sum rule, we turn next to the energies
of the dimer configurations. First, for both types of dimer [(os/st) and
(ss/ot)], all states with equal numbers of each bond type are degenerate,
subject to equal solutions of the frustration problem. Next, if
frustration is neglected, it is clear from Eqs.~(\ref{ehdt2},\ref{ehds2}),
(\ref{ehdt1},\ref{ehds1}), and (\ref{ehdt0},\ref{ehds0}), that the AF and
AO energy values for the three bond types (obtained by substituting $-
\frac14$ for ${\vec S}_i \cdot {\vec S}_j$ and ${\vec T}_i \cdot {\vec
T}_j$) are respectively $- \frac12 J$, $- \frac{3}{8} J$ and $- \frac14
J$, which, when taken together with the sum rule, suggest a very large
degeneracy of dimer covering energies.

\begin{figure}[t!]
\mbox{\includegraphics[width=4.1cm]{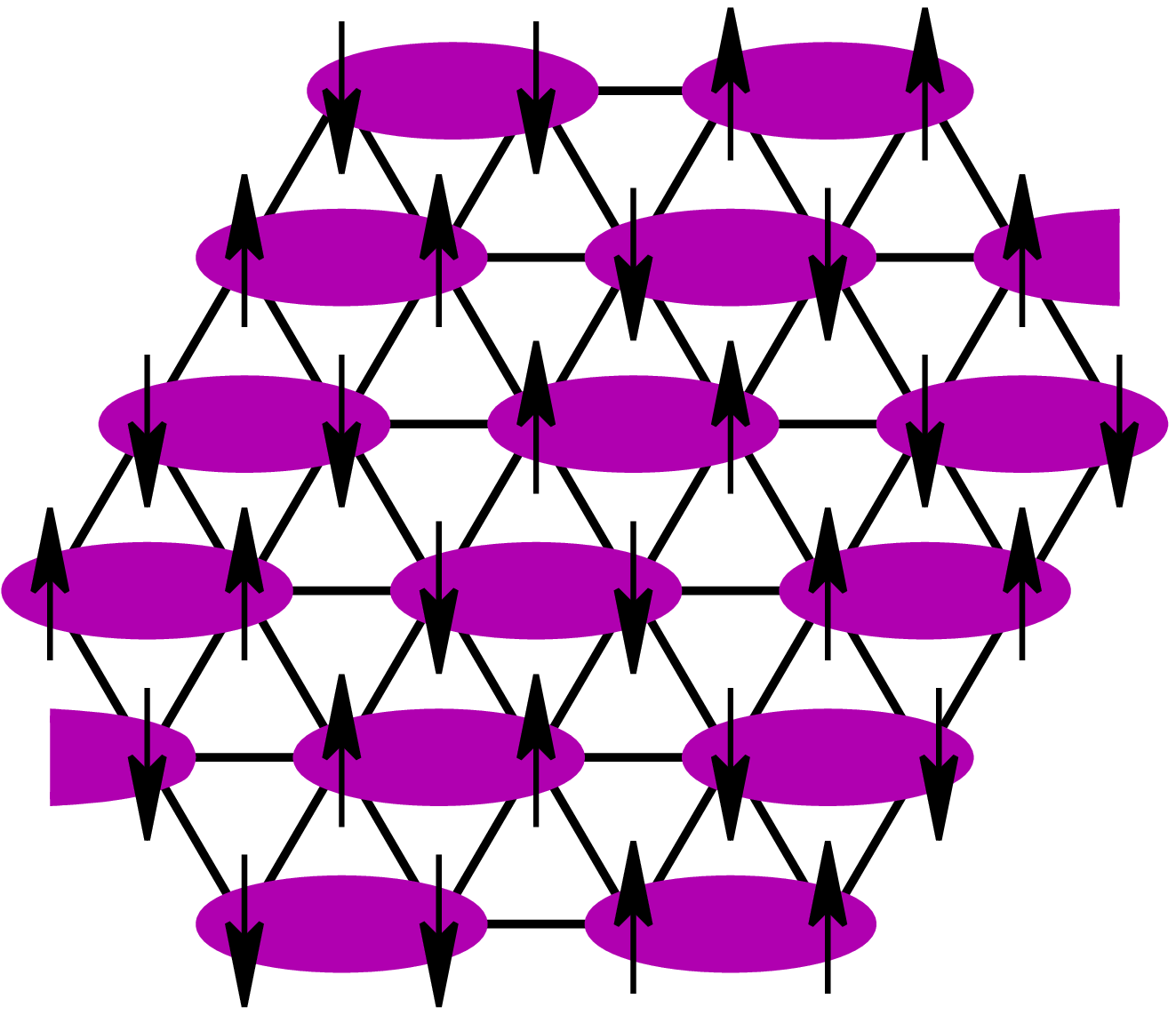}}
\hskip 0.2cm
\mbox{\includegraphics[width=4.1cm]{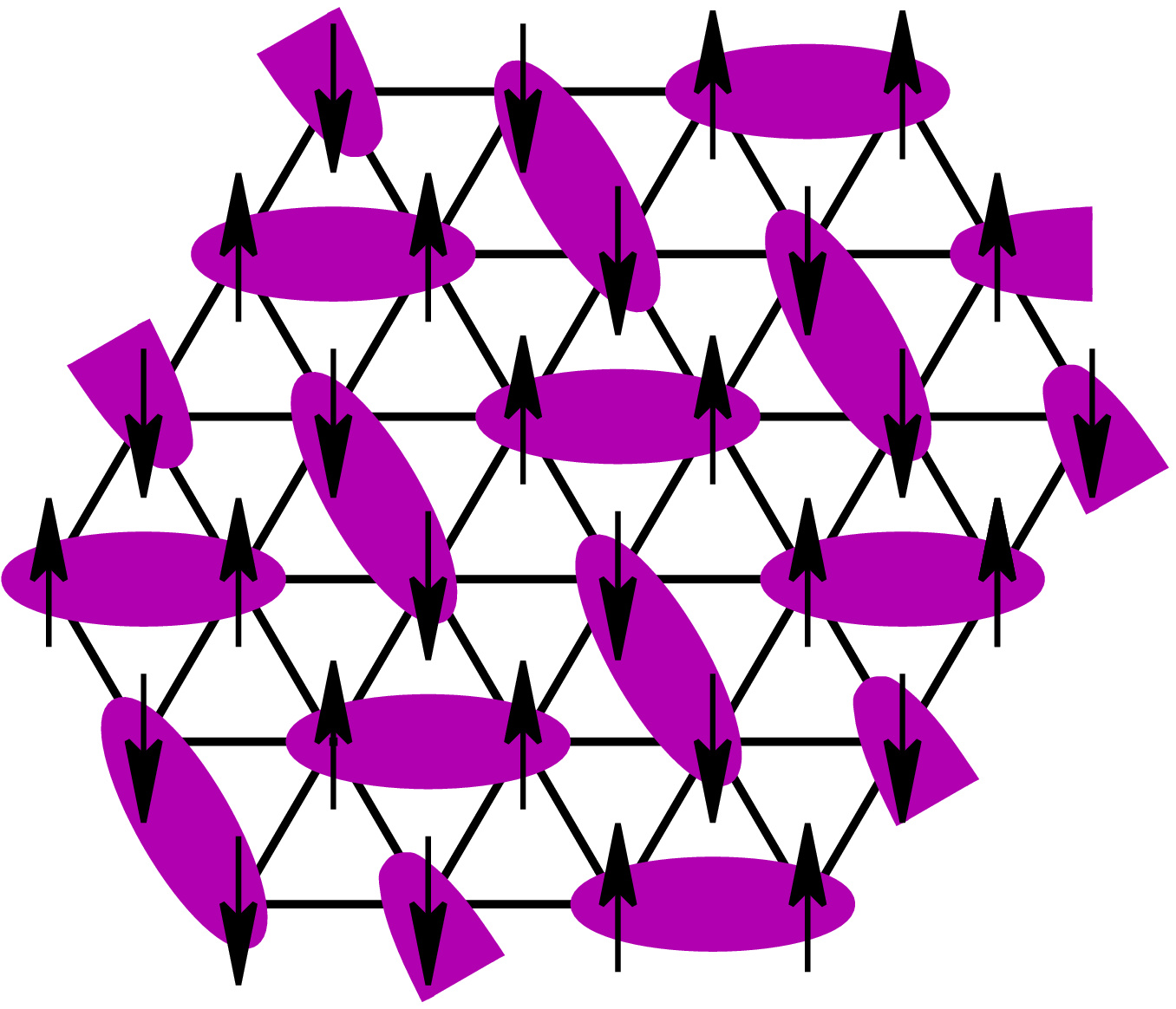}}
\centerline{(a) \hskip 3.5cm  (b)}
\caption{(Color online)
Spin or orbital configurations (black arrows) within (a) linear
and (b) zig--zag orbital-- or spin--singlet dimer coverings of
the triangular lattice. The number of frustrated interdimer
bonds is reduced to 1/6 of the total, and all are of type
(\protect{\ref{fig:bonds}}c). This figure emphasizes that for
the spin--orbital model, dimer singlet formation does not exhaust
the available degrees of freedom.}
\label{pdca}
\end{figure}

Returning to the question of frustration, a covering of minimal energy
is one which both minimizes the number of FM or FO bonds, and ensures
that they fall on bonds of type (\ref{fig:bonds}c); both criteria are
equally important. For the dimer covering (\ref{pdc}a), with maximal
aligned bonds, it is possible by using the spin (for (os/st) dimers)
or orbital (for (ss/ot) dimers) configuration represented by the
arrows in Fig.~\ref{pdca}(a) to make the number of frustrated (FM/FO)
interdimer bonds equal to 1/6 of the total. Bearing in mind that the
1/6 of bonds covered by dimers are also FM/FO, and that at least 1/3
of bonds on the triangular lattice must be frustrated for collinear
spins, this number is an absolute minimum. [Here we do not consider
the possibility of non--collinear order of the non--singlet degree
of freedom.] Further, for this configuration one observes that all
of the FM/FO bonds already fall on bonds of type (\ref{fig:bonds}c),
providing an optimal case with energy
\begin{eqnarray}
\label{enedim0}
E_{\rm dim}(0) & = & - J \left( \frac{1}{6} + \frac{1}{6} \cdot
\frac12 + \frac12 \cdot \frac14 + \frac{1}{6} \cdot \frac{3}{16}
\right) \nonumber \\ & = & - \frac{13}{32} J
\end{eqnarray}
at $\eta = 0$. This value constitutes a basic bound which
demonstrates that a simple, static dimer covering has lower energy
than any long--range--ordered spin or orbital state discussed in
Sec.~\ref{sec:mfa} in this limit ($\alpha = 0$) of the model.

It remains to establish the degeneracy of the ground--state
manifold of such coverings, and we provide only a qualitative
discussion using further examples. If alternate four--site (dimer
pair) clusters in Fig.~\ref{pdc}(a) are rotated to give the
covering of Fig.~\ref{pdc}(b), the minimal frustration is spoiled:
by analogy with Fig.~\ref{pdca}, it is easy to show that, if only
1/6 of the bonds are to be frustrated, then they are of type
(\ref{fig:bonds}b), and otherwise 1/3 of the bonds are frustrated
if all are to be of type (\ref{fig:bonds}c). On the periodic
12--site cluster [Fig.~\ref{pdc}(c)], one may place three
four--site clusters in each of the possible orientations, which as
above removes all bonds of type (\ref{fig:bonds}a) and maximizes
those of type (\ref{fig:bonds}b). Within this cluster it is
possible to have only four frustrated interdimer bonds out of 18,
while between the clusters there is again an arrangement of the
spin or orbital arrows ({\it cf.}~Fig.~\ref{pdca}) with only six
FM or FO bonds out of 24, for a net total of 1/6 frustrated
interdimer bonds, of which half are of type (\ref{fig:bonds}b).
The covering of Fig.~\ref{pdc}(d) represents an extension of the
procedure of enlarging unit cells and removing four--site
plaquettes, which demonstrates that it remains possible in the
limit of no type--(\ref{fig:bonds}a) bonds to reduce frustration
to 1/6 of the bonds, and to bonds of type (\ref{fig:bonds}c)
[Fig.~\ref{pdca}(b)], whence the energy of the covering is again
$E_{\rm dim}(0) = - \frac{13}{32}$ (\ref{enedim0}). Thus it is
safe to conclude that, for the static--dimer problem, the
ground--state manifold for $\alpha  = 0$ consists of a significant
number of degenerate coverings. We do not pursue these
considerations further because of degeneracy lifting by dimer
resonance processes, and because the energetic differences between
static dimer configurations are likely to be dwarfed by the
contributions from dimer resonance, the topic to which we turn in
Sec.~V.

\subsection{ Direct exchange model }
\label{sec:dimdex}

The very strong preference for bond spin singlets (the factor of 4
in Eq.~(\ref{Hd})] suggests that dimer states will also be competitive
in this limit, even though only 1/6 of the bonds may redeem an
energy of $- J$. Following the considerations and terminology of the
previous subsection, we note (i) that $\langle{\vec S}_i \cdot {\vec
S}_j \rangle = 0$ on interdimer bonds and (ii) that in this case,
interdimer bonds have energy $- \frac14 J$ at $\eta = 0$ for types
(\ref{fig:bonds}a) and (\ref{fig:bonds}b), and $0$ for type
(\ref{fig:bonds}c). Because any state with a maximal number (1/6)
of type--(\ref{fig:bonds}a) bonds must have only bonds of type
(\ref{fig:bonds}c) for the other 2/3 [states (\ref{pdc}a)], such a
state is manifestly less favorable at $\alpha = 1$ than those of type
(\ref{pdc}b)--(\ref{pdc}d), where there are no aligned pairs of dimers.
In this latter case, the full calculation gives
\begin{equation}
\label{enedimeta}
E_{\rm dim}(1) = - \frac{1}{144} J (9 r_1 + 19 r_2 + 8 r_3),
\end{equation}
and $E_{\rm dim}(1) = -\frac14 J$ for $\eta = 0$. This energy does
now exceed that available from the formation of Heisenberg spin chains
in one of the three lattice directions (Sec.~\ref{sec:mfdex}), which
gave the value $E_{\rm 1DAF}(1) = - 0.231 J$.

At the level of these calculations, the manifold of degenerate states
with this energy is very large, and its counting is a problem which
will not be undertaken here. We will show in Sec.~V that, precisely
in this limit, no dimer resonance processes occur and the static dimer
coverings do already constitute a basis for the description of the
ground state. The question of fluctuations leading to the selection of
a particular linear combination of these states which is of lowest
energy, {\it i.e.}~of a type of order--by--disorder mechanism, is
addressed in Ref.~\onlinecite{Jac07}.

At finite values of the Hund exchange, this type of state will come
into competition with the simple avoided--blocking states which gain,
with a FM spin state, an energy
\begin{equation}
\label{enedimfm}
E_{\rm FM}(1) = - \frac{1}{6} J r_1,
\end{equation}
as 2/3 of the bonds contribute with an energy of $-\frac{1}{4} J r_1$.
The critical value of $\eta$ required to drive the transition from the
low--spin dimerized state to the FM state is found to be
\begin{equation}
\label{etac}
\eta_c = 0.1589.
\end{equation}

\subsection{ Mixed model }
\label{sec:dimmex}

Because both of the endpoints, $\alpha = 0$ and $\alpha = 1$, favor
dimerized states over states of long--ranged order, it is natural to
expect that a dimer state will provide a lower energy also at $\alpha
 = 0.5$. However, we remind the reader that there are no intermediate
dimer bases, and caution there is no strong reason to expect one or
other of the limiting dimer states to be favored close to $\alpha = 0.5$.
By inspection, the energy of an $\alpha > 0$ state can be obtained by
direct addition of the diagonal interdimer bond contributions in an
(ss/ot) or (os/st) dimer state, which is established by pure
off--diagonal hopping, because no site occupancies arise which allow
mixed processes. For the same reason, no interdimer terms impede a
calculation of the energy of an $\alpha < 1$ state by summing the
off--diagonal interdimer bond contributions in a spin--singlet dimer
state stabilized by purely diagonal processes. We will not analyze
the static dimer solutions for the intermediate regime in great
detail, and provide only a crude estimate of the $\alpha = 0.5$
 energy by averaging over both results at the limits of their
applicability. We will make no attempt here to exclude other forms
of disordered state at $\alpha = 0.5$, and return to this question
in Sec.~V.

\begin{table}[b!]
\caption{
Additional interdimer bond energies at $\alpha = 0.5$ due respectively
to (i) diagonal hopping occurring in a state (designated by $\alpha = 0$)
stabilized by off--diagonal processes and (ii) off--diagonal hopping in a
state ($\alpha = 1$) stabilized by diagonal processes.}
\begin{ruledtabular}
\begin{tabular}{ccccc}
bond & (\ref{fig:bonds}a) & (\ref{fig:bonds}b) & (\ref{fig:bonds}c) \cr
\colrule
$\alpha = 0$, (os/st), AF & 0 & $- \frac{1}{16} (r_1 + r_2)$ &
$- \frac{1}{8}(r_1 + r_2)$ \cr
$\alpha = 0$, (os/st), FM & 0 & $- \frac{1}{8} r_1 $ & $- \frac14 r_1$ \cr
$\alpha = 0$, (ss/ot), AO & 0 & $- \frac{1}{32} (3 r_1 + r_2)$ & $
- \frac{1}{16}(3 r_1 + r_2)$ \cr
$\alpha = 0$, (ss/ot), FO & 0 & $- \frac{1}{32} (3 r_1 + r_2)$ & $
- \frac{1}{16}(3 r_1 + r_2)$ \cr
$\alpha = 1$, $\parallel$ dimers & 0 & $- \frac{1}{16} (3 r_1 + r_2)$ & $
- \frac{1}{8}(3 r_1 + r_2)$ \cr
$\alpha = 1$, non--$\parallel$ dimers & 0 & $- \frac{1}{16} (3 r_1 + r_2)$
& $- \frac{1}{12} (2 r_2 + r_3)$ \cr
\end{tabular}
\end{ruledtabular}
\label{tab:int}
\end{table}

For each type of bond it is straightforward to compute the energy gained
from interdimer hopping processes of the type not constituting the dimer
state, and the results are shown in Table II. The first four lines give
the energies per bond from diagonal hopping processes occurring on the
bonds of the different $\alpha = 0$ dimer states, and conversely for the
final two lines. It is clear that the occupations of type (\ref{fig:bonds}a)
bonds preclude any hopping of the opposite type. For $\alpha = 0$ dimer
configurations, the interdimer diagonal hopping on (\ref{fig:bonds}b)
bonds is always of avoided--blocking type, while on (\ref{fig:bonds}c)
bonds a blocking can occur, and like the other terms is evaluated using
$\langle {\vec S}_i \cdot {\vec S}_j \rangle$. For $\alpha = 1$,
off--diagonal hopping on the interdimer bonds is evaluated with
$\langle {\vec S}_i \cdot {\vec S}_j \rangle = 0$ between the spin
singlets: all processes on (\ref{fig:bonds}b) bonds are those for one
active orbital; complications arise only for (\ref{fig:bonds}c) bonds,
where an interdimer bond between parallel dimers has two active AO
orbitals, while one between dimers which are not parallel has two
active FO orbitals.

At $\alpha = 0.5$, the energy of an (os/st) or (ss/ot) dimer state
augmented by diagonal hopping processes is minimized by states
(\ref{pdc}a) and (\ref{pdc}d): the interdimer bond contributions of all
coverings in Fig.~\ref{pdc} are equal, despite the different type
counts, so only the $\alpha = 0$ energy is decisive. At $\eta = 0$,
\begin{eqnarray}
E_o^{(\ref{pdc}a)} (0.5) & = & - \frac12 \left( \frac{13}{32}
+ \frac{2}{3} \! \cdot \! \frac{1}{4} \right) J
\; = \; - \frac{55}{192}\, J, \;\;\;\;\;\; \\
E_o^{(\ref{pdc}d)} (0.5) & = & - \frac12 \left( \frac{13}{32}
+ \frac{1}{3} \! \cdot \! \frac{1}{8} + \frac12 \! \cdot \!
\frac{1}{4} \right) J \nonumber \\ & = & - \frac{55}{192}\, J.
\label{ede5bodd}
\end{eqnarray}
The energy of a spin--singlet dimer state augmented by off--diagonal
hopping is minimal in states (\ref{pdc}b) and, curiously, (\ref{pdc}a):
although the latter has explicitly a worse ground--state energy than the
other states shown, the effect of the additional hopping is strong, not
least because all interdimer type--(\ref{pdc}c) bonds are between parallel
dimers. Thus at $\eta = 0$,
\begin{eqnarray}
E_d^{(\ref{pdc}a)} (0.5) & = & - \frac12 \left( \frac{5}{24} + \frac12
\cdot \frac{2}{3} \right) J \;\; = \;\; - \frac{13}{48} J, \nonumber \\
E_d^{(\ref{pdc}b)} (0.5) & = & - \frac12 \left( \frac14 + \frac13 \cdot
\frac14 + \frac12 \cdot \frac{2}{3} \cdot \frac12 + \frac12 \cdot \frac13
\cdot \frac14 \right) J \nonumber \\ & = & - \frac{13}{48} J.
\label{ede5bdod}
\end{eqnarray}
Despite the fact that these are two completely different expansions, it is
worth noting that the two sets of numbers are rather similar, which occurs
because the significantly inferior energy of the $\alpha = 1$ ground state
is compensated by the significantly greater interdimer bond energies
available from off--diagonal hopping processes. However, this result also
implies that no special combinations of diagonal and off--diagonal dimers
can be expected to yield additional interdimer energies beyond this value.

Taking the covering (\ref{pdc}a) as representative of the lowest
available energy, but bearing in mind that many other states lie very
close to this value, an average over the two approaches yields
\begin{equation}
E_{\rm dim}^{(\ref{pdc}a)} (0.5) = - \frac{107}{384} J
\label{ejunk}
\end{equation}
at $\eta = 0$. This number is no longer lower than the value obtained in
Sec.~IIID for fully ordered states gaining energy from mixed processes,
raising the possibility that non--dimer--based phases may be competitive
in the intermediate regime, where neither of the limiting types of dimer
state alone is expected to be particularly suitable. However, we will
not investigate this question more systematically here, and caution that
the approximations made both in Sec.~IIID and here make it difficult to
draw a definitive conclusion.

\subsection{ Summary }
\label{sec:dimsum}

The results of this section make it clear that static dimer states, while
showing the same energetic trend, are considerably more favorable than
any long--range--ordered states (Sec.~III) over most of the phase diagram.
As a function of $\alpha$, the dimer energy increases monotonically from
$- \frac{13}{32} J$ to $- \frac14 J$, and both end--point values also
lie below the results obtained for quasi--1D spin--disordered states in
Sec.~III. We stress that the results of this section are provisional in
the sense that we have not performed a systematic exploration of all
possible dimer coverings, but rather have focused on a small number of
examples illustrative of the limiting cases in terms of interdimer bond
types. More importantly, we have considered only static dimer coverings
with effective interdimer interactions: the kinetic energy contributions
due to dimer resonance processes for all values of $\alpha < 1$ are
missing in this type of calculation. For this reason, we have also
refrained from investigating higher--order processes, which may select
particular dimer states from a manifold of static coverings degenerate
at the level of the current considerations. Gaining some insight into
the magnitude and effects of resonance contributions is the subject of
the following section.

\section{Exact diagonalization}
\label{sec:ed}

\subsection{Clusters and correlation functions}
\label{sec:cluster}

In this Section we present results obtained for small systems by
full exact diagonalization (ED). Because each site has two spin
and three orbital states, the dimension of the Hilbert space
increases with cluster size as $6^N$, where $N$ is the number of
sites. As a consequence, we focus here only on systems with $N =$
2, 3, and 4 sites: all three clusters can be considered as two--,
three-- or four--site segments of an extended triangular lattice,
connected with periodic boundary conditions. For the single bond
and triangle this only alters the bond energies by a factor of two,
a rescaling not performed here, but for the four--site system it is
easy to see that the intercluster bonds ensure that the system
connectivity is tetrahedral. We will also compare some of the
single--bond and tetrahedron results with those for a four--site
chain. Other accessible cluster sizes ($N =$ 5 and 6) yield awkward
shapes which disguise the intrinsic system properties. Indeed we will
emphasize throughout this Section those features of our very small
clusters which can be taken to be generic, and those which are
shape--specific.

Given the clear tendency to dimerization illustrated in
Secs.~\ref{sec:mfa} and \ref{sec:dim}, it is to be expected
that spin correlation lengths in all regimes of $\alpha$ are
very small. To the extent that the behavior of the model for any
parameter set is driven by local physics, the cluster results
should be highly instructive for such trends as dimer formation,
relative roles of diagonal and off--diagonal hopping, dimer
resonance processes, lifting of degeneracies both in the orbital
sector and between states of (os/st) and (ss/ot) dimers, and the
importance of joint spin--orbital correlations. However, generic
features of extended systems which cannot be accessed in small
clusters are those concerning questions of high system degeneracy
and subtle selection effects favoring specific states.

We will compute and discuss the cluster energies, degeneracies,
site occupations, bond hopping probabilities in diagonal and
off--diagonal channels (discussed in Sec.~VC), and the spin,
orbital, and spin--orbital (four--operator) correlation functions.
All of these quantities will be calculated for representative values
of $\alpha$ and $\eta$ covering the full phase diagram, and each
contains important information of direct relevance to the local
physics properties listed in the previous paragraph. Although the
systems we study are perforce rather small, we will show that one
may recognize in them a number of general trends valid also in the
thermodynamic limit.

We introduce here the three correlation functions, which for a bond
$\langle ij\rangle$ oriented along axis $\gamma$ are given respectively
by
\begin{eqnarray}
\label{ss} S_{ij}\!& \equiv & \frac{1}{d} \; \sum_n \big\langle n
\big| {\vec S}_i
\cdot {\vec S}_j \big| n \big\rangle, \\
\label{tt} T_{ij}\!& \equiv & \frac{1}{d} \; \sum_n \big\langle n
\big| {\vec T}_{i
\gamma} \cdot {\vec T}_{j\gamma} \big| n \big\rangle, \\
\label{ct} C_{ij}\!& \equiv & \! \frac{1}{d} \; \sum_n \big\langle
n \big| ({\vec S}_i \! \cdot \! {\vec S}_j) ({\vec T}_{i \gamma}
\! \cdot \! {\vec T}_{j \gamma}) \big| n \big\rangle \nonumber \\
&-&\!\frac{1}{d^2} \sum_n \big\langle n \big| {\vec S}_i \! \cdot
\! {\vec S}_j \big| n \big\rangle \sum_m \big\langle m \big| {\vec
T}_{i\gamma} \! \cdot \! {\vec T}_{j\gamma} \big| m \big\rangle,
\end{eqnarray}
where $d$ is the degeneracy of the ground state. The definitions of the
spin ($S_{ij}$) and orbital ($T_{ij}$) correlation functions are standard,
and we have included explicitly all of the quantum states $\{|n\rangle\}$
which belong to the ground--state manifold. The correlation function
$C_{ij}$ (\ref{ct}) contains information about spin--orbital entanglement,
as defined in Sec.~I: it represents the difference between the average over
the complete spin--orbital operators and the product of the averages over
the spin and orbital parts taken separately. It is formulated in such a
way that $C_{ij} = 0$ means the mean--field decoupling of spin and orbital
operators on every bonds is exact, and both subsystems may be treated
independently from each other. Such exact factorizability is
found\cite{Ole06} in the high--spin states at large $\eta$; its
breaking, and hence the need to handle coupled spin and orbital
correlations in a significantly more sophisticated manner, is what
is meant by ``entanglement'' in this context.

\subsection{Single bond}
\label{sec:bond}

\begin{figure}[t!]
\includegraphics[width=7.5cm]{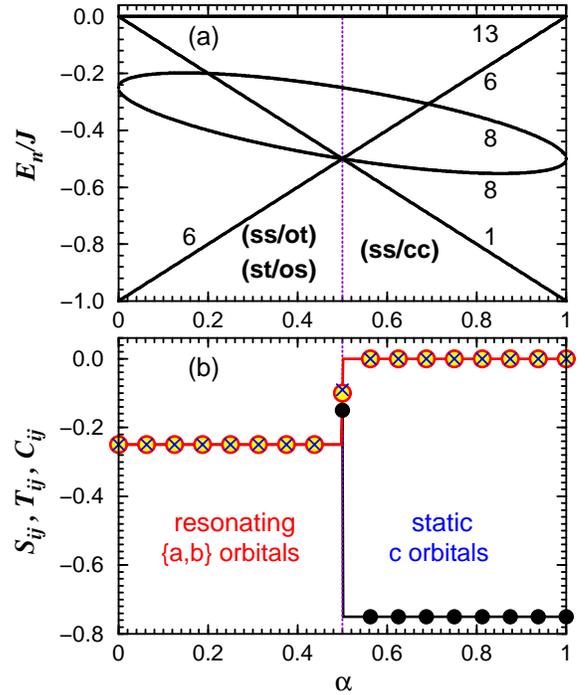}
\caption{(Color online)
Evolution of the properties of a single bond $\gamma \equiv c$ as a
function of $\alpha$ at $\eta = 0$:
(a) energy spectrum (solid lines) with degeneracies as shown;
(b) spin ($S_{ij}$, filled circles), orbital ($T_{ij}$, empty circles),
and spin--orbital ($C_{ij}$, $\times$) correlations: $S_{ij} = T_{ij}
 = C_{ij} = -0.25$ for $\alpha < 0.5$, while $T_{ij} = C_{ij} = 0$ for
$\alpha > 0.5$.
The ground--state energy $E_0$ is $-J$ for both the superexchange
($\alpha = 0$) and direct--exchange ($\alpha = 1$) limits, and its
increase between these is a result of the scaling convention. The
transition between the two regimes occurs by a level crossing at
$\alpha = 0.5$.
For $\alpha < 0.5$, the two types of dimer wave function [(ss/ot) and
(os/st)] are degenerate ($d = 6$) for resonating orbital configurations
$\{ab\}$, while at $\alpha > 0.5$, the nondegenerate spin singlet is
supported by occupation of $c$ orbitals at both sites [(ss/cc)]. }
\label{fig:bond}
\end{figure}

We consider first a single bond oriented along the $c$--axis
(Fig.~\ref{fig:bond}). In the superexchange limit the active
orbitals are $a$ and $b$, while for direct exchange only the
$c$ orbitals contribute in Eq.~(\ref{som}). As discussed in
Sec.~\ref{sec:dimsex}, a single bond gives energy $-J$ in the
superexchange model ($\alpha = 0$) [Fig.~\ref{fig:bond}(a)],
where the ground state has degeneracy $d = 6$ at $\eta = 0$,
from the two triply degenerate wave functions (ss/ot) and (os/st).
At finite $\eta$, the latter is favored as it permits a greater
energy gain from excitations to the lowest triplet state in the
$d^2$ configuration [Eqs.~(\ref{eosst}) and (\ref{essot})].

Although orbital fluctuations which appear in the mixed exchange
terms in Eq.~(\ref{Hm}) may in principle contribute at $\alpha > 0$,
one finds that the wave function remains precisely that for $\alpha
 = 0$, {\it i.e.} (ss/ot) degenerate with (os/st), all the way to
$\alpha = 0.5$. Thus for the parameter choice specified in Sec.~II,
the ground--state energy increases to a maximum of $E_0 = -0.5 J$
here [Fig.~\ref{fig:bond}(a)]. The degeneracy $d = 6$ is retained
throughout the regime $\alpha < 0.5$, and only at $\alpha = 0.5$
do several additional states join the manifold, causing the degeneracy
to increase to $d = 15$. For the entire regime $\alpha \in (0.5,1]$,
the ground state is a static orbital configuration with $c$ orbitals
occupied at both sites to support the spin singlet, and $d = 1$.
The evolution of the spectrum with $\alpha$ demonstrates not only
that superexchange and direct exchange are physically distinct, unable
to contribute at the same time, but that the two limiting wave
functions are extremely robust, their stability quenching all mixed
fluctuations for a single bond. In this situation it is not the
ground--state energy but the higher first excitation energy which
reveals the additional quantum mechanical degrees of freedom active
at $\alpha = 0$ compared to $\alpha = 1$ [Fig.~\ref{fig:bond}(a)].

The spin, orbital, and composite spin--orbital correlation functions
defined in Eqs.~(\ref{ss})--(\ref{ct}) give more insight into the nature
of the single--bond correlations. The degeneracy of wavefunctions (ss/ot)
and (os/st) for $0 \leq \alpha < 0.5$ leads to equal spin and orbital
correlation functions, as shown in Fig.~\ref{fig:bond}(b), and averaging
over the different states gives $S_{ij} = T_{ij} = - \frac14$. As a
singlet for one quantity is matched by a triplet for the other, the
two sectors are strongly correlated, and indeed $C_{ij} = - \frac14$,
indicating an entangled ground state. However, a considerably more
detailed analysis is possible. Each of the six individual states
$\{ |n \rangle \}$ within the ground manifold has the expectation value
$\big\langle n \big| ({\vec S}_i \! \cdot \! {\vec S}_j) ({\vec T}_{i
\gamma} \! \cdot \! {\vec T}_{j \gamma}) \big| n \big\rangle \nonumber
 = - \frac{3}{16}$, which we assert is the minimum possible when the
spin and pseudospin are the quantum numbers of only two electrons.
It is clear that if the operator in $C_{ij}$ is evaluated for any
one of these states alone, the result is zero. Entanglement arises
mathematically because of the product of averages in the second
term of Eq.~(\ref{ct}), and physically because the ground state is
a resonant superposition of a number of degenerate states. We
emphasize that the resulting value, $C_{ij} = - \frac14$, is the
minimum obtainable in this type of model, reflecting the maximum
possible entanglement. We will show in Sec.~VE that this value is
also reproduced for the Hamiltonian of Eq.~(\ref{som}) on a linear
four--site cluster, whose geometry ensures that the system is at
the SU(4) point of the 1D SU(2)$\otimes$SU(2) model.\cite{Ole06}

By contrast, for $\alpha > 0.5$ those states favored by superexchange
become excited, and the spin--singlet ground state has $S_{ij} = -
\frac{3}{4}$. The orbital configuration is characterized by $\langle
n_{ic} n_{jc} \rangle = 1$, a rigid order which quenches all orbital
fluctuations (indeed, the orbital pseudospin variables ${\vec T}_{i
\gamma}$ are zero). Thus the spin and orbital parts are trivially
decoupled, giving $C_{ij} = 0$. Finally, at the transition point
$\alpha = 0.5$, averaging over all 15 degenerate states yields
$S_{ij} = -0.15$, $T_{ij} = - 0.10$, and $C_{ij} = - 0.09$. In
summary, the very strong tendency to dimer formation in the two limits
$\alpha = 0$ and $\alpha = 1$ precludes any contribution from mixed
terms on a single bond, leading to a very simple interpretation of the
ground--state properties for all parameters.

\subsection{Triangular cluster}
\label{sec:tri}

We turn next to the triangle, which has one bond in each of the
lattice directions $a$, $b$, and $c$. Unlike the case of the single
bond, here the spin--orbital interactions are strongly frustrated, in
a manner deeper than and qualitatively different from the Heisenberg
spin Hamiltonian. Not only can interactions on all three bonds not
be satisfied at the same time, but also the actual form of these
interactions changes as a function of the occupied orbitals. The
triangle is sufficient to prove (numerically and analytically) the
inequivalence in general of the original model and the model after
local transformation, for frustration reasons discussed in Sec.
\ref{sec:sex}.

We begin with the observation that the results to follow are
interpreted most directly in terms of resonant dimer states on the
triangle. This fact is potentially surprising, given that the number
of sites is odd and dimer formation must always exclude one of them,
but emphasizes the strong tendencies to dimer formation in all
parameter regimes of the model. For their interpretation we use a
VB ansatz where it is assumed that one bond is occupied by an
optimal dimer state, minimizing its energy, and the final state of
the system is determined by the contributions of the other two bonds.
This ansatz is perforce only static, and breaks the symmetry at a
crude level, but enables one to understand clearly the effects of
the resonance processes captured by the numerical studies in
restoring symmetries and lowering the total energy.

Considering first the VB ansatz for the superexchange model, the
energy $-J$ may be gained only on a single bond, in one of two ways.
For the bond spin state to be a singlet ($S = 0$, (ss/ot) wave function),
two different active orbitals are occupied at both sites in one of the
orbital triplet states. The other two bonds lower the total energy when
the third site has an electron of the third orbital color, each gaining
an energy of $- 0.25J$ due to the orbital interactions in Eq.~(\ref{Hs}).
The energy of the triangle is then $E_{\rm VB} (0) =
- 0.5J$ per bond, and the cluster has a low--spin ($S = \frac12$) ground
state with degeneracy $d = 6$ from the combination of the orbital triplet
and the spin state of the third electron. We stress that the location
($a$, $b$, or $c$ bond) of the spin singlet does not contribute to
the degeneracy because the three VB states are mixed within the
ground state by the contributing off--dimer hopping processes. The
same considerations applied to an (os/st) dimer on one of the bonds
of the triangle shows that there is no color and spin state of the third
electron which allows both non--dimer bonds to gain the energy $- 0.25J$
simultaneously, so the cluster has a higher energy of $- \frac{5}{12}J$
per bond. Thus the VB ansatz illustrates a lifting of the degeneracy
between the two types singlet state, the physical origin of which lies
in the permitted off--dimer fluctuation processes, and this will be
borne out in the calculations below. However, the net spin state of
the cluster has little effect on the estimated energy of the (os/st)
case, and its high--spin version ($S = \frac{3}{2}$) will be a strong
candidate for the ground state at higher values of $\eta$. In the
direct--exchange limit ($\alpha = 1$), the VB ansatz for spin singlets
again returns an energy $E_{\rm VB} (1) = - \frac{5}{12}J$, also because
only one non--dimer bond can contribute. Here the off--dimer processes
are restricted to the third electron, which has arbitrary color and spin,
and cannot mix the three VB states, whence the degeneracy is $d = 12$.

\begin{figure}[t!]
\includegraphics[width=7.7cm]{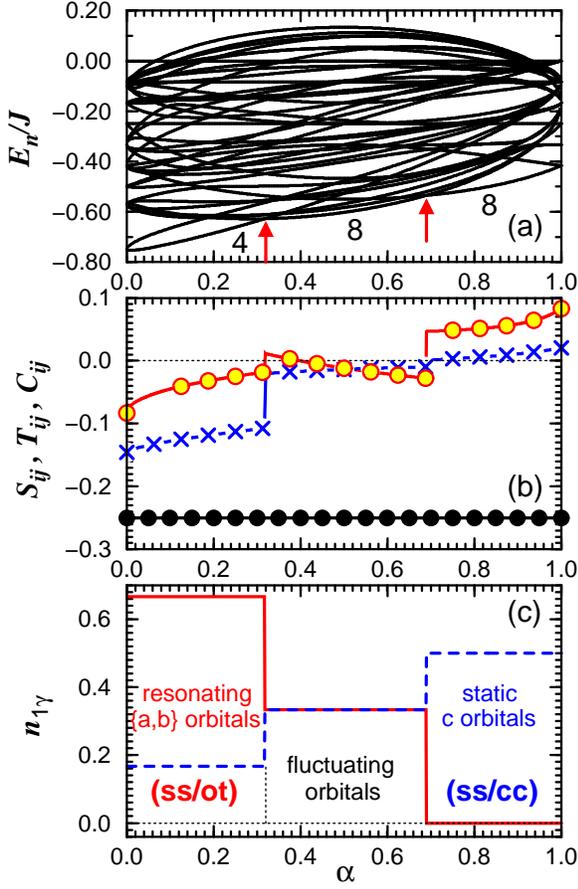}
\caption{(Color online)
(a) Energy spectrum per bond for a triangular cluster as a function
of $\alpha$ for $\eta = 0$. Ground--state degeneracies are as indicated,
with $d = 6$ at $\alpha = 0$ and $d = 12$ at $\alpha = 1$. The arrows
mark two transitions in the nature of the (low--spin) ground state,
which are further characterized in panels (b) and (c). (b) Spin
($S_{ij}$, filled circles), orbital ($T_{ij}$, empty circles), and
spin--orbital ($C_{ij}$, $\times$) correlation functions on the $c$
bond. (c) Average electron densities in the $t_{2g}$ orbitals at site 1
[Figs.~\ref{fig:hops}(b,c)], showing $n_{1b}$ (solid line) and $n_{1a}
 = n_{1c}$ (dashed). The orbital labels are shown for a $c$ bond. All
three panels show clearly a superexchange regime for $\alpha < 0.32$,
a direct--exchange regime for $\alpha > 0.69$, and an intermediate
regime ($0.32 < \alpha < 0.69$). A full description is presented in
the text.}
\label{fig:tri}
\end{figure}

With this framework in mind, we turn to a description of the
numerical calculations at all values of $\alpha$, beginning with
the most important results: at $\alpha = 0$ the degeneracy is $d =
6$, and hence VB resonance is confirmed, yielding an energy very
much lower than the static estimate, at $E_0 = - 0.75J$ per bond
[Fig.~\ref{fig:tri}(a)]. Thus strong orbital dynamics and positional
resonance effects operate in the ground--state manifold. These break
the (ss/ot)/(os/st) symmetry, but act to restore other symmetries
broken in the VB ansatz. At $\alpha = 1$, the energy and degeneracy
from the VB ansatz are exact, showing that the orbital sector is
classical andintroduces no resonance effects.

\begin{figure}[t!]
\includegraphics[width=7.7cm]{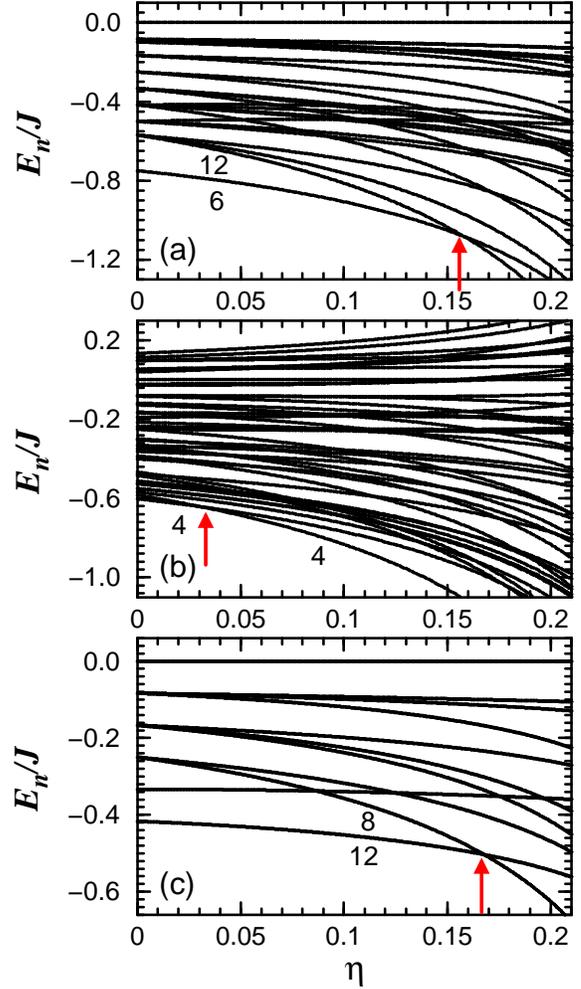}
\caption{(Color online) Energy spectra for a triangular cluster as
a function of Hund exchange $\eta$. Energies are quoted per bond,
and shown for: (a) $\alpha = 0$, (b) $\alpha = 0.5$, and (c)
$\alpha = 1$. The arrows indicate transitions at $\eta_c$ from the
low--spin ($S = 1/2$) to the high--spin ($S = 3/2$) ground state.
The numbers in all panels give degeneracies for the two lowest
states for $\eta<\eta_c$ and $\eta>\eta_c$, respectively. }
\label{fig:spe3}
\end{figure}

Figure \ref{fig:tri}(a) shows the complete spectrum of the
triangular cluster for all ratios of superexchange to direct
exchange, and in the absence of Hund coupling. Frustration of
spin--orbital interactions is manifest in rather dense energy
spectra away from the symmetric points, and in a ground--state
energy per bond significantly higher than the minimal value $-J$.
At $\alpha = 0$ the spectrum is rather broad, with a significant
number of states of relatively low degeneracy due to the strong
fluctuations and consequent mixing of VB states in this regime.
However, even in this case the ground state is well separated from
the first excited state. As emphasized above, the ground--state
energy, $E_0 (0) = - 0.75 J$, is quite remarkable, demonstrating a
very strong energy gain from dimer resonance processes. By
contrast, the value $E_0 (1) = - \frac{5}{12} J$ per bond found at
$\alpha = 1$ is exactly equal to that deduced from the VB ansatz,
demonstrating that this wave function is exact. Here the excited
states have high degeneracies, mostly of orbital origin, and thus
the spectrum shows wide gaps between these manifolds of states;
this effect is more clearly visible in Fig.~\ref{fig:spe3}(c). The
degeneracies shown in Fig.~\ref{fig:tri}(a) are discussed below.
In the intermediate regime, many of the degeneracies at the
end--points are lifted, leading to a very dense spectrum. The two
transitions at $\alpha = 0.32$ and $\alpha = 0.69$ appear as clear
level--crossings: the intermediate ground state is a highly
excited state in both of the limits ($\alpha = 0,1$), reinforcing
the physical picture of a very different type of wave function
dominated by orbital fluctuations and, as we discuss next, with
little overt dimer character.

The correlation functions for any one bond of the triangle are shown
in Fig.~\ref{fig:tri}(a). That $S_{ij}$ is constant for all $\alpha$
can be understood in the dimer ansatz by averaging over the three
configurations with one (ss/ot) or (ss) bond and one 'decoupled' spin
on the third site, which gives $S_{ij} = - \frac{1}{4}$ everywhere.
The orbital and spin--orbital correlation functions show a continuous
evolution accompanied by discontinuous changes at two transitions, where
the nature of the ground state is altered. The orbital correlation
function $T_{ij} = - \frac{1}{12}$ at $\alpha = 0$ may be understood as
an average over the orbital triplet ($+ \frac14$) and the two non--dimer
bonds (each $- \frac14$). When $\alpha$ increases, this value is weakened
by orbital fluctuations, and undergoes a transition at $\alpha = 0.32$ to
a regime where orbital fluctuations dominate, and $T_{ij}$ is close to
zero. Above $\alpha = 0.69$, $T_{ij}$ becomes positive, and approaches
$+ \frac{1}{12}$ as $\alpha \rightarrow 1$, indicating that the wave
function changes to the static--dimer limit. While ${\vec T}_{ic}$
vanishes on the $c$ bond here, the cluster average has a finite value
due to the contribution $T_{ij} = \frac14$ from the active non--singlet
bond.

The spin--orbital correlation function $C_{ij}$ also marks clearly the
three different regimes of $\alpha$. When $\alpha < 0.32$, $C_{ij}$
has a significant negative value [Fig.~\ref{fig:tri}(b)] whose primary
contributions are given by the four--operator component $\langle ({\vec
S}_i\!\cdot\! {\vec S}_j) ({\vec T}_{i\gamma} \! \cdot \! {\vec
T}_{j\gamma})\rangle$. By contrast, $C_{ij}$ is close to zero in
the intermediate regime, increasing again to positive values for
$\alpha > 0.69$. For all $\alpha > 0.32$, $C_{ij}$ can be shown to
be dominated by the term $- S_{ij} T_{ij}$ in Eq.~(\ref{ct}), while
the four--operator contribution is small, and vanishes as $\alpha
\rightarrow 1$. Thus entanglement, defined as the lack of factorizability
of the spin and orbital sectors, can be finite even for vanishing joint
spin--orbital dynamics.

Further valuable information is contained in the orbital occupancies
at individual sites [Fig.~\ref{fig:tri}(b)], which show clearly the
three different regimes. Although there is always on average one
electron of each orbital color on the cluster, these are not equally
distributed, as each site participates only in two bonds and the
symmetry is broken. A representative site,  labelled $1$ in
Figs.~\ref{fig:hops}(b,c)] has only $a$ and $c$ bonds, and hence the
electron density in the $b$ orbital is expected to differ from the
other two. The values $n_b = 2/3$ and $n_a = n_c = 1/6$ found in
the regime $\alpha < 0.32$ is understood readily as following from
a 1/3 average occupation of $(ab)$ and $(bc)$ orbital triplet states
on the $c$ and $a$ bonds, respectively, and of an $(ac)$ orbital triplet
state on the $b$ bond, which ensures that the electron at site $1$ is
in orbital $b$ [Fig.~\ref{fig:hops}(b)]. By contrast, in the regime
$\alpha > 0.69$, only the two static orbital configurations $(cc)$
and $(bb)$ on the $c$ and $b$ bonds contribute, and $n_a = n_c =
\frac{1}{2}$, while $n_b = 0$; when the system is in the third
possible spin--singlet state, with a $(bb)$ orbital state on the
$b$ bond, the third electron is either $a$ or $c$. Between these
two regimes ($0.32 < \alpha < 0.69$) is an extended phase with equal
average occupancy of all three orbitals at each site, a potentially
surprising result given the broken site symmetry of the cluster. While
this may be interpreted as a restoration of the symmetry of the orbital
sector by strong orbital fluctuations, including those due to terms in
$H_m$ (\ref{Hm}), it does not imply a higher symmetry of the strongly
frustrated interactions at $\alpha = 0.5$.

The spectra as a function of Hund coupling $\eta$ are shown in
Fig.~\ref{fig:spe3} for the $\alpha = 0$ and $\alpha = 1$ limits, and
at $\alpha = 0.5$ to represent the intermediate regime. The lifting of
degeneracies as a function of $\eta$ is a generic feature. States of
higher spin are identifiable by their stronger dependence on $\eta$,
and in all three panels a transition is visible from a low--spin to
a high--spin state. At $\alpha = 0$ [Fig.~\ref{fig:spe3}(a)], the
large low--$\eta$ gap to the next excited state results in the
transition occurring at the rather high value of $\eta_c = 0.158$.
This can be taken as a further indication of the exceptional stability
of the resonance--stabilized ground state in the low--spin sector. The
degeneracy $d = 12$ of the high--spin state is discussed below.

\begin{figure}[t!]
\includegraphics[width=7.7cm]{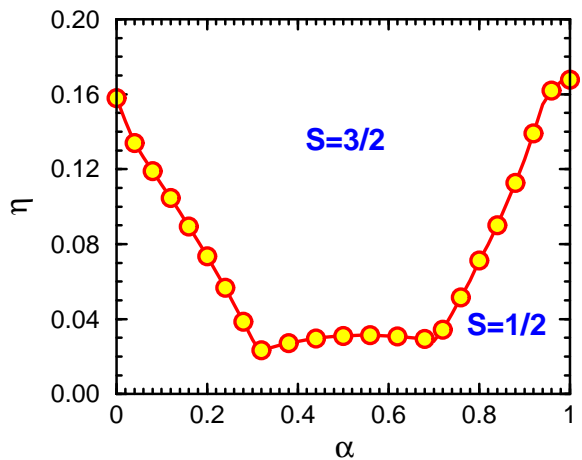}
\caption{(Color online) Phase diagram of the triangular cluster in
the plane $(\alpha,\eta)$. The spin states below and above the
transition line $\eta_c (\alpha)$ are respectively spin doublet
($S = 1/2$) and spin quartet ($S = 3/2$).} \label{phd3}
\end{figure}

The transition to the high--spin state at $\alpha = 1$ also occurs
at a high critical value, $\eta_c = 0.169$
[Fig.~\ref{fig:spe3}(a)], due in this case quite simply to the
lack of competition for the strong singlet states on individual
bonds. Only in the intermediate regime, $0.32 < \alpha < 0.69$,
where we have shown already that the orbital state is quite
different from that in either limit [Fig.~\ref{fig:tri}], is the
transition to the high--spin state much more sensitive to $\eta$.
The orbital fluctuations in this phase occur both in the low--spin
and the high--spin channel, making these very similar in energy,
and the transition occurs for $\alpha = 0.5$ at only $\eta_c =
0.033$ [Fig.~\ref{fig:spe3}(b)]. As expected from the $\alpha = 0$
limit, where fluctuations are also strong, the characteristic
features of this energy spectrum are low degeneracy and a
semicontinuous nature. The location of the high--spin transition
as a function of $\alpha$ may be used to draw a phase diagram for
the triangular cluster, which has the rather symmetric form shown
in Fig.~\ref{phd3}.

Yet more information complementary to that in the energy spectra
and correlation functions can be obtained by considering the
average ``occupation correlations'' for a bond $\langle ij \rangle
\parallel \gamma$,
\begin{eqnarray}
\label{orbipro}
 P & = & \langle n_{i\gamma} n_{j\gamma} \rangle, \\
 Q & = & \langle n_{i\gamma} (1 - n_{j\gamma}) \rangle
     +   \langle (1 - n_{i\gamma}) n_{j\gamma} \rangle, \\
 R & = & \langle (1 - n_{i\gamma}) (1 - n_{j\gamma}) \rangle.
\end{eqnarray}
These probabilities ($P + Q + R = 1$) reflect directly the nature
of the resonance processes contributing to the energy of the cluster
states, in that they show the relative importance of diagonal and
off--diagonal hopping in the ground states, and the evolution of
these contributions with $\alpha$ and $\eta$. We do not present
these quantities in detail here, but only summarize the overall
picture of the ground state whose understanding they help elucidate.

For this summary we return to the VB framework, which accounts for
many of the basic properties illustrated in the numerical results
presented above. Considering first the low--spin states ($\eta =
0$), at $\alpha = 0$ the ground state is given by one (ss/ot) dimer
resonating around the three bonds of the cluster; the third site has
the third color, its hopping gives a large value of $Q = 1/3$ ($R =
2/3$ from the pure superexchange channel) and its spin an addition
twofold degeneracy ($d = 3 \times 2 = 6$); the orbital occupation of
the (ss/ot) dimer is responsible for the net 1/6:1/6:2/3 occupation
distribution. When $\alpha > 0$ the state remains essentially one
with a resonating spin singlet, large $Q$ and dominant $R$, but the
orbital triplet degeneracy is lifted to $2 + 1$ and the ground--state
degeneracy to $d = 2 \times 2$. All quantities, including $P$, $Q$,
and $R$, undergo discontinuous changes at $\alpha \simeq 0.32$, and
in this regime there is no longer strong evidence for an interpretation
in terms of resonating spin singlets: large $Q \simeq 2/3$ and the equal
site occupations suggest the dominance of mixed hopping processes which
are not consistent with either mechanism of singlet formation. The
retention of fourfold degeneracy across this transition is largely
accidental, and stems from twofold spin and orbital contributions.
Only for $\alpha > 0.69$ is a spin--singlet description once again
valid: here $P$ becomes significant, as the resonating singlet is
stabilized by diagonal hopping where the orbital has the bond
color. The third site now has one of two possible colors, its
hopping keeps $Q$ large, and its spin yields another twofold
degeneracy, as do the orbital states, whence the net degeneracy
is $d = 2 \times 2 \times 2$. Only at $\alpha = 1$ does the spin
singlet become static, while the third site still has either of
the other colors, yielding the symmetric result $P = 1/9$, $Q = R
= 4/9$, and degeneracy $d = 12$.

A similar description is possible in the high--spin states at
$\eta > \eta_c$. At $\alpha = 0$ the (os/st) dimer is rendered
static by the fact that hopping to the third site is now excluded
if it has the third color, and so instead this site takes one of
the singlet colors, a twofold degree of freedom which, however, does
not allow singlet motion; as a consequence the orbital occupation
is uniform (1/3:1/3:1/3), the hopping processes include contributions
in the diagonal channel ($P = 1/6$, $Q = 1/3$, $R = 1/2$) and the
degeneracy is $d = 3 \times 4 = 12$.
For $\alpha > 0$ the orbital singlet may again resonate, but the
third site retains one of the singlet colors, orbital degeneracy
is broken and $d = 4$. Once again strong mixed processes dominate
the intermediate regime, in which the spin state is not an
important determining factor. Above $\alpha = 0.69$ the critical
value $\eta_c$ required to overcome spin singlet formation becomes
large again, and the high--spin state is one where avoided--blocking
processes (large $Q$) dominate, while broken orbital degeneracy keeps
$d = 4$. Finally, at $\alpha = 1$ one obtains a pure avoided--blocking
state with orbital configurations $acb$ or $cba$ for the sites
$(1,2,3)$ of Fig.~\ref{fig:hops}(c), and consequent degeneracy
$d = 4 \times 2 = 8$. Thus it is clear that the high--$\eta$ region
is also one yielding interesting orbital models with nontrivial ground
states, some including orbital singlet states.

\subsection{Tetrahedral cluster}
\label{sec:tet}

As in the case of the triangular lattice, interpretation of the
numerical results for the tetrahedral cluster (four--site plaquette
of the triangular lattice) is aided by consideration of the VB ansatz
in the two limits of superexchange and direct--exchange interactions.
The tetrahedral cluster can accommodate exactly two dimers, with all
interdimer bonds of type (\ref{fig:bonds}c), and may thus be expected
to favor dimer--based states by simple geometry. However, because the
considerations and comparisons of this subsection are given only for
this single cluster type, any bias of this sort would not invalidate
the results and trends discussed here.

Because of the different forms and symmetries of the spin and orbital
sectors, there is no possibility of elementary spin--orbital operators,
or of a ground--state wave function which is a net singlet of a higher
symmetry group. The state with two orbital singlets on one pair of bonds,
two spin singlets on a second pair and pure interdimer bonds on the third
pair does exist, but is not competitive: the energy cost for removing the
orbital singlets from the spin state maximizing their energy is by no
means compensated by the energy gain from having two spin singlet bonds
in an orbital state which also does not maximize their energy. This
result may be taken as a further indication for the stability of dimers
only in the forms (os/st) or (ss/ot) in this model, and states of shared
orbital and spin singlets are not considered further here. We return to
this point in the following subsection, in the context of the four--site
chain.

We discuss only the energies of the VB wave functions at $\eta =
0$. The minimal values obtainable for $\langle {\vec S}_i \cdot
{\vec S}_j \rangle$ and $\langle {\vec T}_{i\gamma} \cdot {\vec
T}_{j\gamma} \rangle$ on the interdimer bonds is $- 1/4$,
corresponding to the AF/AO order. Thus at $\alpha = 0$ the energy
per bond is
\begin{equation}
\label{enetet0}
   E_{os/st}(0) = E_{ss/ot}(0) = - \frac12 J,
\end{equation}
with the degeneracy of the (ss/ot) and (os/st) wave functions
restored as for the single bond. In the limit of direct exchange, the
VB wave function consists of spin singlets with two active orbitals of
the bond. The geometry of the cluster precludes these orbitals from
being active on any of the interdimer bonds, as a result of which the
energy per bond at $\eta = 0$ is
\begin{equation}
\label{enetet1}
E(1) = - \frac{1}{3}J,
\end{equation}
and the ground state has degeneracy $d = 3$.

The most important results for the tetrahedron, which we discuss
in detail in the remainder of the subsection, are the following.
At $\alpha = 0$, the exact ground state energy is $E_0 = -
0.5833J$: while not as large as in the case of the triangle (Sec.
\ref{sec:tri}), the resonance energy contribution is very
significant also for an even number of cluster sites. The
degeneracy of the numerical ground state, $d = 6$, has its origin
in only one of the (ss/ot) or (os/st) wave functions (below),
demonstrating again that there is no sense in which the quantum
fluctuations in the spin and orbital sectors are symmetrical, and
that the VB ansatz is capturing the essence of the local physics
only at a very crude level. At $\alpha = 1$, as also for the
triangular cluster, the numerical results confirm not only the
energy given by the VB ansatz but every detail (degeneracies,
occupations, correlations) of this state.

\begin{figure}[t!]
\includegraphics[width=7.7cm]{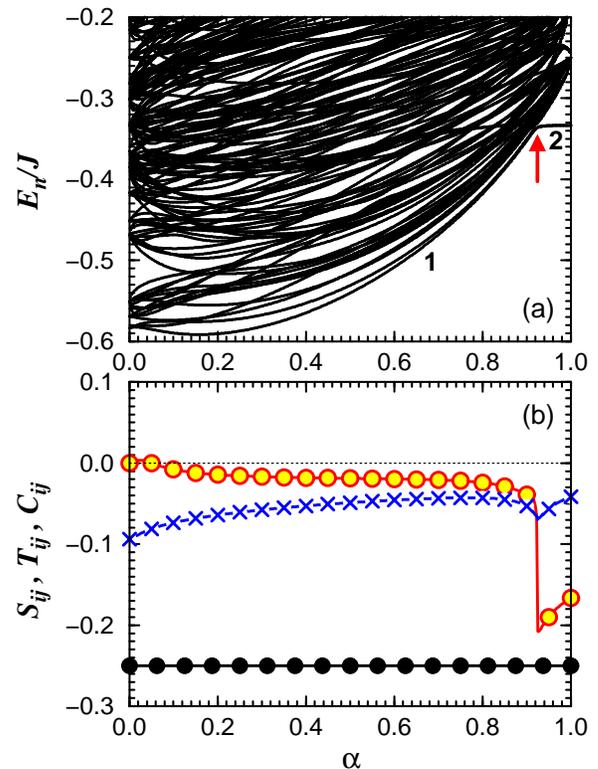}
\caption{(Color online)
(a) Energy spectrum per bond for a tetrahedral cluster as a function
of $\alpha$ for $\eta = 0$. Ground--state degeneracies are as indicated,
with $d = 6$ at $\alpha = 0$ and $d = 150$ at $\alpha = 1$. The arrow
marks a transition in the nature of the (low--spin) ground state. (b)
Spin ($S_{ij}$, filled circles), orbital ($T_{ij}$, empty circles), and
spin--orbital ($C_{ij}$, $\times$) correlation functions on the $c$ bond
of the tetrahedral cluster as functions of $\alpha$ for $\eta = 0$. }
\label{tetcor}
\end{figure}

We begin the systematic presentation of results by discussing the
energy spectra at $\eta = 0$ [Fig.~\ref{tetcor}(a)]. As soon as the
degeneracies of the superexchange limit ($\alpha = 0$) are broken,
the spectrum becomes very dense, and remains so across almost the
complete phase diagram until a level--crossing at $\alpha_c = 0.92$.
The ground--state energy for all intermediate values of $\alpha$
interpolates smoothly towards the transition, showing an initial
decrease not observed in the triangle: for the tetrahedron, mixed
hopping terms make a significant contribution, leading to an overall
energy minimum around $\alpha = 0.15$. The dominance of these terms
is indicated by both the extremely high value of $\alpha_c$ and the
steepness of the low--$\alpha$ curve where the transition to the
static VB phase is finally reached.

The bond correlation functions shown in Fig.~\ref{tetcor}(b) illustrate
the effects of corrections to the VB ansatz. The spin correlations
always have the constant value $S_{ij} = - \frac14$, which is the most
important indication of the breaking of symmetry between (ss/ot) and
(os/st) sectors at low $\alpha$: this value is an average over the
spin--singlet result $- 3/4$ (on two bonds) and four bonds with value
0, and thus it is clear that (ss/ot) dimers afford more resonance energy.
However, the proximity of (os/st) states suggests that a low value of
$\eta_c$, the critical Hund coupling for the transition to the high--spin
state, is to be expected (below).

The orbital correlations average to zero at $\alpha = 0$, a non--trivial
result whose origin lies in the breaking of nine--fold degeneracy within
the orbital sector, and remain close to this value until the transition
at $\alpha_c$. It is worth noting here that $T_{ij} = 0$ implies a higher
frustration in the orbital sector than would be obtained in the spin sector
for an (os/st) state ($S_{ij} = - \frac{1}{12}$), which is due to the
complex direction--dependence of the orbital degrees of freedom. This
phase is maintained across much of the phase diagram, with only small
changes to the correlation functions, the negative value of $T_{ij}$
reflecting an easing of orbital frustration. The lack of a phase transition
throughout the region in which mixed processes are also important suggests
that a dimer--based schematic picture of the ground state remains
appropriate for the four--site system, with only quantitative evolution
as a function of $\alpha$ until $\alpha_c = 0.92$. At $\alpha = 1$, the
result $T_{ij} = - \frac{1}{6}$ is the consequence of $c$--orbital
operators on the interdimer $a$ and $b$ bonds.

Significant spin--orbital correlations, $C_{ij}\simeq - 0.1$ at $\alpha
 = 0$ [Fig.~\ref{tetcor}(b)], are found to be due exclusively to the
four--operator term at low $\alpha$. While these negative contributions
drop steadily through most of the regime $\alpha < \alpha_c$, signifying
a gradual decoupling of orbitals and spins as the static limit ($\alpha
 = 1$) is approached, near $\alpha_c$ the negative value of $C_{ij}$ is
again enhanced by the contribution $- S_{ij} T_{ij}$ due to the interdimer
bonds. Thus, as for the triangle (Sec.~\ref{sec:tri}), the entanglement is
finite, complete factorization is not possible, and a finite value $C_{ij}
 = - \frac{1}{24}$ is found even at $\alpha = 1$. We note here that on the
tetrahedron there is little information in the orbital occupations, which
are constant ($n_{\gamma} = \frac13$) over the entire phase diagram,
demonstrating only the symmetry of this cluster geometry, and are
therefore not shown.

\begin{figure}[t!]
\includegraphics[width=7.7cm]{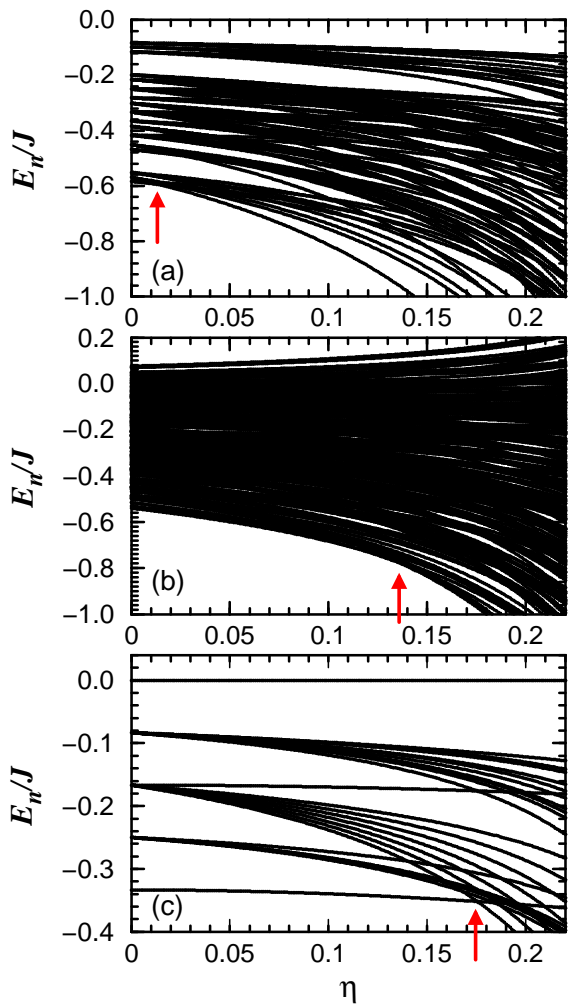}
\caption{(Color online)
Energy spectra for a tetrahedral cluster as a function of Hund
exchange $\eta$. Energies are quoted per bond, and shown for:
(a) $\alpha = 0$,
(b) $\alpha = 0.5$, and
(c) $\alpha = 1$. The
arrows indicate transitions from the low--spin ($S = 0$) to
the high--spin ($S = 2$) ground state. }
\label{fig:spe4}
\end{figure}

The spectra as a function of Hund coupling $\eta$ are shown for the
three parameter choices $\alpha = 0$, 0.5, and 1 in Fig.~\ref{fig:spe4}.
Once again, the spectra become very dense away from $\eta = 0$. At
$\alpha = 0$ [Fig.~\ref{fig:spe4}(a)] high--spin states are found also
in the low--energy sector, as a consequence of the near--degeneracy
of (ss/ot) and (os/st) states, and the high--spin transition occurs
at a very low value of $\eta_c$ [Fig.~\ref{fig:spe4}(a)]. The
direct--exchange limit is both qualitatively and quantitatively
different, because the quantum fluctuations and the corresponding
energy gains are limited to the spin sector, making the low--spin
states considerably more stable and giving $\eta_c = 0.175$
[Fig.~\ref{fig:spe4}(c)]. The spin excitation gap decreases gradually
with increasing $\eta$, but until just below $\eta_c$, for all values
of $\alpha$, the spin excitation is to $S = 1$ states. However, these
triplet states are never the ground state in the entire regime of
$\eta$, a single transition always occurring directly into an $S = 2$
state. In the intermediate regime represented by $\alpha = 0.5$, the
energy spectrum is so dense that individual states are difficult to
follow (a more systematic analysis of the spectra in different
subspaces of $S^z$ is not presented here). The high--spin transition
occurs at the relatively high value $\eta_c = 0.136$, due mainly to
the large energy gains in the low--spin sector from mixed exchange.
Further evidence for the importance of the orbital excitations in
${\cal H}_m$ (\ref{Hm}) can be found in the broadening of the spectrum
which leads to the occurrence of quantum states with weakly positive
energies: for both superexchange and direct--exchange processes, the
Hamiltonians are constructed as products of projection operators with
negative coefficients, so positive energies are excluded.

\begin{figure}[t!]
\includegraphics[width=7.7cm]{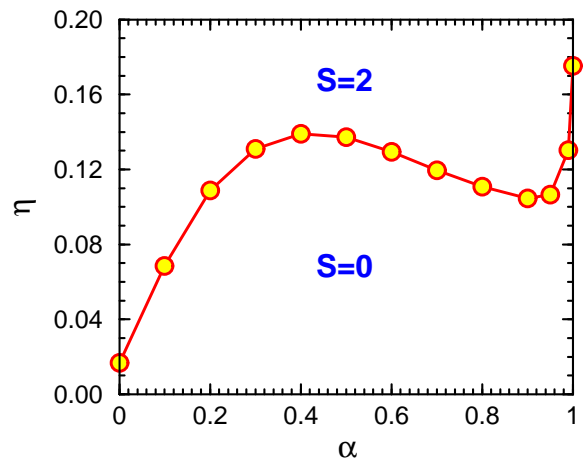}
\caption{(Color online)
Phase diagram of the tetrahedral cluster in the plane $(\alpha,\eta)$.
As for the triangular cluster, the spin states below and above the
line $\eta_c (\alpha)$ are respectively singlet ($S = 0$) and quintet
($S = 2$), with no intermediate triplet phase.}
\label{phd4}
\end{figure}

The low-- to high--spin transition points at all values of $\alpha$
can be collected to give the full phase diagram of the tetrahedron
shown in Fig.~\ref{phd4}. As shown above, in the superexchange limit
the high--spin state lies very close to the low--spin ground state,
and the transition to an $S = 2$ spin quintet occurs at $\eta_c = 0.017$.
We comment here that this high--spin state is in no sense classical or
trivial, being based on orbital singlets which are stabilized by strong
orbital fluctuations, and emphasize again that the high--spin sector
also contains a manifold of rich problems in orbital physics, which we
will not consider further here. The near degeneracy of (ss/ot) and
(os/st) states is further lifted in the presence of the mixed terms in
${\cal H}_m$, raising $\eta_c$ to values on the order of 0.12 across
the bulk of the phase diagram. For no choice of parameters is a spin
triplet state found at intermediate values of $\eta$. The reentrant
behavior close to $\alpha = 0.5$ is an indication of the importance
of mixed terms in stabilizing a low--spin state, the tetrahedral
geometry providing one of the few examples we have found of anything
other than a direct competition, and hence an interpolation, between
the two limiting cases. The rapid upturn in the limit of $\alpha \to
1$ reflects the anomalous stability of the static VB states in the
direct--exchange limit. The very strong asymmetry of the transition
line in Fig.~\ref{phd4} contrasts sharply with the near--symmetry
about $\alpha = 0.5$ observed for the triangle (Fig.~\ref{phd3}),
and shows directly the differences between those features of the phase
diagram which are universal and those which are effects of even or
odd cluster sizes in a dimer--based system.

We close our discussion of the tetrahedral cluster with a brief
discussion of degeneracies and summary of the picture provided by
the VB ansatz with additional resonance. For the orbital occupation
correlations and degeneracies, we begin with the low--spin sector
($\eta = 0$). At $\alpha = 0$ one has two (ss/ot) VBs resonating
around the 6 bonds of the cluster, a state characterized by $P = 1/6$,
$Q = 1/3$, and $R = 1/2$; however, a mixing of the orbital triplet
states lowers the degeneracy from 9 to $d = 6$. For $\alpha > 0$ the
state is the same, with slow evolution of $P < 1/6$, $Q > 1/2$, and
$R > 1/3$, but now mixed hopping terms break all orbital degeneracies,
giving $d = 1$. Only when $\alpha > 0.92$ is the ground state more
accurately characterized as one based on spin singlets of the bond
color, with significant values of $P$ and the restoration of an orbital
degeneracy $d = 2$. As $\alpha \rightarrow 1$, the diagonal hopping
component is strengthened ($P \rightarrow 1/3$) as the pair of
bond--colored spin singlets resonates, until at $\alpha = 1$ they
become static and the degeneracy is $d = 3$.

For the high--spin states in the regime $\eta > \eta_c$, at $\alpha
 = 0$ one has two resonating (os/st) VBs, with the hopping channels
unchanged and only the spin degeneracy $d = 5$. This state is not
altered qualitatively for any $\alpha < 0.92$, a transition value
independent of $\eta$. For $0.92 < \alpha < 1$, orbital correlations
are strongly suppressed and the state is characterized by hopping
processes largely of the avoided--blocking type (one active orbital,
$Q$ dominant), still with $d = 5$. Finally, $\alpha = 1$ represents
the limit of a pure avoided--blocking state ($P = 0$, $Q = 2/3$,
$R = 1/3$), where the degeneracy jumps to 150, a number which can
be understood as 5 (spin degeneracy) $\times$ [6 (number of two--color
states with no bonds requiring spin singlets) + 24 (number of
three--color states with no bonds requiring spin singlets)].

\subsection{Four--site chain}
\label{sec:chain}

As a fourth and final case, we present results from a linear four--site
cluster. While not directly relevant to the study of the triangular
lattice, this system offers further valuable insight into the intrinsic
physics of the spin--orbital model. The cluster is oriented along the
$c$--axis with periodic boundary conditions. As for the single bond
(Sec.~VB), only the $a$ and $b$ orbitals contribute at $\alpha = 0$,
where indeed one finds average electron densities per site $n_{ia} =
n_{ib} = \frac12$, and $n_{ic} = 0$. Likewise, at $\alpha = 1$ only
the $c$ orbitals are occupied, with $n_{ic} = 1$, a result dictated
by the spin singlet correlations, which are fully developed only for
complete orbital occupation.

\begin{figure}[t!]
\includegraphics[width=7.5cm]{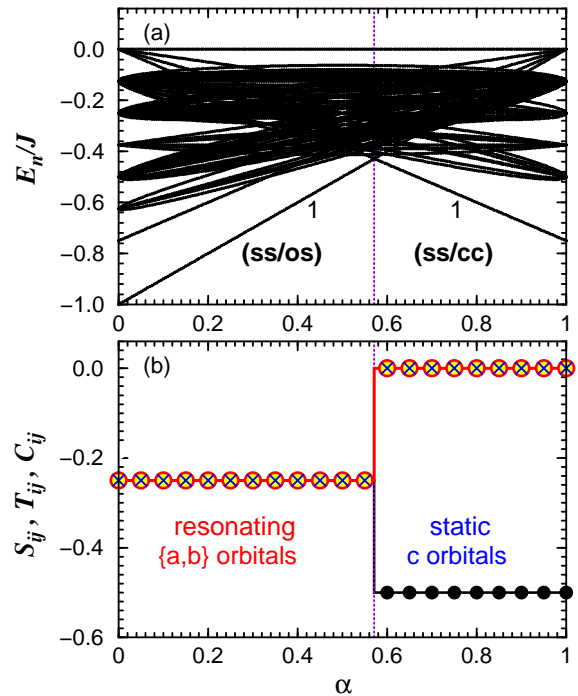}
\caption{(Color online)
Evolution of the properties of the four--site chain as a function of
$\alpha$ at $\eta = 0$: (a) energy spectrum and (b) spin ($S_{ij}$,
filled circles), orbital ($T_{ij}$, empty circles), and spin--orbital
($C_{ij}$, $\times$) correlation functions. Both panels show a
transition occurring at a level crossing at $\alpha = 4/7$. In panel
(a), the labels show a nondegenerate ground state ($d = 1$) in both
regimes, which has prediminantly spin singlet character at $\alpha
 > 0.571$, but both spin and orbital singlet components at $\alpha <
0.571$. In panel (b), $S_{ij} = T_{ij} = C_{ij} = - \frac14$ for
$\alpha < 0.571$ due to a resonating $(ab)$ orbital configuration,
while $T_{ij} = C_{ij} = 0$ for $\alpha > 0.571$ as a consequence of
the static $c$ orbital configuration.}
\label{fig:chain}
\end{figure}

The energy per bond for the four--site chain in the superexchange limit
is again $-J$, as for a single bond [Fig.~\ref{fig:chain}(a)]: somewhat
surprisingly, the bonds do not "disturb" each other, and joint
spin--orbital fluctuations extend over the entire chain. However, in
contrast to a single bond, this behavior is due to only one quantum
state, the SU(4) singlet. In this geometry, only one SU(2) orbital
subsector is selected, and the resulting SU(2)$\otimes$SU(2) system
is located precisely at the SU(4) point of the Hamiltonian.\cite{riqa}
Thus, exactly as in the SU(4) chain, all spin, orbital and spin--orbital
correlation functions are equal, $S_{ij} = T_{ij} = C_{ij} = - 0.25$,
as shown in Fig.~\ref{fig:chain}(b). For $S_{ij}$ and $T_{ij}$, this
result may be understood as an average over equal probabilities of
singlet and triplet states on each bond. In more detail, the condition
set on the correlation functions by SU(4) symmetry\cite{Fri99} is
$\frac{4}{3} \langle ({\vec S}_i \! \cdot \! {\vec S}_j) ({\vec T}_{ic}
\! \cdot \! {\vec T}_{jc}) \rangle = S_{ij} = T_{ij}$, an equality
also obeyed by the single bond (Sec.~VB). The product of $S_{ij}$ and
$T_{ij}$ in its definition ensures the identity for $C_{ij}$. The
unique ground state is nevertheless a linear superposition of states
expressed in the spin and orbital bases, and has not only finite but
maximal entanglement. This state persists, with a perfectly linear
$\alpha$--dependence, all the way to $\alpha = 1$, but ceases to be
the ground state at $\alpha = \frac{4}{7}$ [Fig.~\ref{fig:chain}(a)],
where there is a level--crossing with the $\alpha = 1$ ground state
(also perfectly linear). This latter state has a completely different,
fluctuation--free orbital configuration, with pure $c$--orbital
occupation at every site, and gains energy solely in the direct--exchange
channel. The spins and orbitals are decoupled, $T_{ij}$ and $C_{ij}$
vanish, and the spin state has $S_{ij} = - 0.50$: this result can be
understood as an equal average over bond states with ${\vec S}_i \cdot
{\vec S}_j = - \frac{3}{4}$ and $- \frac14$, and matches that obtained
for the four--site AF Heisenberg model with a resonating VB (RVB)
ground state.\cite{Faz99} The energy at $\alpha = 1$, $E_0 = - 0.75J$
[Fig.~\ref{fig:chain}(a)], is given directly by including the constant
term, $- \frac14 J$ per bond, in the definition of the Hamiltonian
(\ref{Hd}).

The results for the linear four--site cluster demonstrate again the
competition between superexchange and direct exchange. The orbital
fluctuations arising due to the mixed exchange term, ${\cal H}_m$
(\ref{Hm}), are responsible for removing the high degeneracies of
the eigenenergies in the limits $\alpha = 0$ and $\alpha = 1$
[Fig.~\ref{fig:chain}(a)]. In fact the spectrum of the excited states
is quasi--continuous in the regime around $\alpha = 0.5$, but has a
finite spin and orbital gap everywhere other than the quantum critical
point at $\alpha = \frac{4}{7}$.

These chain results raise a further possibility for the spontaneous
formation at $\alpha = 0$ of a 1D state not discussed in Sec.~III.
A set of (for example) $c$--axis chains, with only $a$ and $b$ orbitals
occupied in the pseudospin sector, would create exactly the 1D SU(4)
model, and would therefore redeem an energy $E = - \frac{3}{4} J$ per
bond from the formation of linear, four--site spin--orbital singlets.
The energy of the triangular lattice would receive a further, constant
contribution from the cross--chain bonds, which was calculated in
Eq.~(\ref{e1c1d}) for general $\eta$, and hence would be given at
$\eta = 0$ by
\begin{equation}
\label{e1dsu4}
E_{\rm 1D}^{SU(4)}(0) = - \frac13 \cdot \frac{3}{4} J
                        - \frac{1}{6} J \; = \; - \frac{5}{12} J.
\end{equation}
This energy represents a new minimum compared with all of the results
in Sec.~III. That it was obtained from a melting of both spin and orbital
order confirms the conclusion that ordered phases are inherently unstable
in this class of model, being unable to provide sufficient energy to
compete with the kinetic energy gains available through resonance
processes. That its value is now lower than that obtained for a static,
2D dimer covering (Sec.~IV) is not of any quantitative significance,
given the results of Sec.~V confirming the importance of the positional
resonance of dimers.

\subsection{Summary}
\label{sec:edsum}

To summarize, we have shown in this section the results of exact
numerical diagonalization calculations performed on small
clusters. Detailed analysis of ground--state energies,
degeneracies, site occupancies and a number of correlation
functions can be used to extract valuable information about the
local physics of the model across the full regime of parameters.
Essentially all of the quantities considered show strong local
correlations and the dominance of quantum fluctuations of the
shortest range, with ready explanations in terms of resonating
dimer states.

\begin{figure}[t!]
\includegraphics[width=8.0cm]{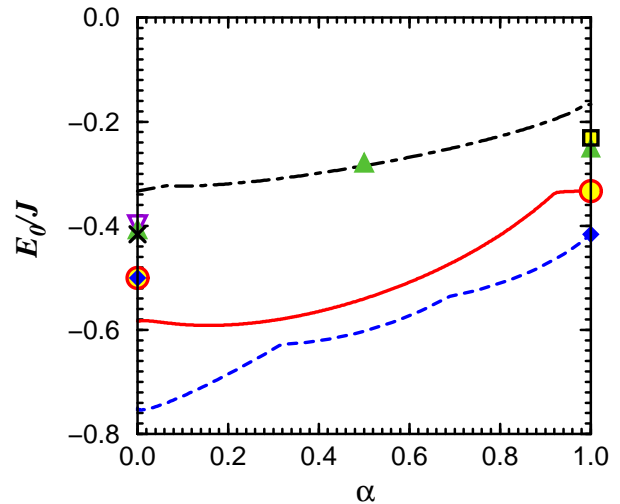}
\caption{(Color online) Ground--state energy per bond as a
function of $\alpha$, obtained with $\eta = 0$ for a triangular
cluster with 3 bonds (blue, dashed line), and a tetrahedral
cluster with 6 bonds (red, solid line). For comparison, the
energies obtained from the VB ansatz in the limiting cases $\alpha
 = 0$ and $\alpha = 1$ are shown for the triangular cluster (blue,
diamonds) and tetrahedral cluster (red, yellow--filled, open
circles); at $\alpha = 0$ both VB energies are the same, while at
$\alpha = 1$ they match the exact solutions. Green,
upward--pointing triangles show the static--dimer results of
Sec.~IV for the extended system, and the black, dot--dashed line
the lowest energy per bond obtained for fully spin and orbitally
ordered phases in Sec.~III. The violet, downward--pointing
triangle shows the energy of the orbitally ordered but
spin--disordered Heisenberg--chain state at $\alpha = 0$
[Eq.~(\ref{e1c1d})] and the open, yellow--filled square that of
the analogous state at $\alpha = 1$ [Eq.~(\ref{af2a1})], while the
cross shows the energy of the spin-- and orbitally disordered,
SU(4)--chain state [Eq.~(\ref{e1dsu4})]. } \label{fig:eneall}
\end{figure}

We draw particular attention to the extremely low ground--state
energy of the triangular cluster, which shows large gains from
dimer resonance. The tetrahedral cluster also has a very
significant resonance contribution, although more of its
ground--state energy is captured at the level of a static dimer
model. Such a VB ansatz provides the essential framework for the
understanding of all the results obtained, even for systems with
odd site numbers. The energies and their evolution with $\alpha$
contain some quantitative contrasts between even-- and odd--site
systems, allowing further insight concerning the range over which
the qualitative features of the cluster results extend.

Focusing in detail upon these energies, Fig.~\ref{fig:eneall}
summarizes the exact diagonalization results at zero Hund
coupling, and provides a comparison not only with the VB ansatz,
but with all of the other results obtained in Secs.~III--V. From
bottom to top are shown: the exact cluster energies including all
physical processes; the cluster VB ansatz, showing the importance
of dimer resonance energy; the static VB ansatz for extended
systems, suggesting by comparison with clusters the effects of
resonance; the energies of ``melted'' states with 1D spin (and
orbital) correlations; the optimal energy of states with full,
long--ranged spin and orbital order.

Returning to the cluster results, their degeneracies can be
understood precisely, and demonstrate the restoration of various
symmetries due to resonance processes. We provide a complete
explanation for all the correlation functions computed, and use
these to quantify the entanglement as a function of $\alpha$,
$\eta$ and the system size. There is a high--spin transition as a
function of $\eta$ for all values of $\alpha$, which sets the
basic phase diagram and establishes a new set of disentangled
orbital models at high $\eta$.

The extrapolation of the cluster results to states of extended
systems, some approximations for which are shown in
Fig.~\ref{fig:eneall}, is not straightforward, and cannot be
expected to include any information relevant to subtle selection
effects within highly degenerate manifolds of states. However,
with the exception of the static--dimer regime around $\alpha =
1$, our calculations suggest that nothing subtle is happening in
this model over the bulk of the phase diagram, where the physics
is driven by large energetic contributions from strong, local
resonance processes.

\section{Rhombic, honeycomb, and kagome lattices}
\label{sec:rhk}

In Sec.~I we alluded to the question of different sources of
frustration in complex systems such as the spin--orbital model
of Eq.~(\ref{som}). More specifically, this refers to the
relative effects of pure geometrical frustration, as understood
for AF spin interactions, and of interaction frustration of the
type which can arise in spin--orbital models even on bipartite
lattices.\cite{Fei97} Because the interaction frustration depends
in a complex manner on system geometry, no simple separation of
these contributions exists. In this section we alter the lattice
geometry to obtain some qualitative results with a bearing on this
separation, by considering the same spin--orbital model on the three
simple lattice geometries which can be obtained from the triangular
lattice by the removal of active bonds or sites.

\begin{figure}[t!]
\mbox{\includegraphics[width=3.5cm]{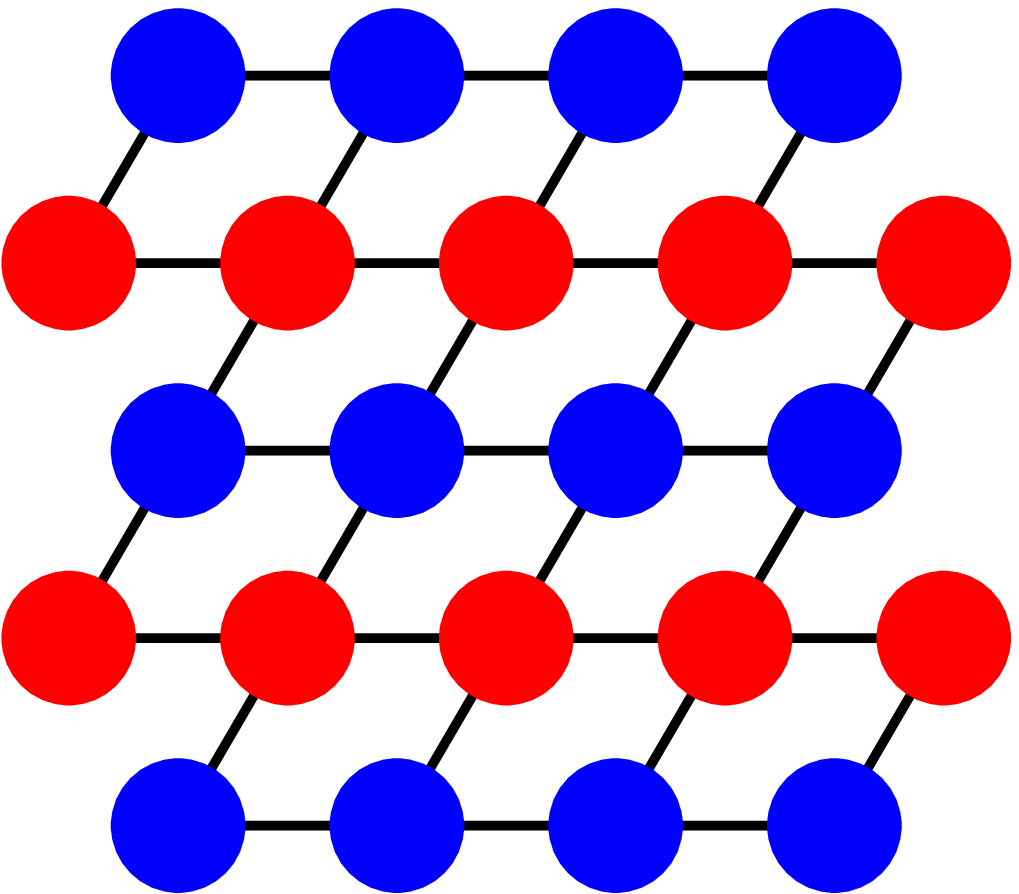}}
\hskip 0.2cm
\mbox{\includegraphics[width=4.3cm]{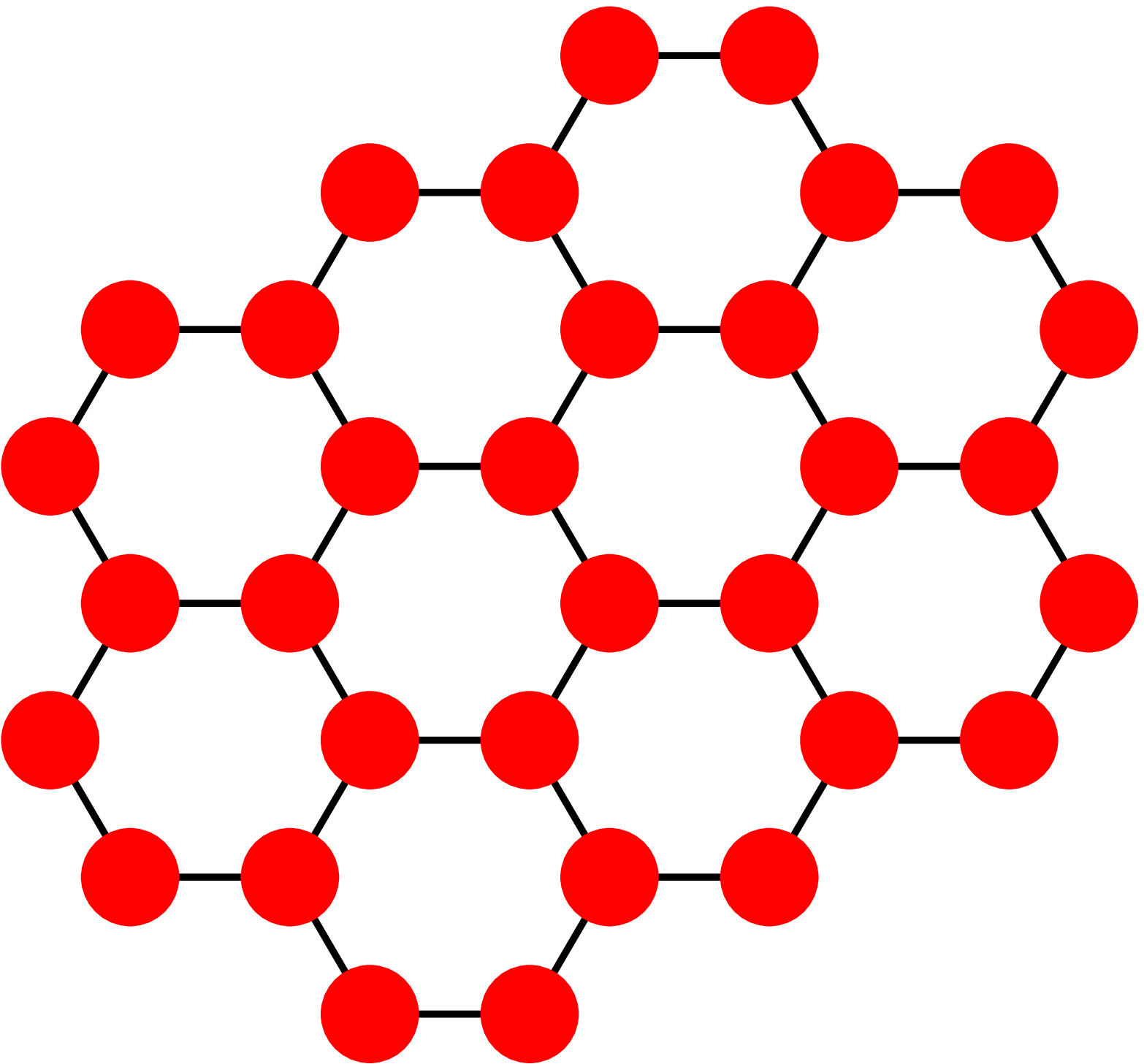}}
\vskip 0.2cm
\centerline{(a) \hskip 4.0cm (b)}
\vskip 0.2cm
\includegraphics[width=4.7cm]{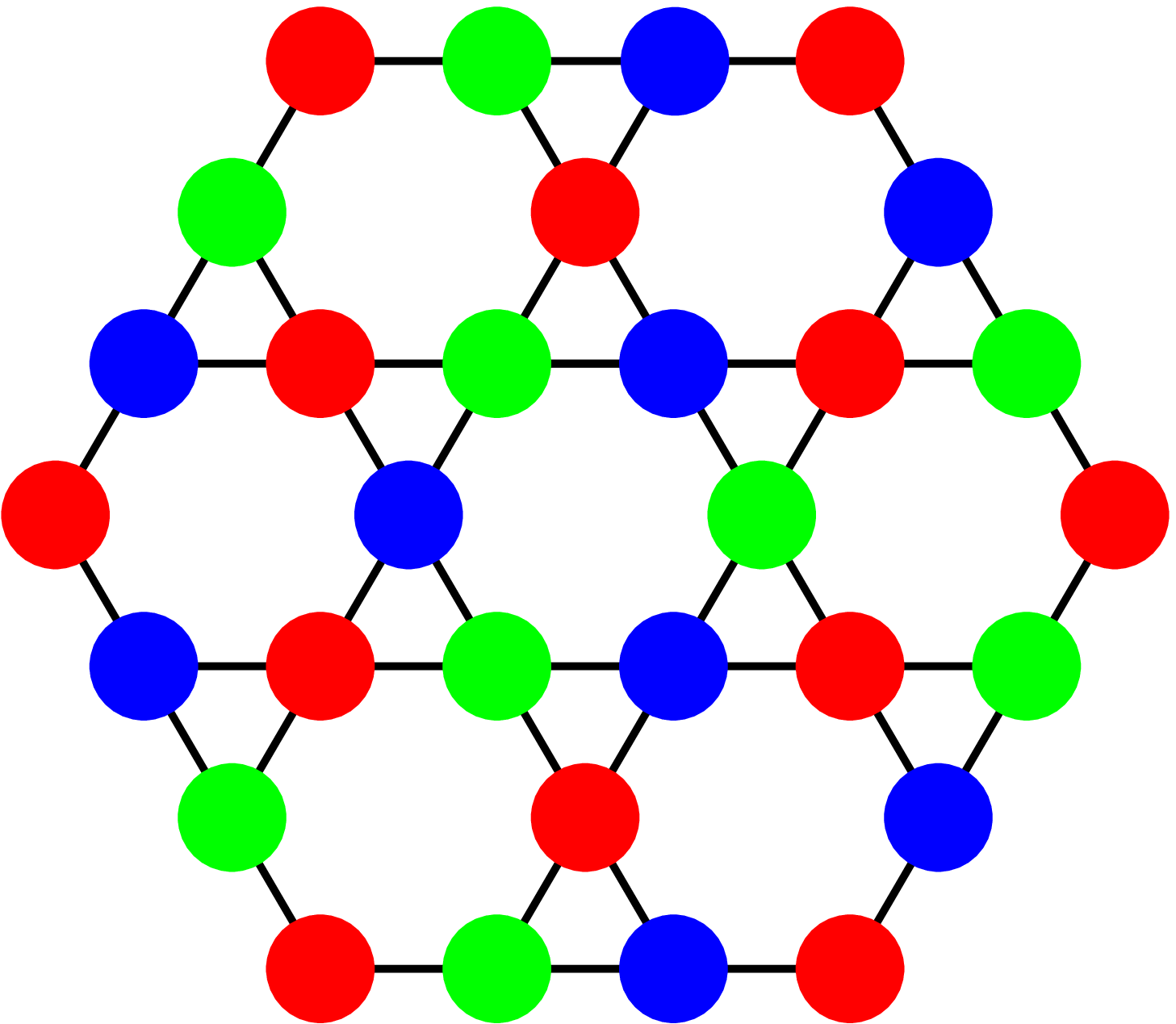}
\centerline{(c)}
\caption{(Color online)
(a) Rhombic lattice, showing a two--color orbitally ordered state.
(b) Honeycomb lattice, showing a one--color orbitally ordered state.
(c) Kagome lattice, showing a three--color orbitally ordered state.}
\label{rhk1}
\end{figure}

The geometries we discuss are rhombic, obtained by removing all bonds
in one of the three triangular lattice directions [Fig.~\ref{rhk1}(a)],
honeycomb, or hexagonal, obtained by removing every third lattice site
[Fig.~\ref{rhk1}(b)], and kagome, obtained by removing every fourth
lattice site in a 2$\times$2 pattern [Fig.~\ref{rhk1}(c)]. Simple
geometrical frustration is removed in the rhombic and honeycomb cases,
but for Heisenberg spin interactions the kagome geometry is generally
recognized (from the ground--state degeneracy of both classical and
quantum problems) to be even more frustrated than the triangular lattice.
We consider only the $\alpha = 0$ and $\alpha = 1$ limits of the model,
and $\eta = 0$. We discuss the results for long--range--ordered states
(Sec.~\ref{sec:mfa}) and for static dimer states (Sec.~\ref{sec:dim})
for all three lattice geometries. Here we do not enter into numerical
calculations on small clusters, and comment only on those systems for
which exact diagonalization may be expected to yield valuable
information not accessible by analytical considerations.

\subsection{Rhombic lattice}
\label{sec:rho}

While the connectivity of this geometry is precisely that of the square
lattice, we refer to it here as rhombic to emphasize the importance of
the bond angles of the chemical structure in maintaining the degeneracy
of the $t_{2g}$ orbitals and in determining the nature of the exchange
interactions. It is worth noting that the spin--orbital model (\ref{som})
on this lattice may be realized in Sr$_2$VO$_4$ (below). In the absence
of geometrical frustration, the spin problem created by imposing any
fixed orbital configuration selected from Sec.~\ref{sec:mfa}
(Figs.~\ref{fig:pure} and \ref{fig:mix}) is generally rather easy to
solve. Further, at $\eta = 0$ both FM and AF, and by extension AFF,
spin states have equal energies, leading to a high spin degeneracy.

Following Sec.~\ref{sec:mfa}, the $\alpha = 0$ energies for the
majority of the orbitally ordered states of Fig.~\ref{fig:pure} are
\begin{equation}
E_{\rm lro}^{\rm rh}(0) = - \frac12 J
\end{equation}
per bond at $\eta = 0$ for a number of possible spin configurations,
whose degeneracy is lifted (in favor of FM lines or planes) at finite
$\eta$. Indeed, the only exceptions to this rule occur for the
three--color state [Fig.~\ref{fig:pure}(d)] and for orientations of
the other states which preclude hopping in one of the two lattice
directions, whose triangular symmetry properties are broken by the
missing bond. As noted in Sec.~\ref{sec:mfa}, for superpositions is
it the exception rather than the rule for all hopping processes to be
maximized, but on the rhombic lattice this is possible for the states
in Fig.~\ref{fig:mix}(a) and some orientations of those in
Figs.~\ref{fig:mix}(d) and \ref{fig:mix}(e).

For $\alpha = 1$, the energy limit even on the triangular lattice
was set rather by the number of active bonds than by the problem of
minimizing their frustration. Similar to the $\alpha = 0$ case, all
states where the active hopping direction is one of the two lattice
directions, plus in this case state (3d), can redeem the maximum
energy available,
\begin{equation}
E_{\rm lro}^{\rm rh}(1) = - \frac14 J
\end{equation}
at $\eta = 0$, which is simply the avoided--blocking energy, for a
large number of possible spin configurations. Finite Hund exchange
favors FM spin states.

\begin{figure}[t!]
\mbox{\includegraphics[width=4.1cm]{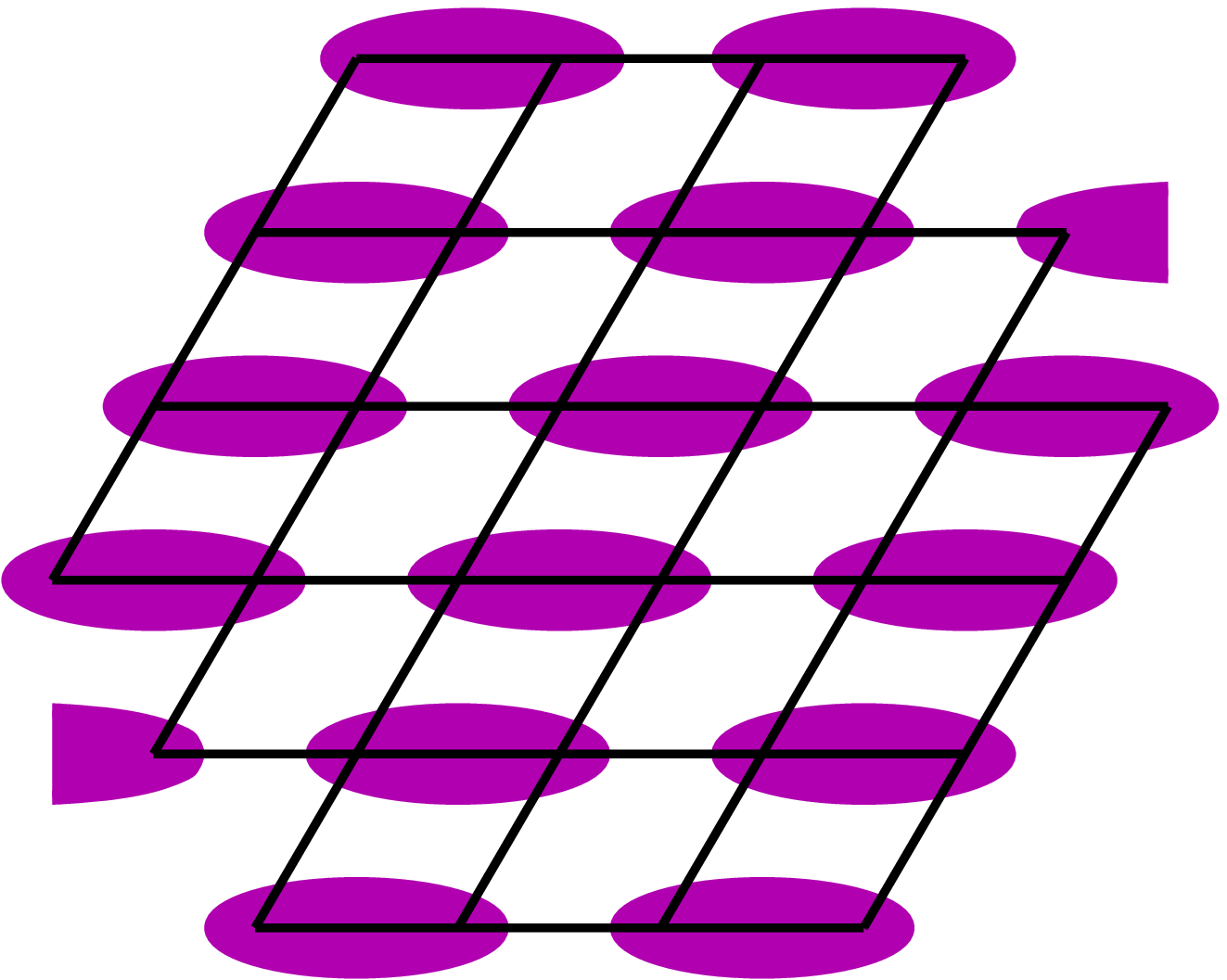}}
\hskip 0.2cm
\mbox{\includegraphics[width=4.1cm]{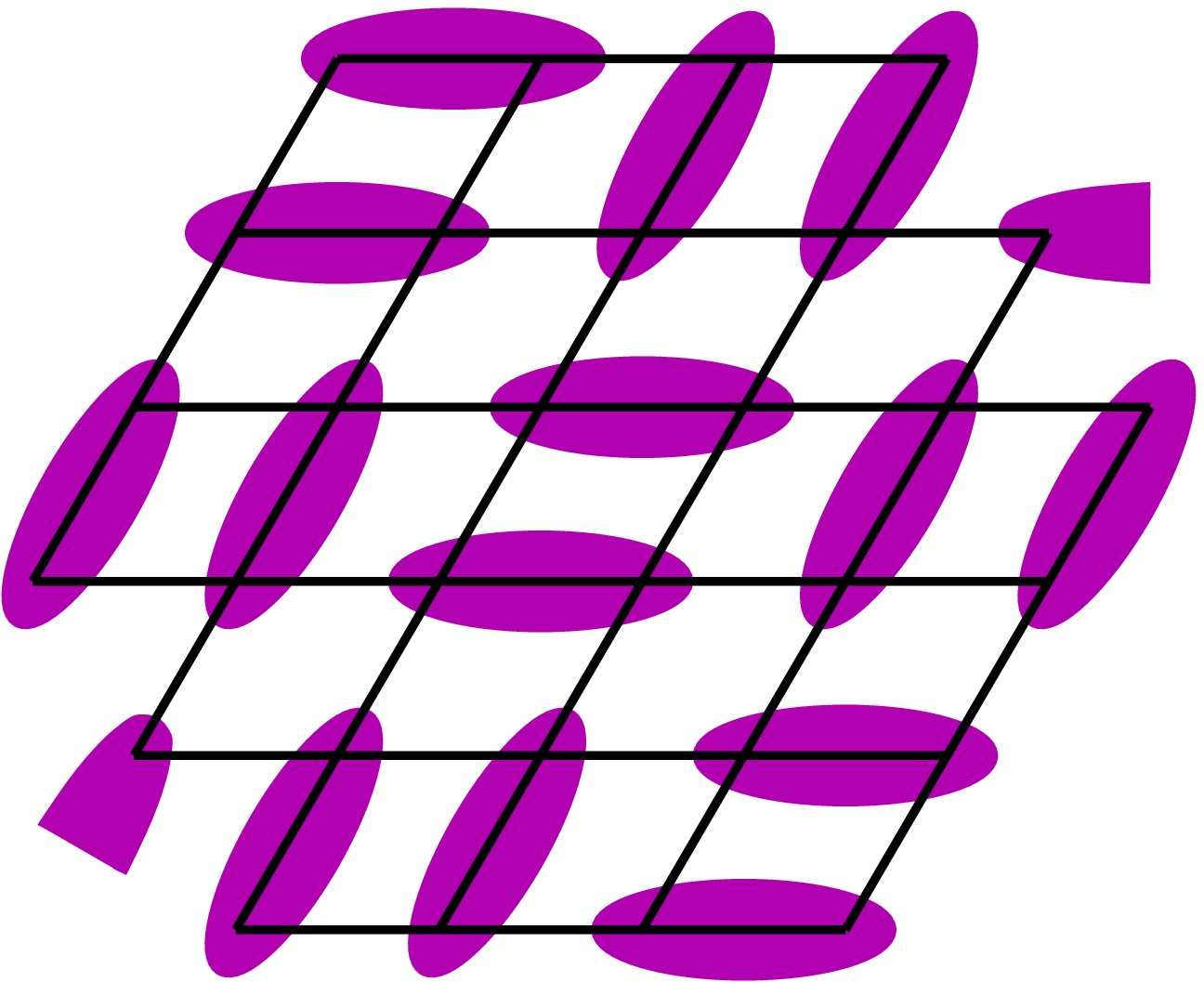}}
\centerline{(a) \hskip 3.6cm (b)}
\caption{(Color online)
Rhombic lattice with (a) columnar and (b) plaquette dimer coverings.}
\label{rhk2}
\end{figure}

Turning to dimerized states, the calculation of the energy of any
given dimer covering proceeds as in Sec.~\ref{sec:dim}, namely by
counting for each the respective numbers of bonds of types
(\ref{fig:bonds}a), (\ref{fig:bonds}b), and (\ref{fig:bonds}c)
[Fig.~\ref{fig:bonds}]. For the rhombic lattice, lack of geometrical
frustration means that all interdimer bonds can be chosen to be AF/AO.
The two most regular dimer coverings of the rhombic lattice with small
unit cells may be designated as ``columnar'' [Fig.~\ref{rhk2}(a)] and
``plaquette'' [Fig.~\ref{rhk2}(b)]. In both cases, $1/4$ of the bonds
are the dimers, and by inspection $1/4$ of the interdimer bonds in
the columnar state are of type (\ref{fig:bonds}a), while the remainder
are (\ref{fig:bonds}c); by contrast, the plaquette state has no
type--(\ref{fig:bonds}a) bonds, $1/2$ type--(\ref{fig:bonds}b) bonds,
and the remainder are of type (\ref{fig:bonds}c). For $\alpha = 0$,
the energies are
\begin{eqnarray}
E_{\rm dc}^{\rm rh}(0)\! & = & - \frac14 J - \frac14 \cdot \frac12 J
 - \frac12 \cdot \frac14 J \; = \; - \frac12 J, \nonumber \\
E_{\rm dp}^{\rm rh}(0)\! & = & - \frac14 J - \frac12 \cdot \frac{3}{8} J
 - \frac14 \cdot \frac{1}{4} J \; = \; - \frac12 J
\end{eqnarray}
at $\eta = 0$, both for (ss/ot) and for (os/st) dimers. The degeneracy
of these two limiting cases, in the sense of maximal and minimal
numbers of types--(\ref{fig:bonds}a) and --(\ref{fig:bonds}b) bonds,
suggests a degeneracy of all dimer coverings at this level of analytical
sophistication. Further, all of these dimer coverings are degenerate with
all of the unfrustrated ordered states at $\eta = 0$. The selection of
a true ground state from this large manifold of static states
(order--by--disorder) would hinge on higher--order processes, but
these considerations are likely to be rendered irrelevant by dimer
resonance (Sec.~V).

For the spin--singlet dimer states at $\alpha = 1$ one finds
\begin{eqnarray}
E_{\rm dc}^{\rm rh}(1)\! & = & - \frac14 J - \frac14 \cdot \frac14 J
 - \frac12 \cdot 0 J \; = \; - \frac{5}{16} J, \nonumber \\
E_{\rm dp}^{\rm rh}(1)\! & = & - \frac14 J - \frac12 \cdot \frac14 J
 - \frac{1}{4} \cdot 0 J \; = \; - \frac{3}{8} J,
\end{eqnarray}
at $\eta = 0$, and thus that, as for the triangular lattice, the energy
is minimized by dimer configurations excluding linear interdimer bonds.
This remains a large manifold of dimer coverings, whose energy is
manifestly lower than any of the possible orbitally ordered states
in this limit of the model, and within which order--by--disorder is
expected to operate (Sec.~V).\cite{Jac07}

The considerations of this subsection, extended to finite values of
$\eta$, may be relevant in the understanding of experimental results
for Sr$_2$VO$_4$. These suggest weak FM order,\cite{Noz91} accompanied
by an AO order\cite{Tok05} which could be interpreted as arising from
the formation of dimer pairs. When the oxygen octahedra distort, the
threefold degeneracy of the $t_{2g}$ orbitals is lifted, to give a
model containing only two degenerate orbitals, $d_{yz}$ and $d_{xz}$.
This leads to a situation with Ising--like superexchange interactions
and quasi--1D hole propagation in an effective $t$--$J$ model.\cite{Dag08}

\subsection{Honeycomb lattice}
\label{sec:hex}

The situation for the honeycomb lattice is very similar to that for
the rhombic case. Again the absence of geometrical frustration makes
it possible to obtain the minimal energy for a number of orbital
orderings, with a high spin degeneracy at $\eta = 0$. For pure
superexchange interactions, once again
\begin{equation}
E_{\rm lro}^{\rm h}(0) = - \frac12 J
\end{equation}
per bond, while in the direct--exchange limit
\begin{equation}
E_{\rm lro}^{\rm h}(1) = - \frac14 J,
\end{equation}
both at $\eta = 0$, for the same physical reasons as above.

\begin{figure}[t!]
\includegraphics[width=6cm]{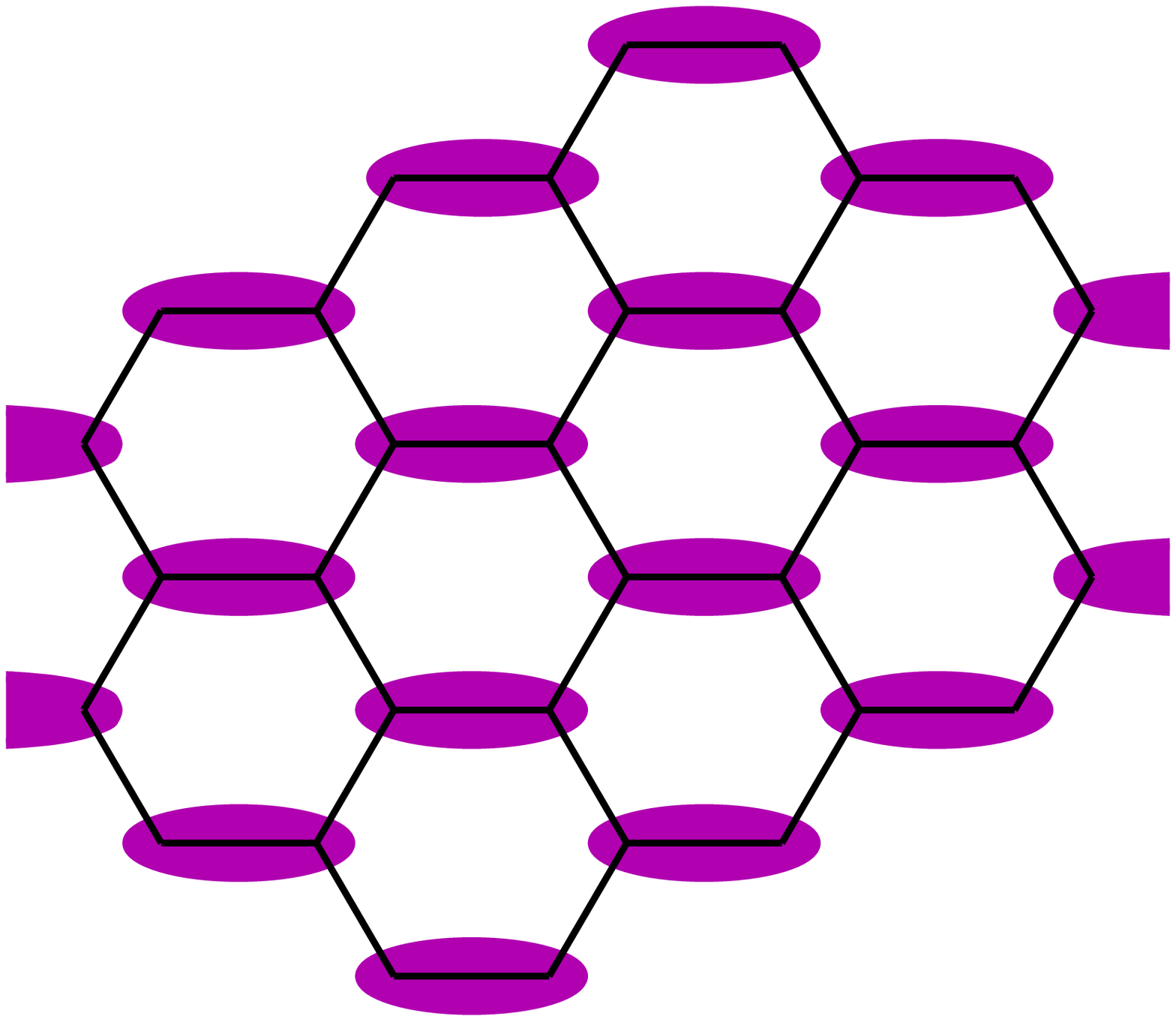}
\centerline{(a)}
\vskip .2cm
\includegraphics[width=6cm]{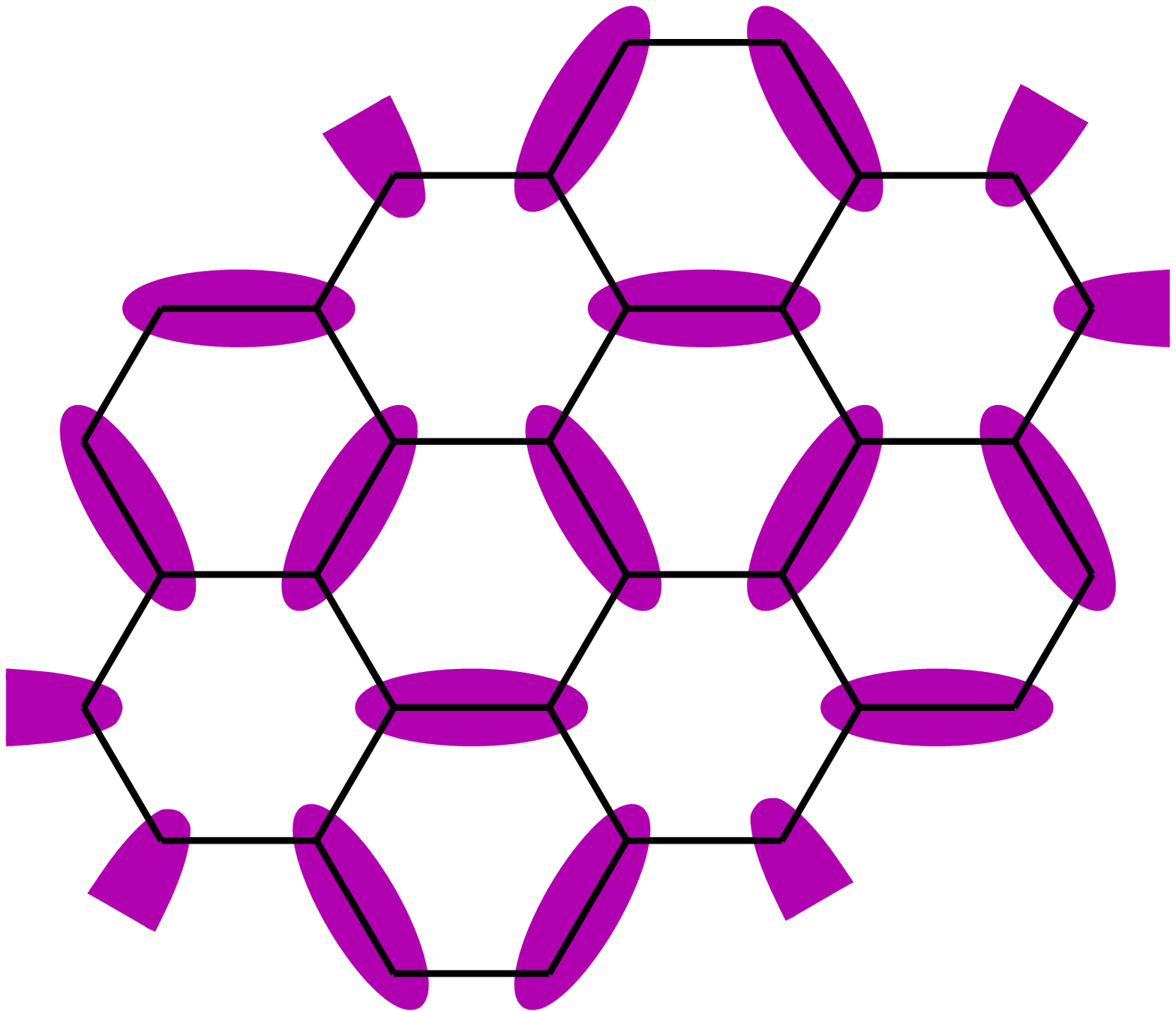}
\centerline{(b)}
\caption{(Color online)
Honeycomb lattice with (a) columnar and (b) three--way dimer coverings.}
\label{rhk3}
\end{figure}

For dimer states, on the honeycomb lattice all interdimer bonds
are by definition of type (\ref{fig:bonds}c), and again can be made
AF/AO because frustration is absent, so the energies of all dimer
coverings are {\it de facto\/} identical. By way of demonstration,
the two simplest regular configurations, which we label ``columnar''
and ``three--way'', are shown in Fig.~\ref{rhk3}, and, from the fact
that now $1/3$ of the bonds contain dimers, their energies are
\begin{eqnarray}
E_{\rm dc}^{\rm h}(0) & = & - \frac13 J - \frac{2}{3} \cdot \frac14 J
\; = \; - \frac12 J, \nonumber \\
E_{\rm d3}^{\rm h}(0) & = & - \frac13 J - \frac{2}{3} \cdot \frac14 J
\; = \; - \frac12 J,
\end{eqnarray}
per bond at $\alpha = 0 = \eta$. Thus static dimer states are again
degenerate with unfrustrated ordered states in the superexchange limit,
and detailed consideration of kinetic processes would be required to
deduce the lowest total energy. In this context, the dimer coverings
shown in Fig.~\ref{rhk3} exemplify two limits about which little
kinetic energy can be gained from resonance (Fig.~\ref{rhk3}(a),
where large numbers of dimers must be involved in any given process)
and in which kinetic energy gains from processes involving short loops
[the three dimers around 2/3 of the hexagons, Fig.~\ref{rhk3}(b)] are
maximized.

At $\alpha = 1$, only the dimer energy is redeemed, and this on $1/3$
of the bonds, so
\begin{equation}
E_{\rm d}^{\rm h}(1) = - \frac13 J
\end{equation}
at $\eta = 0$ for a large manifold of coverings. This energy is
once again significantly better than any of the possible ordered
states, a result which can be ascribed to the low connectivity.
That the ground state of the extended system in this limit for
both the rhombic and honeycomb lattices involves a selection from
a large number of nearly degenerate states suggests that numerical
calculations on small clusters would not be helpful in resolving
detailed questions about its nature. The same model for the honeycomb
geometry in the $\alpha = 1$ limit has been discussed for the $S = 1$
compound Li$_2$RuO$_3$,\cite{Jac08} where the authors invoked the
lattice coupling, in the form of a structural dimerization driven by
the formation of spin singlets, to select the true ground state.

\subsection{Kagome lattice}
\label{sec:kag}

The kagome lattice occupies something of a special place among
frustrated spin systems\cite{Diep} as one of the most highly
degenerate and intractable problems in existence, for both classical
and quantum spins, and even with only nearest--neighbor Heisenberg
interactions. Interest in this geometry has been maintained by the
discovery of a number of kagome spin systems, and has risen sharply
with the recent synthesis of a true $S = 1/2$ kagome material,
ZnCu$_3$(OH)$_6$Cl$_2$.\cite{rsnbn} Preliminary local--probe
experiments\cite{rinbsn,rmea} show a state of no magnetic order
and no apparent spin gap, whose low--energy spin excitations have
been interpreted\cite{rhea} as evidence for an exotic spin--liquid
phase. Both experimentally and theoretically, kagome systems of higher
spins ($S = 3/2$ and 5/2) are found to have flat bands of magnetic
excitations, reflecting the very high degeneracy of the spin
sector.\cite{rmgnyhlnl} While no kagome materials are yet known
with both spin and orbital degrees of freedom, Maekawa and
coworkers\cite{Kos03,Kha04} have considered the itinerant electron
system on the triangular lattice for $\alpha = 0$ (actually for the
motion of holes in Na$_x$CoO$_2$), demonstrating that the combination
of orbital, hopping selection, and geometry leads to any one hole
being excluded from every fourth site, and thus moving on a system
of four interpenetrating kagome lattices.

\begin{figure}[t!]
\mbox{\includegraphics[width=4.1cm]{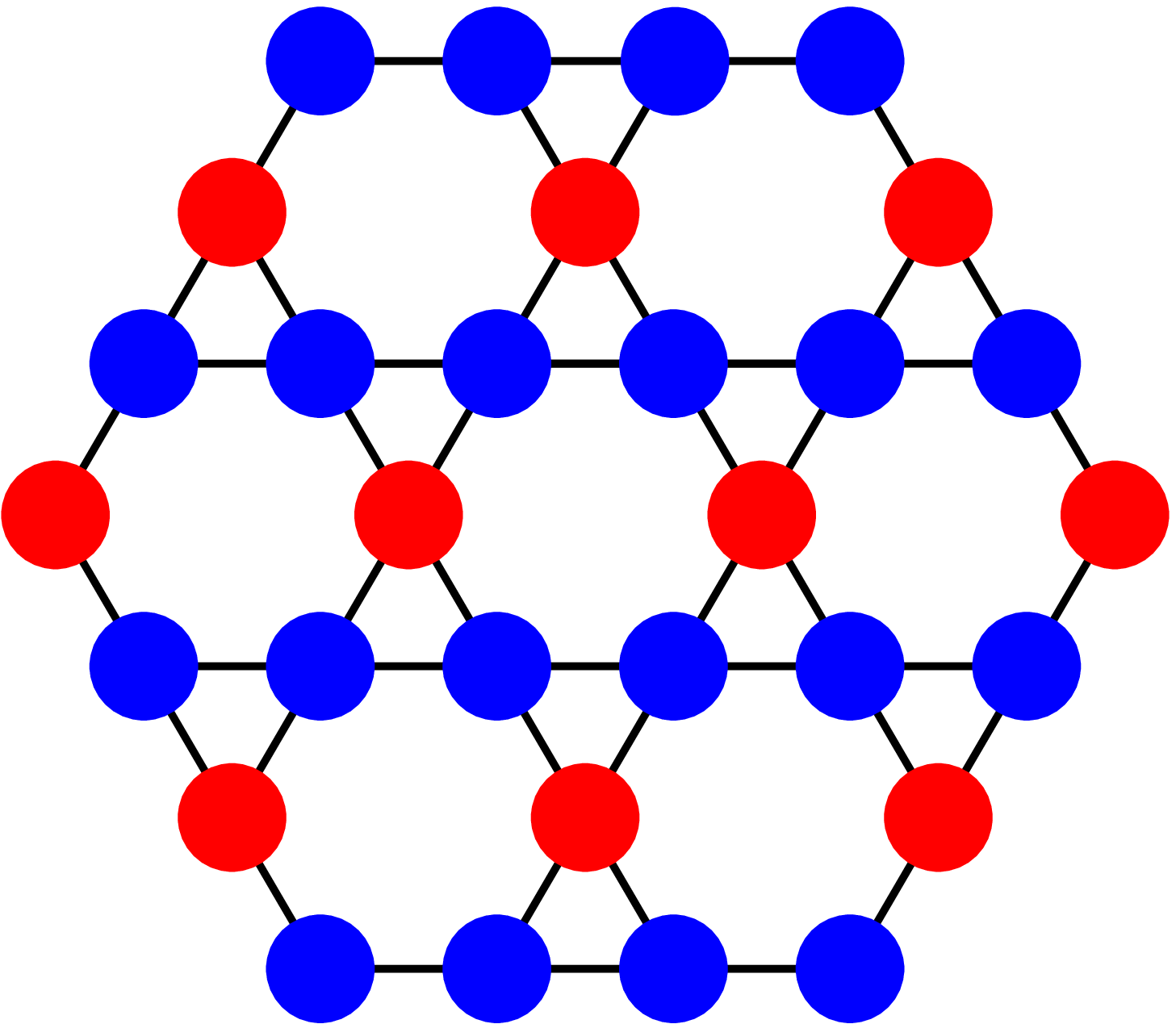}}
\hskip 0.2cm
\mbox{\includegraphics[width=4.1cm]{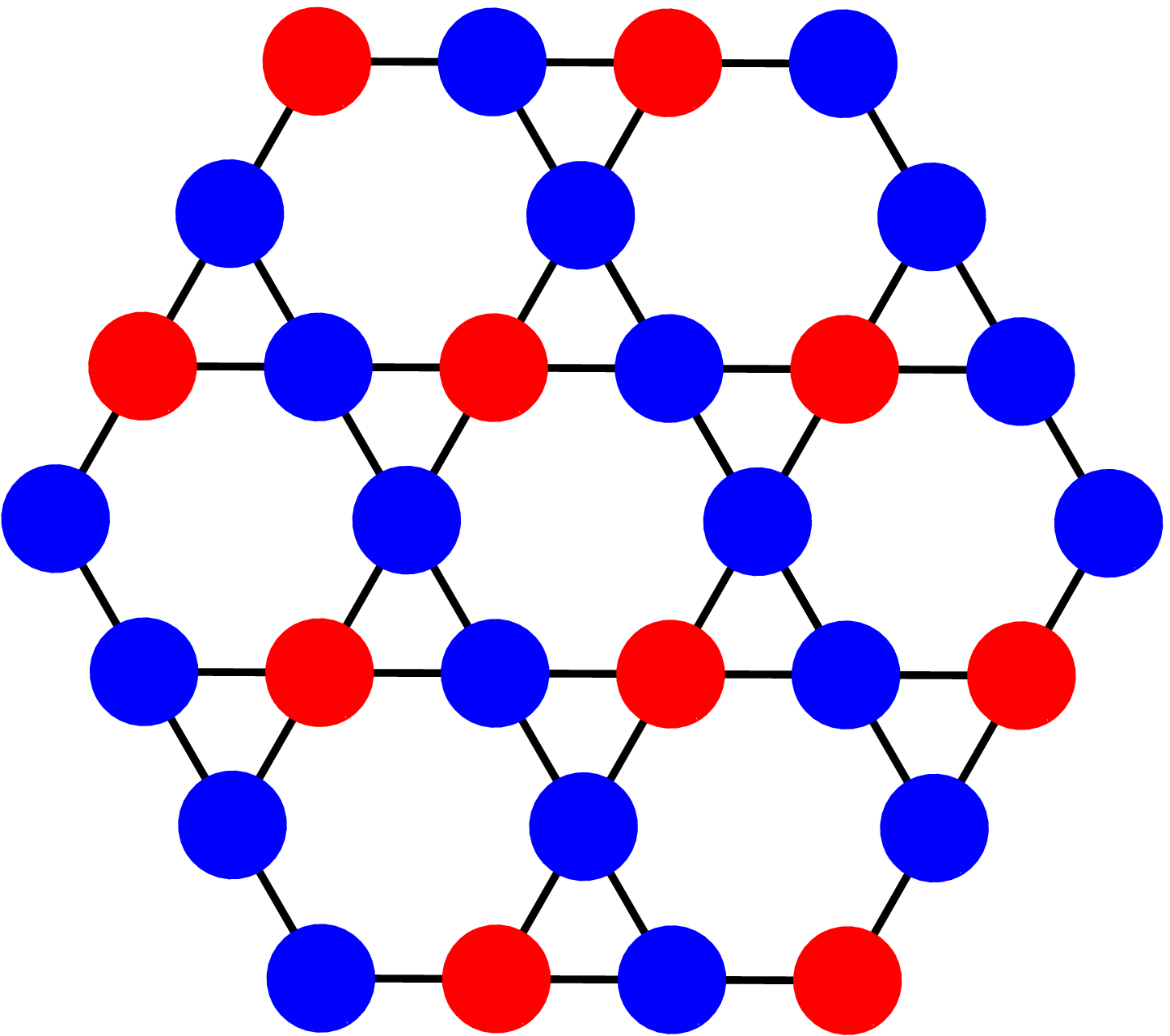}}
\centerline{(a) \hskip 3.6cm (b)}
\caption{(Color online)
Kagome lattice with unequally weighted two--color states oriented
(a) with and (b) against the lattice direction corresponding to the
majority orbital color. }
\label{rhk4}
\end{figure}

Considering first the energies per bond for states of long--ranged
spin and orbital order, in a number of cases the values for the
kagome lattice are identical to those of the triangular lattice.
This is easy to show by inspection for the one--color state (3a), and
for the superposition states (4a), (4b), and (4c), where bonds of all
types are removed in equal number. However, for the less symmetrical
orbital color configurations a more detailed analysis of the type
performed in Sec.~\ref{sec:mfa} is required, and yields provocative
results. The two simple possibilities for ordered two--color states
with a single color per site are shown in Fig.~\ref{rhk4}, and differ
only in the orientation of the continuous lines (the majority color)
relative to the active orbitals. These can be considered as the
kagome--lattice analogs of states (3b) and (3c), as well as of
(3e) and (3f).

When the lines of $c$--orbitals are aligned with the $c$--axis
[Fig.~\ref{rhk4}(a)], this direction is inactive at $\alpha = 0$,
and only the other two directions contribute, one with two active
FO orbitals, mandating an AF spin state to give energy $- \frac12 J$
per bond, and the other with energy $- \frac14 J$ and no strong spin
preference, whence
\begin{equation}
E_{\rm (k3b)} (0) = - \frac14 J
\end{equation}
at $\eta = 0$ for sets of unfrustrated AF chains. By contrast,
when the lines of $c$--orbitals fall along the $b$--direction
[Fig.~\ref{rhk4}(b)], the $\alpha = 0$ problem contains one FO and
one AO line each with two active orbitals, and one line with one
active orbital. Only the first requires AF spin alignment, while
the other two lines are not frustrating, with the result that an
energy
\begin{equation}
E_{\rm (k3c)} (0) = - \frac{5}{12} J
\end{equation}
can be obtained. This value is lower than that on the triangular
lattice, showing that for the class of models under consideration,
where not all hopping channels are active in all directions, a
system of lower connectivity can lead to frustration relief even
when its geometry remains purely that of connected triangles.

\begin{figure}[t!]
\mbox{\includegraphics[width=4.1cm]{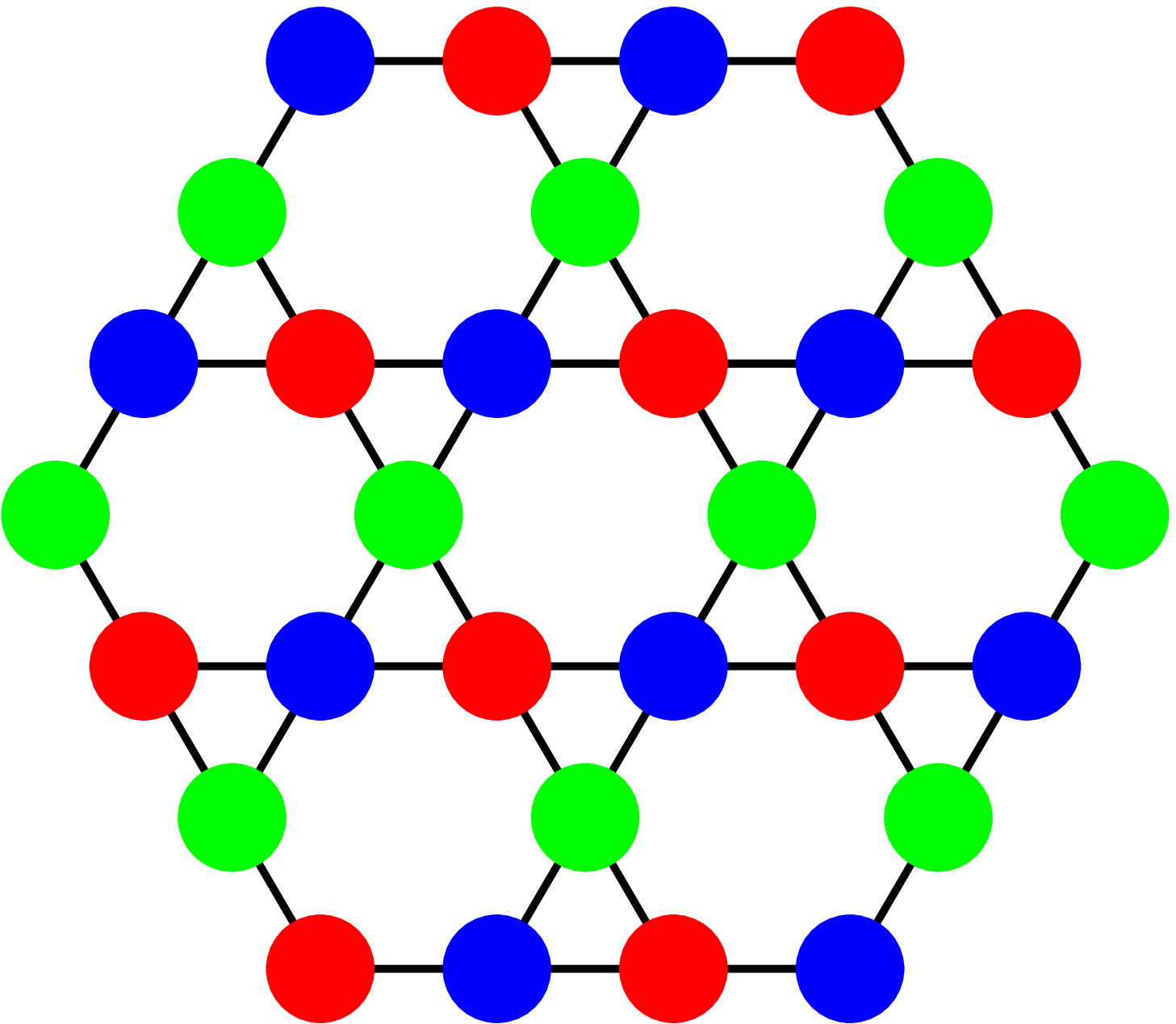}}
\hskip 0.2cm
\mbox{\includegraphics[width=4.1cm]{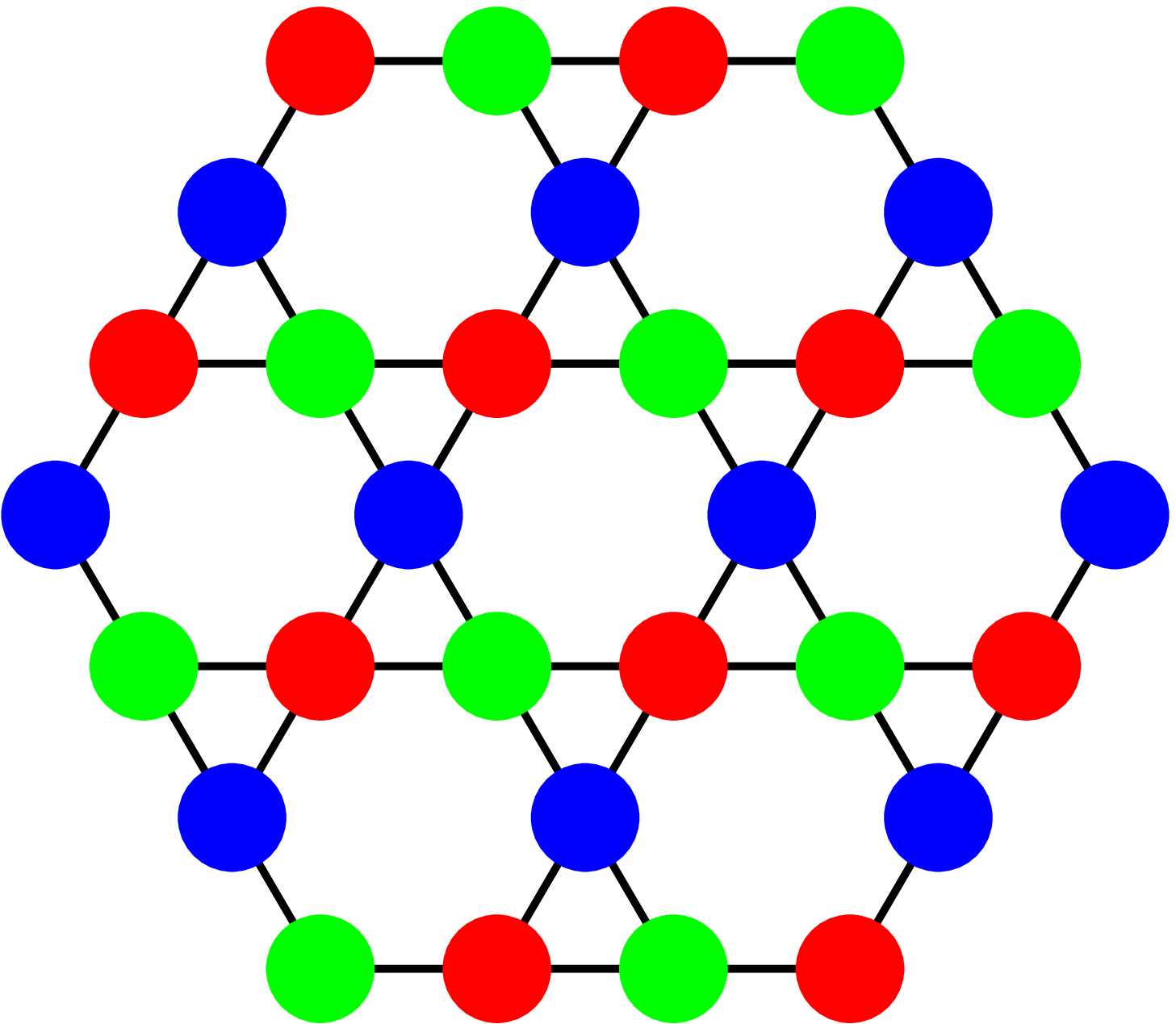}}
\centerline{(a) \hskip 3.6cm (b)}
\caption{(Color online)
Kagome lattice with two different, equally weighted three--color states:
(a) two--color lines oriented such that only one superexchange channel,
plus the direct exchange channel, is active on every bond. (b) two--color
lines oriented such that all superexchange channels are active, but no
direct exchange channels. }
\label{rhk5}
\end{figure}

With this result in mind, we consider again the possibilities
offered by different three--color states, specifically those shown in
Fig.~\ref{rhk5}. With reference to the superexchange problem, the state
in Fig.~\ref{rhk1}(c), which by analogy with (3d) we denote as (k3d),
contains only a small number of remnant triangles and isolated bonds
still with two active orbitals. However, the state (k3d1), shown in
Fig.~\ref{rhk5}(a) is that which ensures that no such bonds remain, and
every single bond of the lattice has one active superexchange channel.
The state (k3d2) in Fig.~\ref{rhk5}(b) is that in which every single
bond of the lattice has two active (FO) superexchange channels: this
possibility can be realized for the kagome geometry, at the cost of
creating a frustrated magnetic problem requiring a 120$^{\circ}$ spin
state to minimize the energy,
\begin{eqnarray}
E_{\rm (k3d)}  (0) & = & - \frac{5}{16} J, \\
E_{\rm (k3d1)} (0) & = & - \frac14 J, \\
E_{\rm (k3d2)} (0) & = & - \frac{3}{8} J.
\end{eqnarray}
Thus one finds that lower energies than the value $- \frac13 J$ per
bond, which was the lower bound for fully (orbitally and spin--)ordered
states on the triangular lattice, are again possible for three--color
ordered states. However, the residual spin frustration means that the
lowest ordered--state energy on the kagome lattice is given by the
unfrustrated, two--color AFF state, $E_{\rm (k3c)} (0) = - \frac{5}{12}
J$.

We present briefly the energies of the same states at $\alpha = 1$,
where only a maximum of one hopping channel per bond can be active,
and as noted above this is generally a stricter energetic limit than
any frustration constraints. The results at $\eta = 0$ are
\begin{equation}
E_{\rm (k3b)} (1) = - \frac14 J
\end{equation}
for an AFF state gaining most of its energy from the $c$--axis
chains, and
\begin{equation}
E_{\rm (k3c)} (1) = - \frac{1}{12} J
\end{equation}
due to the dearth of active orbitals in this orientation. Similarly,
by counting active orbitals in the three--color states,
\begin{eqnarray}
E_{\rm (k3d)}  (1) & = & - \frac{1}{6} J, \\
E_{\rm (k3d1)} (1) & = & - \frac{1}{4} J, \\
E_{\rm (k3d2)} (1) & = & 0,
\end{eqnarray}
and it is the state of Fig.~\ref{rhk5}(a) which achieves the
unfrustrated value $ - \frac{1}{4} J$ by permitting one active
hopping channel on every bond of the kagome lattice.

We will not discuss the orbital superposition states which are the
analogs of (4d) and (4e), noting only that these present again two
different possibilities on the kagome lattice, depending on the
orientation of the majority lines. Even with the frustration relief
offered by this geometry for the type of model under consideration,
superposition states contain too many hopping channels for all to
be satisfied simultaneously, and it is not possible to equal the
energy values found respectively for the configurations in
Figs.~\ref{rhk5}(a) and (b) at $\alpha = 1$ and $\alpha = 0$.

\begin{figure}[t!]
\includegraphics[width=6.0cm]{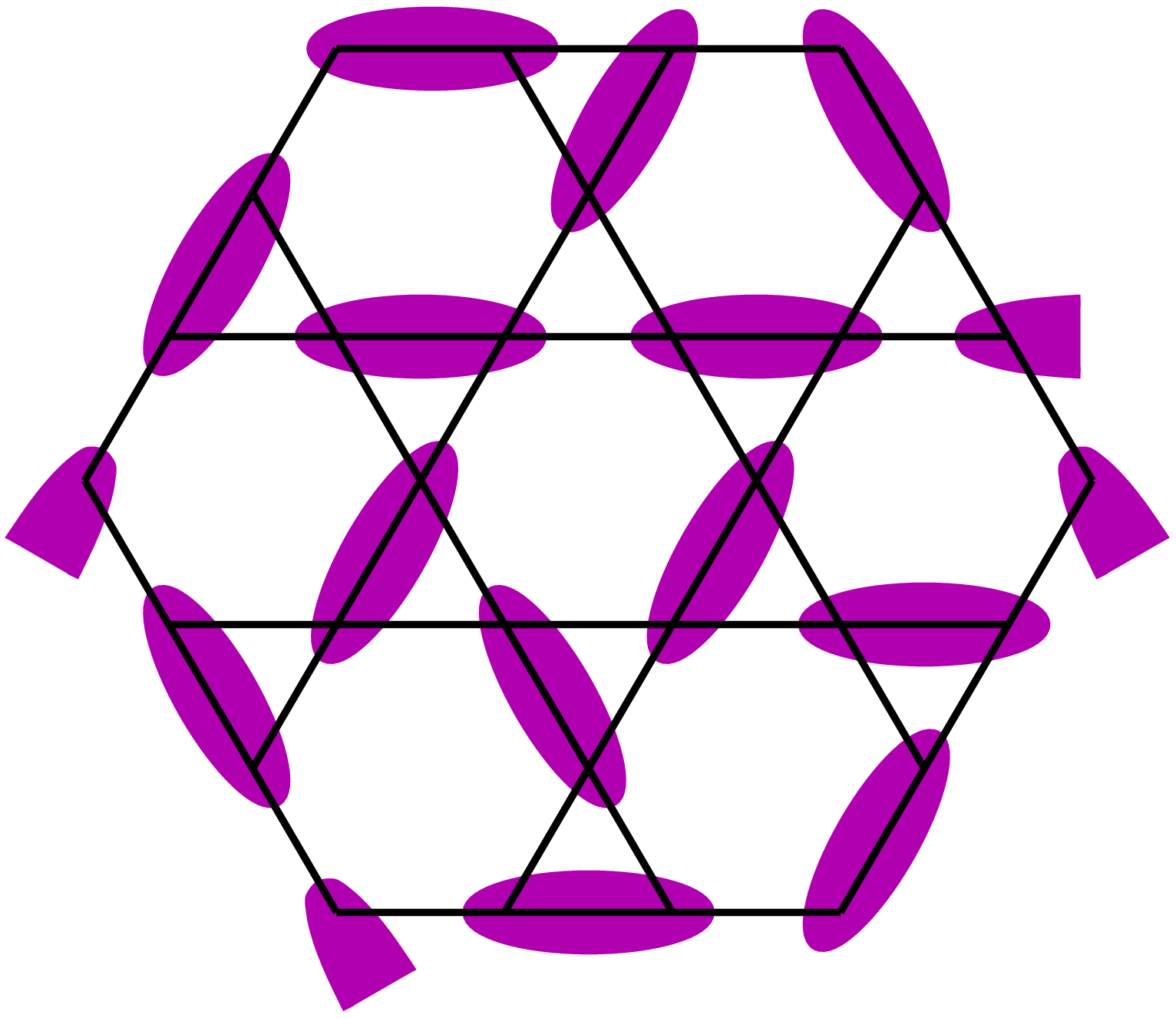}
\centerline{(a)}
\vskip .2cm
\includegraphics[width=5.4cm]{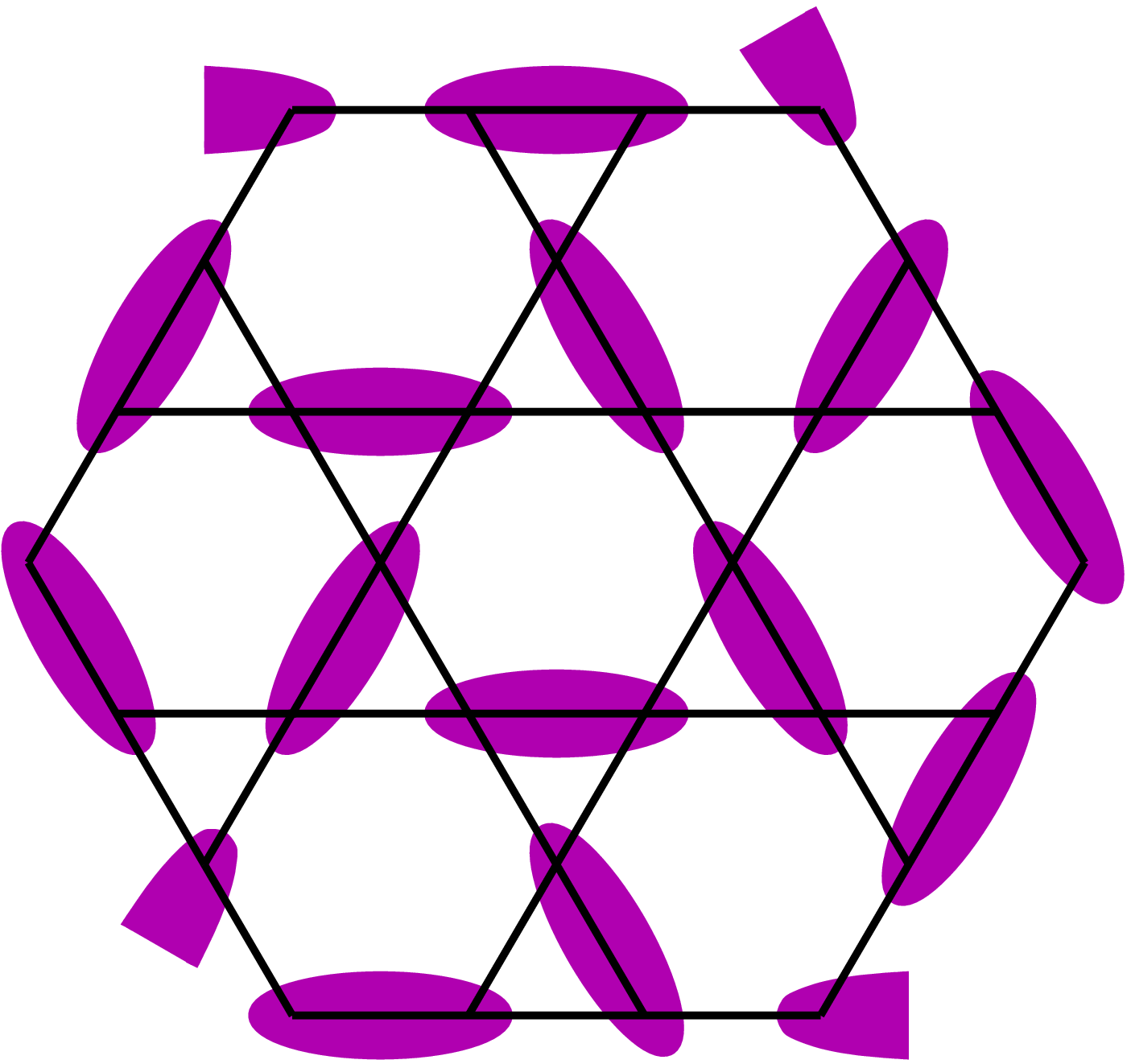}
\centerline{(b)}
\caption{(Color online)
Kagome lattice with two different dimer coverings, (a) and (b). In
both examples, only two of the twelve triangles shown explicitly on the
cluster are ``defective'' (contain no dimer), but the reader may notice
that many of the next twelve triangles adjoining the boundary must
also be so.}
\label{rhk6}
\end{figure}

It remains to consider dimer states on the kagome lattice, as these
have been of equal or lower energy for every case analyzed so far.
The set of nearest--neighbor dimer coverings of the kagome
lattice is large, and for the $S = 1/2$ Heisenberg model in this
geometry the spin singlet manifold has been proposed as the basis
for an RVB description.\cite{rmm} Two dimer coverings degenerate
at the level of the current treatment are shown in Fig.~\ref{rhk6}.

Dimer coverings of the kagome lattice have the property that $3/4$
of the triangles contain one dimer. In this case, the other bonds of
the triangle are interdimer bonds, one of which is of type (7b) while
the other is of type (7c). The other 1/4 of the triangles, known\cite{re}
as ``defect triangles'', have no dimers, and their three bonds are
either all of type (7b), with probability 1/4, or one each of types
(7a), (7b), and (7c), with probability 3/4. The frustration of the
system is contained in the problem of minimizing the number of FM/FO
interdimer bonds; this exercise is complex and no solution is known,
so only an upper bound will be estimated here.

The bonds of a defect triangle connect three different dimers, and
so one (or all three) must be FM/FO. A hexagon of the kagome lattice
with no dimers on its bonds is surrounded by six non--defective
triangles, one with one dimer by one defective neighbor, with two
dimers two, and a hexagon with three dimers shares its non--dimer
bonds with three defect triangles. Hexagons with odd dimer numbers
must create a FM/FO bond between at least one pair of dimers, and
it is reasonable to place this bond on the defect triangle(s)
where an energy cost is already incurred. We note immediately that the
cost of reversing the type--(7a) bond, $\frac14 J$ (Sec.~IVA), exceeds
that of reversing both interdimer bonds of a non--defective triangle,
which is $\frac{1}{8} J + \frac{1}{16} J$. As a consequence, we take
this cost, which is equal to that of reversing both a non--defective
triangle and the weakest bond of the defect triangle, to be an upper
bound on the effect of frustration. The net energy of a dimer state
for $\alpha = 0 = \eta$ is then estimated to be
\begin{eqnarray}
E_{\rm kd} (0)\!\! & = &
- \frac{3}{4} \! \cdot \! \frac{1}{3} J - \frac{3}{4} \! \cdot \!
\frac{1}{3} J \left( \frac{3}{8} + \frac{1}{4} \right) \nonumber
\\ & & - \frac{1}{4} J \left[ \frac14 \! \left( \frac{2}{3} \! \cdot
\! \frac{3}{8} \! + \! \frac{1}{3} \! \cdot \! \frac14 \right) \! +
\! \frac{3}{4} \! \cdot \! \frac{1}{3} \! \left(\frac{1}{4} \! + \!
\frac{3}{8} \! + \! \frac{1}{4} \right) \! \right] \nonumber
\\ & = & \! - \frac{209}{384} J \; \simeq \; - \frac{13}{24} J.
\end{eqnarray}
This is a very large number for the kagome lattice, exceeding even
the value $- \frac12 J$ per bond (which, however, is of no special
significance here). Thus we find that dimer states in this type of
model are strongly favored, gaining a very much higher energy than
even the best ordered states. Qualitatively, the dimer energy shares
with the ordered--state energy the feature that it is considerably
better than anything obtainable for the triangular lattice. This
implies that the reduced connectivity of the lattice geometry for
a model where the orbital degeneracy provides a number of mutually
exclusive hopping channels makes it easier to find states where
every remaining bond can support a favorable hopping process
without strong frustration.

Applying all of the above geometrical considerations to the
direct--exchange model ($\alpha = 1$), where there is no frustration
problem between the spin singlets, one finds
\begin{eqnarray}
E_{\rm kd} (1) & = & - \frac{3}{4} \cdot \frac{1}{3} J - \frac{3}{4}
 \cdot \frac{1}{3} J \left[ \frac{1}{4} + 0\right] \nonumber
\\ & & - \frac{1}{4} \cdot \frac{1}{3} J \left[ \frac{1}{4} \cdot
\frac{3}{4} + \frac{1}{4} \cdot \frac12 \right] \nonumber
\\ & = & - \frac{21}{64} J
\end{eqnarray}
at $\eta = 0$. Once again this energy is significantly lower
than the value $E_{\rm dim}(1) = - \frac14 J$ obtained for the
triangular lattice in Eq.~(\ref{enedimeta}), demonstrating that
the multichannel spin--orbital model of the type considered here
is less frustrated in the kagome geometry.

We comment in closing that the dimer energies we have estimated are
only those of static VB configurations, and, away from $\alpha = 1$,
the possibility remains of a significant resonance energy gain from
quantum fluctuations between these states ({\it cf.}~Sec.~V). Numerical
calculations on small clusters of sufficient size (here at least 6
sites for a unit cell) would be helpful in this frustrated case.

To summarize this section, the spin--orbital model on bipartite
lattices appears to present competing ordered and dimerized states
with the prospect of high degeneracies. Among ``frustrated'' systems
(in the sense of being non--bipartite), the kagome lattice provides
an example where geometrical and orbital frustration effects cancel
partially, affording favorable dimerized solutions. Thus, while it
is possible to ascribe some of the frustration effects we have studied
in the triangular lattice to a purely geometrical origin, for more
complex models it is in general necessary to extend the concept of
``geometrical frustration'' beyond that applicable to pure spin
systems.

\section{Discussion and summary}
\label{sec:summa}

We have considered a spin--orbital model representative of a
strongly interacting $3d^1$ electron system with the cubic
structural symmetry of edge--sharing metal--oxygen octahedra,
conditions which lead to a triangular lattice of magnetic
interactions between sites with unbroken, threefold orbital
degeneracy. We have elucidated the qualitative phase diagram,
which turns out to be very rich, in the physical parameter
space presented by the ratio ($\alpha$) of superexchange to
direct--exchange interactions and the Hund exchange ($\eta$).

Despite the strong changes in the fundamental nature of the
model Hamiltonian as a function of $\alpha$ and $\eta$, a number of
generic features persist throughout the phase diagram. With the
exception of the ferromagnetic phases at high $\eta$, which
effectively suppresses quantum spin fluctuations (below), there
is no long--ranged magnetic or orbital order anywhere within the
entire parameter regime. This shows a profound degree of frustration
whose origin lies both in the geometry and in the properties of the
spin--orbital coupling; a qualitative evaluation of these respective
contributions is discussed below.

All of the phases of the model show a strong preference for the
formation of dimers. This can be demonstrated in a simple, static
valence--bond (VB) ansatz, and is reinforced by the results of numerical
calculations. The static ansatz is already an exact description of
the direct--exchange limit, $\alpha = 1$, and gives the best analytic
framework for understanding the properties of much of the remainder
of the phase diagram. The most striking single numerical result is
the prevalence of VB states even on a triangular cluster, and the
underlying feature reinforced by all of the calculations is the
very large additional ``kinetic'' contribution to the ground--state
energy arising from the resonance of VBs due to quantum fluctuations.
It is this resonance which drives symmetry restoration in some or all
of the spin, orbital, and translational sectors over large regions of
the parameter space. The sole exception to dimerization is found at
high $\eta$ and around $\alpha = 1$, where the only mechanism for
virtual hopping is the adoption of orbital configurations which
permit one orbital to be active (``avoided blocking'').

The ``most exotic'' region of the phase diagram is that at small
$\alpha$ and $\eta$, and this we have assigned tentatively as an
orbital liquid. In this regime, quantum fluctuations are at their
strongest and most symmetrical, and every indication obtained from
energetic considerations of extended systems, and from microscopic
calculations of a range of local quantities on small clusters,
suggests a highly resonant, symmetry--restored phase. While this
orbital liquid is in all probability (again from the same indicators)
based on resonating dimers, an issue we discuss in full below, we
cannot exclude fully the possibility of a type of one--dimensional
physics: short, fluctuating segments of frustration--decoupled spin
or orbital chains, whose character persists despite the high site
coordination. It should be stressed here that the point $(\alpha,
\eta)$ = (0,0) is not in any sense a parent phase for exotic
states in the rest of the phase diagram: mixed and direct exchange
processes are qualitatively different elements, which introduce
different classes of frustrated model at finite $\alpha$. While
the matter is somewhat semantic, we comment only that one cannot
argue for the point $\alpha = 0.5$ being ``more exotic'' than
$\alpha = 0$ despite having the maximal number of equally weighted
hopping channels, because it does not possess any additional
symmetries which mandate qualitative changes to the general picture.
In this sense, the limit $\alpha = 1$ serves as a valuable fixed
point which is understood completely, and yet is still dominated
by the purely quantum mechanical concept of singlet formation.

One indicator which can be employed to quantify ``how exotic'' a phase
may be is the entanglement of spin and orbital degrees of freedom. We
define entanglement as the deviation of the spin and orbital sectors
from the factorized limit in which their fluctuations can be treated
separately. We compute a spin--orbital correlation function and use
it to measure entanglement, finding that this is significant over the
whole phase diagram. Qualitatively, entanglement is maximal around the
superexchange limit, which is dominated by dimers where singlet formation
forces the other sector to adopt a local triplet state. However, for
particular clusters and dimer configurations, the high symmetry may
allow less entangled possibilities to intervene exactly at $\alpha = 0$.
The direct--exchange limit, $\alpha = 1$, provides additional insight
into the entanglement definition: the four--operator spin--orbital
correlation function vanishes, reflecting the clear decoupling of the
two sets of degrees of freedom at this point, but the finite product
of separate spin and orbital correlation functions violates the
factorizability condition.

This preponderance of evidence for quantum states based on robust,
strongly resonating dimers implies further that the (spin and orbital)
liquid phase is gapped. Such a state would have only short--ranged
correlation functions. However, these gapped states are part of a
low--energy manifold, and for the extended system we have shown that
this consists quite generally of large numbers of (nearly) degenerate
states. The availability of arbitrary dimer rearrangements at no energy
cost has been suggested to be sufficient for the deconfinement of
elementary $S = 1/2$ (and by analogy $T = 1/2$) excitations with
fractional statistics.\cite{rdmnm} However, the spinons (orbitons)
are massive in such a model, in contrast to the properties of
algebraic liquid phases.\cite{rhsf}

A low--spin to high--spin transition, occurring as a function of
$\eta$, is present for all values of $\alpha$. The quantitative
estimation of $\eta_c$ in the extended system remains a problem
for a more sophisticated analysis. At the qualitative level, large
$\eta$ can be considered to suppress quantum spin fluctuations by
promoting parallel--spin (ferromagnetic) intermediate states on
the magnetic ions. However, even when this sector is quenched,
the orbital degrees of freedom remain frustrated, and contain
non--trivial problems in orbital dynamics. In the superexchange
(low--$\alpha$, high--$\eta$) region, frustration is resolved by
the formation of orbital singlet (spin triplet) dimers, whose
resonance minimizes the ground--state energy. The frustration in
the direct--exchange (high--$\alpha$, high--$\eta$) region is
resolved by avoided--blocking orbital configurations, and
order--by--disorder effects are responsible for the selection
of the true ground state from a degenerate manifold of possibilities;
this is the only part of the phase diagram not displaying dimer physics.
Thus the ferromagnetic orbital models in both limits exhibit a behavior
quite different from that of systems with only $S = 1/2$ spin degrees
of freedom on the triangular lattice.

We have commented on both geometry and spin--orbital interactions as
the origin of frustration in the models under consideration. However,
a statement such as ``on the triangular lattice, geometrical frustration
enhances interaction frustration for spin--orbital models'' must be
qualified carefully. We have obtained anecdotal evidence concerning
such an assertion in Sec.~VI by considering other lattice geometries,
and find that indeed the same model on an unfrustrated geometry appears
capable of supporting ordered states; however, the interplay of the two
effects is far from direct, as the kagome lattice presents a case where
dimer formation acts to reduce the net frustation. Quite generally,
spin--orbital models contain in principle more channels which can be
used for relieving frustration, but the exact nature of the coupling
of spin and orbital sectors may result in the opposite effect. Specific
data characterizing mutual frustration can be obtained from the spin and
orbital correlations computed on small clusters: as shown in Sec.~V, for
the triangular lattice there are indeed regimes where, for example, the
effective orbital interactions enforced by the spin sector make the
orbital sector more frustrated (higher $T_{ij}$) than would be the
analogous pure spin problem (measured by $S_{ij}$), and conversely.

We comment briefly on other approaches which might be employed
to obtain more insight into the states of the extended system, with
a view to establishing more definitively the nature and properties of
the candidate orbital liquid phase. More advanced numerical techniques
could be used to analyze larger unit cells, but while Lanczos
diagonalization, contractor renormalization\cite{rclm} or other
truncation schemes might afford access to systems two, or even four,
times larger, it seems unlikely that these clusters could provide the
qualitatively different type of data required to resolve the questions
left outstanding in Sec.~V. An alternative, but still non--perturbative
and predominantly unbiased, approach would be the use of variational
wave functions, either formulated generally or in the more specific
projected wave function technique which leads to different types of
flux phase.\cite{rh,rrhlw} Adapting this type of treatment to the
coupled spin and orbital sectors without undue approximation remains
a technical challenge.

Within the realm of effective models which could be obtained by
simplification of the ground--state manifold, we cite only the
possibility motivated by the current results of constructing dimer
models based on (ss/ot) and (os/st) dimers. Dimer models\cite{Rok88}
are in general highly simplified, and there is no systematic
procedure for their derivation from a realistic Hamiltonian, but
they are thought to capture the essential physics of certain classes
of dimerized systems. Because QDM Hamiltonians provide exact solutions,
and in some cases genuine examples of exotica long sought in spin systems,
including the RVB phase and deconfined spinon excitations, they represent
a valuable intermediate step in understanding how such phenomena may
emerge in real systems. Here we have found (i) a very strong tendency
to dimer formation, (ii) a large semi--classical degeneracy of basis
states formed from these dimers, and (iii) that resonance processes
even at the four--site plaquette scale provide a very significant
energetic contribution. From the final observation alone, a minimal
QDM, meaning only exchange of parallel dimers of all three directions
and on all possible plaquette units, would already be expected to
contain the most significant corrections to the VB energy. At this
point we emphasize that, because of the change of SU(2) orbital
sector with lattice direction, our 2D models are not close to the
SU(4) point where four--site plaquette formation, and hence very
probably a crystallization, would be expected.\cite{rpmfm} From the
results of Secs.~IV and V, a rather more likely phase of the QDM
would be one with complete plaquette resonance through all three
colors, and without breaking of translational symmetry.

Rigorous proof of a liquid phase, such as that represented by an RVB
state, is more complex, and as noted in Sec.~I it requires satisfying
both energetic and topological criteria. Following the prescription
in Ref.~\onlinecite{Mil07}, three conditions must be obeyed: (i) a
propensity for dimer formation, (ii) a highly degenerate manifold of
basis states from which the RVB ground state may be constructed, and
(iii) a mapping of the system to a liquid phase of a QDM. Criteria
(i) and (ii) match closely the labels in the previous paragraph, and
both dimer formation and high degeneracy have been demonstrated
extensively here. The energetic part of criterion (iii) also appears
to be obeyed here: static dimers have an energy ($V$), and allowing
their location and orientation to change gains more ($t$). The regime
$V/t < 1$ of the triangular--lattice QDM is the RVB phase demonstrated
in Ref.~\onlinecite{Moe01}, whose properties include short--range
correlation functions and gapped, deconfined spinons. This mapping
also contains the criterion of togological degeneracy, and could in
principle be partially circumvented by a direct demonstration. However,
no suitable numerical studies are available of non--simply connected
systems, and so here we can present only plausibility arguments based
on the high degeneracy and spatial topology of the dimer systems
analyzed in Secs.~IV and V. It is safe to conclude that the
threefold--degenerate $t_{2g}$ orbital system on the triangular
lattice is one of best candidates yet for a true spin--orbital RVB
phase.

In closing, spin--orbital models have become a frontier of intense
current interest for both experimental and theoretical studies of
novel magnetic and electronic states emerging as a consequence of
intrinsic frustration. Our model has close parallels to, and yet
crucial differences from, similar studies of manganites (cubic
systems of $e_g$ orbitals), LiNiO$_2$ (triangular, $e_g$),
YTiO$_3$ and CaVO$_3$ (cubic, $t_{2g}$), and many other
transition--metal oxides, appearing in some respects to be the
most frustrated yet discussed. One of its key properties, arising
from the extreme (geometrical and interaction--driven) frustration,
is that ordered states become entirely uncompetitive compared to
the resonance energy gained by maximizing quantum (spin and orbital)
fluctuations. In the orbital sector, the restoration of symmetry by
orbital fluctuations makes the model a strong candidate to display
an orbital liquid phase. Because this liquid is based on robust
dimer states, the mechanism for its formation is very likely to
be spin--orbital RVB physics.

\acknowledgments

We thank G. Khaliullin and K. Penc for helpful discussions, and J.
Chaloupka for technical assistance. A.~M.~Ole\'s acknowledges
support by the Foundation for Polish Science (FNP) and by the
Polish Ministry of Science and Education under Project No.~N202
068 32/1481.

\end{document}